\documentclass[preprint,aps]{revtex4}
\usepackage{graphicx,slashed,color}
\begin{document}
%\preprint{}
\draft
\title{Electroweak Chiral Lagrangian for $W'$ Boson}

\author{Shun-Zhi Wang$^1$,  Shao-Zhou Jiang$^2$, Feng-Jun Ge$^3$, Qing Wang$^4$}

\address{Department of Physics, Tsinghua University, Beijing 100084,
P.R.China$^1$\footnote{wsz04@mails.tsinghua.edu.cn}
$^2$\footnote{jsz@mails.tsinghua.edu.cn}
$^3$\footnote{gfj05@mails.tsinghua.edu.cn}
$^4$\footnote{wangq@mail.tsinghua.edu.cn}\\
    Center for High Energy Physics, Tsinghua University, Beijing 100084, P.R.China$^4$}

\date{May 3, 2008}

\begin{abstract}
 The complete list of electroweak chiral Lagrangian for $W'$, $Z'$
  and a neutral light higgs with symmetry $SU(2)_1\otimes SU(2)_2\otimes U(1)$ is provided.
The bosonic part is accurate up to order of $p^4$, the matter part
involving various fermions representation arrangements includes
dimension three Yukawa type and dimension four gauge type operators.
The universal mixings and masses of gauge boson and fermion are
given. Constraints from mass differences for $K^0-\bar{K}^0$,
$B_d^0-\bar{B}_{d}^0$, $B_s^0-\bar{B}_{s}^0$ systems and indirect CP
violation parameter $|\epsilon_K|$ for $K$ mesons are evaluated.

\bigskip
PACS number(s): 11.10.Ef; 11.10.Lm; 12.15.-y; 12.15.Mm; 12.39.Fe;
12.60.Cn
\end{abstract}
\maketitle \tableofcontents
%%%%%%%%%%%%%%%%%%%%%%%%%%%%%%%%%%%%%%%%%%%%%%
\section{Introduction}

Any electrically charged gauge boson outside of the Standard Model
(SM) is generically denoted $W'$. It is a hypothetical massive
particle of electric charge $\pm 1$ and spin $1$ which always
couples to two different flavors of quarks and (or) leptons, similar
to the $W$ boson (We do not discuss the situation that $W'$ as a
leptoquark gauge boson couples quarks to leptons). $W'$ can be seen
as minimal charged gauge boson extension for SM and is predicted in
various new physics models , such as Left-Right symmetric
models\cite{LRSM1,LRSM2}, Alternate Left-Right
model\cite{AlternateLRSM}, Ununified standard
model\cite{UnunifiedSM}, Non-Commuting Extended
Technicolor\cite{Non-CommutingExtendedTechnicolor}, Little Higgs
models\cite{LH1,LH2,LH3}, models of composite gauge
bosons\cite{Composite4}, Super-symmetric top-flavor
models\cite{SSTopflavor}, Grand Unification\cite{GUT} and
Superstring theories\cite{Sstring1,Sstring2,Sstring3},
Extra-dimensions\cite{ED1,ED2}. Theoretically, unitarity
considerations imply that charged massive vector $W'$s are gauge
bosons associated with some spontaneously broken  non-abelian gauge
symmetry \cite{Unitarity0}. This is true even when it is a composite
particle like the charged techni-$\rho$ in technicolor
theories\cite{TCrho} or a Kaluza-Klein mode in theories where the
$W$ boson propagates in extra dimensions\cite{ED3}. The minimal rank
one non-abelian gauge group is $SU(2)$. Besides $W^{\prime\pm}$, the
group $SU(2)$ demands the existence of extra neutral gauge boson
$Z'$. $W^{\prime\pm}$ and $Z'$ together form a consistent minimal
non-abelian $SU(2)$ gauge group. This gauge group must be completely
spontaneously broken to give $W^{\prime\pm},Z'$ masses through Higgs
mechanism. The breaking mechanism is not known yet which depends on
detail of the model. We can exploit nonlinear realization of the
symmetry to avoid touching upon the details of the breaking
mechanism. This is the $SU(2)$ chiral Lagrangian for
$W^{\prime\pm},Z'$ and three corresponding Goldstone bosons.

Now the new generation hadron collider LHC is going to run and
people are eager expecting the discovery of the new particles. Once
the first new particle shows its signature in the collider
experiment and its spin and parity are evaluated out, the following
work is to check whether it belongs to any of exiting models. In
general, for each kind of possible new particle, there are many
candidate models predicting it and waiting for experiment to check.
It is also possible that the real model our nature chosen is not
presented in this candidate's list. To examine which kind of model
this new particle belongs to and its interactions with those already
discovered particles, we need a phenomenological theory which must
be such general as to include various underlying discovered and
undiscovered candidate models and cover all of its possible
phenomenologies. We call this phenomenological theory the
electroweak chiral Lagrangian (EWCL) for the new particle which
include this new particle and all those already discovered
particles. The symmetry realization of this EWCL should at least
include $SU(2)_L\otimes U(1)_Y$ plus some new part from the new
particle. On the platform of this EWCL, on the one hand, we can
perform model independent phenomenological investigation of the new
particle and fix the corresponding parameters in EWCL from
experiments, on the other hand, we can compute the parameters of
EWCL from concrete underlying models. Through comparison between
parameters from experiments and that from underlying model, we hope
the correct underlying model can be figured out.

In this paper, we are interested in a situation that except
discovered particles in SM, the lowest new particles which are
expected to show up in upcoming collider experiments are
$W^{\prime\pm}$ and $Z'$. According to discussions above, to
describe the corresponding physics phenomenologically, we are lead
to set up a EWCL for $W^{\prime\pm},Z'$ and the symmetry realization
of the theory will be generalized from original  $SU(2)_L\otimes
U(1)_Y$ to $SU(2)_1\otimes SU(2)_2\otimes U(1)$ for which one
$SU(2)$ is for $W^{\prime\pm}$ and $Z'$ and remaining ones are for
SM electro-weak gauge bosons $W^\pm,Z, A$. Naive extension of
conventional unitarity analysis shows that this Lagrangian will
violate unitarity in TeV energy region \cite{Unitarity}, and adding
in theory a neutral Higgs with mass below TeV will kill the
disaster. To keep our theory being unitary at TeV energy region, we
will further include in our theory a neutral Higgs. Thus our EWCL
for $W^{\prime\pm},Z'$ now will include those already discovered
particles, a neutral Higgs, $W^{\prime\pm},Z'$ and corresponding
Goldstone bosons. In fact, without $W^{\prime\pm},Z'$ and
corresponding Goldstone bosons, the EWCL only for a neutral Higgs
boson was already written down in Ref.\cite{WangLiMing} which was a
generalization of original standard
EWCL\cite{longhitano80,Appelquist80,Appelquist93} by adding a
singlet Higgs field to the theory. Now our EWCL can be seen as a
further extension of this generalized EWCL to include in theory
$W^{\prime\pm},Z'$ and corresponding Goldstone bosons. In this work,
we are especially interested in the case that the mass of
$W^{\prime\pm}$ is lighter or roughly same as that of $Z'$. Since if
the mass of $Z'$ is much lighter than that of $W^{\prime\pm}$, the
phenomenological interest will be changed to physics for lighter
$Z'$. The heavier $W^{\prime\pm}$ then can be integrated out
theoretically and we are led to EWCL purely for $Z'$ and neutral
Higgs boson. This EWCL was already discussed by us in another
paper\cite{Z'} in which $Z'$ can be either an element of $SU(2)$
triplet or a remnant of some other underlying dynamics which has
nothing to do with $W'$ and can not be covered in our present
theory. It is shown in Ref.\cite{Z'} that EWCL for $Z'$ is
equivalent to an extended Stueckelberg mechanism for $U(1)$ gauge
boson. From the point of view of Stueckelberg mechanism, our present
EWCL for $W'$ and $Z'$ can be further seen as $SU(2)$ non-abelian
generalization of previous extended $U(1)$ abelian Stueckelberg
mechanism. Due to the passive roles of neutral Higgs and $Z'$, in
this work we focus our attentions mainly on $W'$ and related
physics. For physics related to $W'$, the strongest low energy
phenomenological constraints come from $W-W'$ mixing, $K_L-K_S$ mass
differences and related CP violation parameters. On the platform of
our EWCL, we can explore these constraints in detail, transferring
them to the constraints on parameters of our EWCL and CKM matrix
elements for right hand fermions. We will find that some of these
constraints such as mixings among different particles are universal,
while others are model class dependent. It should be emphasized that
our EWCL will only cover those underlying models which include
massive $W^{\prime\pm},Z'$ and neutral Higgs as lowest new particles
beyond those already discovered particles. For those models which
include new particle with mass lighter than $W'$ or new particle
combining with discovered particle together forms an irreducible
representation of $SU(2)$ group\cite{AlternateLRSM}, our EWCL do not
cover the corresponding physics. We argue for this alternative
situation, a separate EWCL can be built to describe it and this
situation will be investigated elsewhere.

Within the range of our EWCL, a special type of models are
left-right symmetric models\cite{LRSM1,LRSM2} which explore the
possibility of spontaneous parity violation. The EWCL for this kind
models is built up by some of us in Ref.\cite{LRSMEWCLboson} for the
bosonic part and Ref.\cite{LRSMEWCLmatter} for the matter part.
Since we are interested in the general description for $W'$ and $Z'$
physics, it is purpose of this paper to generalize the discussion in
Ref.\cite{LRSMEWCLboson, LRSMEWCLmatter} to cover left-right
non-symmetric models.  For bosonic part of EWCL, no matter which
kind of model involving $W'$ and $Z'$, since gauge bosons and
corresponding Goldstone bosons are all in triplet of $SU(2)_2$
group, their interactions then are fixed as those given in
Ref.\cite{LRSMEWCLboson}. While for matter part, various models
provide at least following different arrangements for fermion
representations\cite{FP0}:
\begin{enumerate}
\item Left-right symmetric (LR)\cite{LRSM1,LRSM2}:~Left hand fermions  belong to doublet of $SU(2)_1$ and singlet
of $SU(2)_2$; Right hand fermions belong to doublet of $SU(2)_2$ and
singlet of $SU(2)_1$.
\item Leptophobic (LP):~Left hand fermions belong to doublet of $SU(2)_1$ and singlet
of $SU(2)_2$; Right hand quarks belong to doublet of $SU(2)_2$ and
singlet of $SU(2)_1$; Right hand leptons belong to singlets of both
$SU(2)$'s.
\item Hadrophobic (HP):~Left hand fermions belong to doublet of $SU(2)_1$ and singlet
of $SU(2)_2$; Right hand leptons belong to doublet of $SU(2)_2$ and
singlet of $SU(2)_1$; Right hand quarks belong to singlets of both
$SU(2)$'s.
\item Fermionphobic (FP)\cite{FP0,LH4,FP}:~Left hand fermions belong to doublet of $SU(2)_1$ and singlet
of $SU(2)_2$; Right hand fermions belong to singlets of both
$SU(2)$'s.
\item Ununified (UN)\cite{UnunifiedSM}:~Left hand leptons belong to doublet of $SU(2)_1$ and singlet
of $SU(2)_2$; Left hand quarks  belong to doublet of $SU(2)_2$ and
singlet of $SU(2)_1$; Right hand fermions belong to singlet of
$SU(2)_1\otimes SU(2)_2$.
\item Non-universal (NU)\cite{NU}:~One or two special family left hand fermions (typical situation is the first two light
families) belong to doublet of $SU(2)_1$ and singlet of $SU(2)_2$;
Remaining left hand fermions belong to doublet of $SU(2)_2$ and
singlet of $SU(2)_1$; Right hand fermions belong to singlet of
$SU(2)_1\otimes SU(2)_2$.
\end{enumerate}
The matter part EWCL given in Ref.\cite{LRSMEWCLmatter} only
involves situation 1 in which although the arrangement of fermion
representations is left-right symmetric, the couplings may or may
not be left-right symmetric. Considering the fact that conventional
EWCL formalism only deals with the system with particles fixed in
some special group representations, the generalization of the
expression to cover different fermion representation arrangements is
not a trivial work.

This paper is organized as follows: Sec.II is the introduction of a
 our EWCL which covers all above situations.
  For the bosonic part we accurate up to order
of $p^4$. For matter part, we limit us in dimension three Yukawa
type and dimension four gauge interaction terms. In Sec.III, we
discuss mixings among $W-W'$ and $A-Z-Z'$ and introduce CKM matrix
to diagonalize fermion mass matrix. Goldstone boson, Higgs boson and
gauge boson couplings to quarks are given in Sec.IV. We build up
effective Hamiltonian for ming of neutral $K$ and $B$ systems in
Sec.V. In Sec.VI, we discuss the constraints on our EWCL for LR and
LP models from mass differences in $K^0-\overline{K}^0$,
$B_d^0-\bar{B}_d^0$, $B_s^0-\bar{B}_s^0$ systems and indirect CP
violation parameter $\epsilon_K$. Sec.VI is the summary.

%%%%%%%%%%%%%%%%%%%%%%%%%%%%%%%%%%%%%%%%%%%%%%%%%%%%%%%%
\section{EWCL in gauge eigenstates}

We first introduce the bosonic part of EWCL which basically is the
same as that for left-right symmetric models given in
Ref.\cite{LRSMEWCLboson}. Let $B_{\mu}$, $W_{1,\mu}^a$,
$W_{2,\mu}^a$ be electroweak gauge fields ($a=1,2,3$) and two by two
unitary unimodular matrices $U_1$ and $U_2$ be corresponding
goldstone boson fields, $h$ be neutral Higgs field which is singlet
of $SU(2)_{\rm 1}\otimes SU(2)_{\rm 2}\otimes U(1)$ group. Consider
covariant derivatives for goldstone fields
$D_{\mu}U_i=\partial_{\mu}U_i+ig_i\frac{\tau^a}{2}W^a_{i,\mu}U_i-igU_i\frac{\tau_3}{2}B_{\mu}$
 and building blocks $X_i^\mu\equiv U_i^\dag(D^\mu U_i)$,
$\overline{W}_{i,\mu\nu}\equiv U_i^\dag g_iW_{i,\mu\nu}U_i$ for
$i=1,2$. The lowest order of chiral Lagrangian is the Higgs
potential $\mathcal{L}_0=-V(h)$ and $p^2$ order of Lagrangian is
\begin{eqnarray}
\mathcal{L}_2&=&\frac{1}{2}(\partial_{\mu}h)^2-\frac{1}{4}f_1^2{\rm
tr}(X_{1,\mu}X_1^{\mu})-\frac{1}{4}f_2^2{\rm
tr}(X_{2,\mu}X_2^{\mu})+\frac{1}{2}\kappa f_1f_2{\rm tr}(X_1^{\mu}
X_2^{\mu})\\
&&+\frac{1}{4}\beta_{1,1}f_1^2[{\rm
tr}(\tau^3X_{1,\mu})]^2+\frac{1}{4}\beta_{2,1}f_2^2[{\rm
tr}(\tau^3X_{2,\mu})]^2+\frac{1}{2}\tilde{\beta}_1f_1f_2[{\rm
tr}(\tau^3X_{1,\mu})][{\rm tr}(\tau^3X_2^{\mu})]\;.\nonumber
\end{eqnarray}
$p^4$ order Lagrangian can be divided into six parts,
\begin{eqnarray}
\mathcal{L}_4=\mathcal{L}_K+\mathcal{L}_1+\mathcal{L}_{H1}+\mathcal{L}_2+\mathcal{L}_{H2}+\mathcal{L}_C
\end{eqnarray}
with kinetic part of $p^4$ order Lagrangian $\mathcal{L}_K$
\begin{eqnarray}
\mathcal{L}_K=-\frac{1}{4}W_{1,\mu\nu}^aW_1^{\mu\nu,a}-\frac{1}{4}W_{2,\mu\nu}^aW_2^{\mu\nu,a}
-\frac{1}{4}B_{\mu\nu}B^{\mu\nu}
\label{LK}
\end{eqnarray}
$\mathcal{L}_i,~i=1,2$ are terms of $p^4$ order Lagrangian which
involve the gauge bosons of first(second) interaction group
$SU(2)_1(SU(2)_2)$ without differential of higgs
\begin{eqnarray}
\mathcal{L}_i&=&\frac{1}{2}\alpha_{i,1}gB_{\mu\nu}{\rm
tr}(\tau^3\overline{W}_i^{\mu\nu}) +i\alpha_{i,2}gB_{\mu\nu}{\rm
tr}(\tau^3X^{\mu}_iX^{\nu}_i)
 +2i\alpha_{i,3}{\rm tr}(\overline{W}_{i,\mu\nu}X^{\mu}_iX^{\nu}_i)
 +\alpha_{i,4}[{\rm tr}(X_{i,\mu}X_{i,\nu})]^2
\nonumber\\
 && +\alpha_{i,5}[{\rm tr}(X_{i,\mu}^2)]^2
 +\alpha_{i,6}{\rm tr}(X_{i,\mu}X_{i,\nu}){\rm tr}(\tau^3X^{\mu}_i){\rm tr}(\tau^3X^{\nu}_i)
 +\alpha_{i,7}{\rm tr}(X_{i,\mu}^2)[{\rm tr}(\tau^3X_{i,\nu})]^2
 \nonumber\\
 &&+\frac{1}{4}\alpha_{i,8}[{\rm tr}(\tau^3\overline{W}_{i,\mu\nu})]^2
 +i\alpha_{i,9}{\rm tr}(\tau^3\overline{W}_{i,\mu\nu}){\rm tr}(\tau^3X_i^{\mu}X_i^{\nu})
 +\frac{1}{2}\alpha_{i,10}[{\rm tr}(\tau^3X_{i,\mu}){\rm tr}(\tau^3X_{i,\nu})]^2
 \nonumber\\
 &&+
 \alpha_{i,11}\epsilon^{\mu\nu\rho\lambda}{\rm tr}(\tau^3X_{i,\mu}){\rm tr}(X_{i,\nu}\overline{W}_{i,\rho\lambda})
 +2\alpha_{i,12}{\rm tr}(\tau^3X_{i,\mu}){\rm tr}(X_{i,\nu}\overline{W}_i^{\mu\nu})
 \nonumber\\
&&+\frac{1}{4}\alpha_{i,13}g\epsilon^{\mu\nu\rho\sigma}\!B_{\mu\nu}{\rm
tr}(\tau^3\overline{W}_{i,\rho\sigma}\!)
+\frac{1}{8}\alpha_{i,14}\epsilon^{\mu\nu\rho\sigma}{\rm
tr}(\tau^3\overline{W}_{i,\mu\nu}){\rm
tr}(\tau^3\overline{W}_{i,\rho\sigma})\;.
\end{eqnarray}
$\mathcal{L}_{Hi},~i=1,2$ are first(second) interaction group part
of $p^4$ order Lagrangian with differential of higgs
\begin{eqnarray}
\mathcal{L}_{Hi}&=&(\partial_{\mu}h)\{\bar{\alpha}_{Hi,1}\mathrm{tr}(\tau^3X_i^\mu)\mathrm{tr}(X_{i,\nu}^2)
+\bar{\alpha}_{Hi,2}\mathrm{tr}(\tau^3X_i^\nu)\mathrm{tr}(X_i^{\mu}X_{i,\nu})+
\bar{\alpha}_{Hi,3}\mathrm{tr}(\tau^3X_i^\nu)\mathrm{tr}(\tau^3X_i^{\mu}X_{i,\nu})\nonumber\\
&&+\bar{\alpha}_{Hi,4}\mathrm{tr}(\tau^3X_i^{\mu})[\mathrm{tr}(\tau^3X_{i,\nu})]^2
+i\bar{\alpha}_{Hi,5}\mathrm{tr}(\tau^3X_{i,\nu})\mathrm{tr}(\tau^3\overline{W}_i^{\mu\nu})
+
ig\bar{\alpha}_{Hi,6}B^{\mu\nu}\mathrm{tr}(\tau^3X_{i,\nu})\nonumber\\
&&+i\bar{\alpha}_{Hi,7}\mathrm{tr}(\tau^3\overline{W}_i^{\mu\nu}X_{i,\nu})
+i\bar{\alpha}_{Hi,8}\mathrm{tr}(\overline{W}_i^{\mu\nu}X_{i,\nu})\}
+(\partial_{\mu}h)(\partial_{\nu}h)[\bar{\alpha}_{Hi,9}\mathrm{tr}(\tau^3X_i^\mu)\mathrm{tr}(\tau^3X_i^\nu)\nonumber\\
&&+ \bar{\alpha}_{Hi,10}\mathrm{tr}(X_i^{\mu}X_i^\nu)]
+(\partial_{\mu}h)^2\{\bar{\alpha}_{Hi,11}[\mathrm{tr}(\tau^3X_{i,\nu})]^2
+ \bar{\alpha}_{Hi,12}\mathrm{tr}(X_{i,\nu}^2)\}\nonumber\\
&&+\bar{\alpha}_{Hi,13}(\partial_{\mu}h)^2(\partial_{\nu}h)\mathrm{tr}(\tau^3X_i^\nu)+\bar{\alpha}_{Hi,14}(\partial_{\mu}h)^4
\end{eqnarray}
The most complex interaction is the crossing part of $p^4$ order
Lagrangian
\begin{eqnarray}
\mathcal{L}_C&=&i\tilde{\alpha}_{2}gB_{\mu\nu}{\rm
tr}(\tau^3X^{\mu}_1X^{\nu}_2)
 +2i\tilde{\alpha}_{3,1}{\rm tr}(\overline{W}_{1,\mu\nu}X^{\mu}_2X^{\nu}_2)
+2i\tilde{\alpha}_{3,2}{\rm
tr}(\overline{W}_{2,\mu\nu}X^{\mu}_1X^{\nu}_1)
 \nonumber\\
&& +2i\tilde{\alpha}_{3,3}{\rm
 tr}(\overline{W}_{1,\mu\nu}X^{\mu}_1X^{\nu}_2)+2i\tilde{\alpha}_{3,4}{\rm
tr}(\overline{W}_{2,\mu\nu}X^{\mu}_2X^{\nu}_1)
 +\tilde{\alpha}_{4,1}{\rm tr}(X_{1,\mu}X_{1,\nu}){\rm tr}(X_2^{\mu}X_2^{\nu})
\nonumber\\
&&+\tilde{\alpha}_{4,2}[{\rm
tr}(X_{1,\mu}X_{2,\nu})]^2+\tilde{\alpha}_{4,3}{\rm
tr}(X_{1,\mu}X_{2,\nu}){\rm
tr}(X_2^{\mu}X_1^{\nu})+\tilde{\alpha}_{4,4}{\rm
tr}(X_{1,\mu}X_{2,\nu}){\rm
tr}(X_2^{\mu}X_2^{\nu})\nonumber\\
 &&+\tilde{\alpha}_{4,5}{\rm
tr}(X_{2,\mu}X_{1,\nu}){\rm tr}(X_1^{\mu}X_1^{\nu})
+\tilde{\alpha}_{5,1}{\rm tr}(X_{1,\mu}^2){\rm tr}(X_{2,\nu}^2)
+\tilde{\alpha}_{5,2}[{\rm
tr}(X_{1,\mu}X_2^{\mu})]^2 \nonumber\\
&&+\tilde{\alpha}_{5,3}{\rm tr}(X_{1,\mu}X_2^{\mu}){\rm
tr}(X_{2,\nu}^2) +\tilde{\alpha}_{5,4}{\rm
tr}(X_{2,\mu}X_1^{\mu}){\rm
tr}(X_{1,\nu}^2)\nonumber\\
&&+\tilde{\alpha}_{6,1}{\rm tr}(X_{1,\mu}X_{1,\nu}){\rm
tr}(\tau^3X^{\mu}_2){\rm
tr}(\tau^3X^{\nu}_2)+\tilde{\alpha}_{6,2}{\rm
tr}(X_{2,\mu}X_{2,\nu}){\rm tr}(\tau^3X^{\mu}_1){\rm
tr}(\tau^3X^{\nu}_1)
\nonumber\\
&&+\tilde{\alpha}_{6,3}{\rm tr}(X_{1,\mu}X_{2,\nu}){\rm
tr}(\tau^3X^{\mu}_1){\rm
tr}(\tau^3X^{\nu}_2)+\tilde{\alpha}_{6,4}{\rm
tr}(X_{1,\mu}X_{2,\nu}){\rm tr}(\tau^3X^{\mu}_2){\rm
tr}(\tau^3X^{\nu}_1)\nonumber\\
&&+\tilde{\alpha}_{6,5}{\rm tr}(X_{1,\mu}X_{2,\nu}){\rm
tr}(\tau^3X^{\mu}_2){\rm
tr}(\tau^3X^{\nu}_2)+\tilde{\alpha}_{6,6}{\rm
tr}(X_{2,\mu}X_{1,\nu}){\rm tr}(\tau^3X^{\mu}_1){\rm
tr}(\tau^3X^{\nu}_1)\nonumber\\
&&+\tilde{\alpha}_{6,7}{\rm tr}(X_{1,\mu}X_{1,\nu}){\rm
tr}(\tau^3X^{\mu}_1){\rm
tr}(\tau^3X^{\nu}_2)+\tilde{\alpha}_{6,8}{\rm
tr}(X_{2,\mu}X_{2,\nu}){\rm tr}(\tau^3X^{\mu}_2){\rm
tr}(\tau^3X^{\nu}_1)
 \nonumber\\
&&+\tilde{\alpha}_{7,1}{\rm tr}(X_{1,\mu}^2)[{\rm
tr}(\tau^3X_{2,\nu})]^2+\tilde{\alpha}_{7,2}{\rm
tr}(X_{2,\mu}^2)[{\rm tr}(\tau^3X_{1,\nu})]^2\nonumber\\
&&+\tilde{\alpha}_{7,3}{\rm tr}(X_{1,\mu}X_2^{\mu}){\rm
tr}(\tau^3X_{1,\nu}){\rm
tr}(\tau^3X_2^{\nu})+\tilde{\alpha}_{7,4}{\rm
tr}(X_{1,\mu}X_2^{\mu})[{\rm tr}(\tau^3X_{2,\nu})]^2\nonumber\\
&& +\tilde{\alpha}_{7,5}{\rm tr}(X_{2,\mu}X_1^{\mu})[{\rm
tr}(\tau^3X_{1,\nu})]^2+\tilde{\alpha}_{7,6}{\rm
tr}(X_{1,\mu}^2){\rm tr}(\tau^3X_{1,\nu}){\rm
tr}(\tau^3X_2^{\nu}) \nonumber\\
 &&+\tilde{\alpha}_{7,7}{\rm
tr}(X_{2,\mu}^2){\rm tr}(\tau^3X_{2,\nu}){\rm
tr}(\tau^3X_1^{\nu})+\frac{1}{4}\tilde{\alpha}_8{\rm
tr}(\tau^3\overline{W}_{1,\mu\nu}){\rm
tr}(\tau^3\overline{W}_2^{\mu\nu})\nonumber\\ &&
+i\tilde{\alpha}_{9,1}{\rm tr}(\tau^3\overline{W}_{1,\mu\nu}){\rm
tr}(\tau^3X_2^{\mu}X_2^{\nu}) +i\tilde{\alpha}_{9,2}{\rm
tr}(\tau^3\overline{W}_{2,\mu\nu}){\rm
tr}(\tau^3X_L^{\mu}X_L^{\nu})\nonumber\\
&& +i\tilde{\alpha}_{9,3}{\rm tr}(\tau^3\overline{W}_{1,\mu\nu}){\rm
tr}(\tau^3X_1^{\mu}X_2^{\nu}) +i\tilde{\alpha}_{9,4}{\rm
tr}(\tau^3\overline{W}_{2,\mu\nu}){\rm tr}(\tau^3X_2^{\mu}X_1^{\nu})
 \nonumber\\
&& +\frac{1}{2}\tilde{\alpha}_{10,1}[{\rm tr}(\tau^3X_{1,\mu}){\rm
tr}(\tau^3X_{2,\nu})]^2+\frac{1}{2}[\tilde{\alpha}_{10,2}[{\rm
tr}(\tau^3X_{1,\mu}){\rm tr}(\tau^3X_2^{\mu})]^2
\nonumber\\
&&+\frac{1}{2}\tilde{\alpha}_{10,3}{\rm tr}(\tau^3X_{1,\mu}){\rm
tr}(\tau^3X_2^{\mu})[{\rm
tr}(\tau^3X_{2,\nu})]^2+\frac{1}{2}\tilde{\alpha}_{10,4}{\rm
tr}(\tau^3X_{2,\mu}){\rm
tr}(\tau^3X_1^{\mu})[{\rm tr}(\tau^3X_{1,\nu})]^2\nonumber\\
&&+ \tilde{\alpha}_{11,1}\epsilon^{\mu\nu\rho\lambda}{\rm
tr}(\tau^3X_{1,\mu}){\rm tr}(X_{2,\nu}\overline{W}_{2,\rho\lambda})
+ \tilde{\alpha}_{11,2}\epsilon^{\mu\nu\rho\lambda}{\rm
tr}(\tau^3X_{2,\mu}){\rm tr}(X_{1,\nu}\overline{W}_{1,\rho\lambda})\nonumber\\
 &&+ \tilde{\alpha}_{11,3}\epsilon^{\mu\nu\rho\lambda}{\rm
tr}(\tau^3X_{1,\mu}){\rm tr}(X_{1,\nu}\overline{W}_{2,\rho\lambda})+
\tilde{\alpha}_{11,4}\epsilon^{\mu\nu\rho\lambda}{\rm
tr}(\tau^3X_{2,\mu}){\rm
tr}(X_{2,\nu}\overline{W}_{1,\rho\lambda})\nonumber\\ &&
+\tilde{\alpha}_{11,5}\epsilon^{\mu\nu\rho\lambda}{\rm
tr}(\tau^3X_{1,\mu}){\rm
tr}(X_{2,\nu}\overline{W}_{1,\rho\lambda})+\tilde{\alpha}_{11,6}\epsilon^{\mu\nu\rho\lambda}{\rm
tr}(\tau^3X_{2,\mu}){\rm tr}(X_{1,\nu}\overline{W}_{2,\rho\lambda})\nonumber\\
&& +2\tilde{\alpha}_{12,1}{\rm tr}(\tau^3X_{1,\mu}){\rm
 tr}(X_{2,\nu}\overline{W}_2^{\mu\nu})+2\tilde{\alpha}_{12,2}{\rm tr}(\tau^3X_{2,\mu}){\rm
tr}(X_{1,\nu}\overline{W}_1^{\mu\nu})\nonumber\\
&& +2\tilde{\alpha}_{12,3}{\rm tr}(\tau^3X_{1,\mu}){\rm
tr}(X_{1,\nu}\overline{W}_2^{\mu\nu})+2\tilde{\alpha}_{12,4}{\rm
tr}(\tau^3X_{2,\mu}){\rm
tr}(X_{2,\nu}\overline{W}_1^{\mu\nu}) \nonumber\\
&&+2\tilde{\alpha}_{12,5}{\rm tr}(\tau^3X_{1,\mu}){\rm
tr}(X_{2,\nu}\overline{W}_1^{\mu\nu})+2\tilde{\alpha}_{12,6}{\rm
tr}(\tau^3X_{2,\mu}){\rm
tr}(X_{1,\nu}\overline{W}_2^{\mu\nu})\nonumber\\
&& +\frac{1}{8}\tilde{\alpha}_{14}\epsilon^{\mu\nu\rho\sigma}{\rm
tr}(\tau^3\overline{W}_{1,\mu\nu}){\rm
tr}(\tau^3\overline{W}_{2,\rho\sigma})
+\tilde{\alpha}_{15}\epsilon^{\mu\nu\rho\sigma}{\rm
tr}(\overline{W}_{1,\mu\nu}\overline{W}_{2,\rho\sigma})\nonumber\\
&&\color{black}+(\partial_{\mu}h)\{\tilde{\alpha}_{H,1,1}\mathrm{tr}(\tau^3X_2^\mu)\mathrm{tr}(X_{1,\nu}^2)
+\tilde{\alpha}_{H,1,2}\mathrm{tr}(\tau^3X_1^\mu)\mathrm{tr}(X_{2,\nu}^2)
\nonumber\\
&&+\tilde{\alpha}_{H,2,1}\mathrm{tr}(\tau^3X_2^\nu)\mathrm{tr}(X_1^{\mu}X_{L,\nu})
+\tilde{\alpha}_{H,2,2}\mathrm{tr}(\tau^3X_1^\nu)\mathrm{tr}(X_2^{\mu}X_{2,\nu})
+\tilde{\alpha}_{H,3,1}\mathrm{tr}(\tau^3X_2^\nu)\mathrm{tr}(\tau^3X_1^{\mu}X_{1,\nu})\nonumber\\
&&+\tilde{\alpha}_{H,3,2}\mathrm{tr}(\tau^3X_1^\nu)\mathrm{tr}(\tau^3X_2^{\mu}X_{2,\nu})
+\tilde{\alpha}_{H,4,1}\mathrm{tr}(\tau^3X_2^{\mu})[\mathrm{tr}(\tau^3X_{1,\nu})]^2\nonumber\\
&&+\tilde{\alpha}_{H,4,2}\mathrm{tr}(\tau^3X_2^{\mu})\mathrm{tr}(\tau^3X_2^{\nu})\mathrm{tr}(\tau^3X_{1,\nu})
+\tilde{\alpha}_{H,4,3}\mathrm{tr}(\tau^3X_1^{\mu})\mathrm{tr}(\tau^3X_2^{\nu})\mathrm{tr}(\tau^3X_{2,\nu})\nonumber\\
&&+\tilde{\alpha}_{H,4,4}\mathrm{tr}(\tau^3X_1^{\mu})\mathrm{tr}(\tau^3X_1^{\nu})\mathrm{tr}(\tau^3X_{2,\nu})
+i\tilde{\alpha}_{H,5,1}\mathrm{tr}(\tau^3X_{2,\nu})\mathrm{tr}(\tau^3\overline{W}_1^{\mu\nu})\nonumber\\
&&+i\tilde{\alpha}_{H,5,2}\mathrm{tr}(\tau^3X_{1,\nu})\mathrm{tr}(\tau^3\overline{W}_2^{\mu\nu})\}
+(\partial_{\mu}h)(\partial_{\nu}h)\tilde{\alpha}_{H,9}\mathrm{tr}(\tau^3X_2^\mu)
\mathrm{tr}(\tau^3X_1^\nu)\nonumber\\
&&+(\partial_{\mu}h)^2\tilde{\alpha}_{H,11}\mathrm{tr}(\tau^3X_{1,\nu})\mathrm{tr}(\tau^3X_2^\nu)]\;.
\end{eqnarray}
Above interaction terms already include all possible $p^4$ order
CP-conserving and CP-violating operators and all $\alpha$
coefficients are functions of higgs field $h$.

Now we come to matter part of EWCL. Except gauge and goldstone
fields introduced in bosonic part EWCL, matter part EWCL further
involves fermions which include SM quarks and leptons (three
generation right hand neutrinos are introduced in our theory, no
other sterile neutrinos are included in). We denote them by left and
right hand quark and lepton doublets $q_{\alpha L,R}$ and $l_{\alpha
L,R}$ with generation index $\alpha$ being summed over the quark and
lepton flavors. The various models defined by the transformation
properties of their fermion contents with respect to the gauge group
are summarized in Table I.

\begin{table}[h]
{\small{\bf TABLE I}. Fermion transformation properties for
different models considered in the text. The numbers in brackets
refer to $SU(2)_1$, $SU(2)_2$ and $U(1)$, respectively. Color
indices are implicit. The right hand neutrinos are not present in
some of original models LP, FP, UN and NU labeled by $-$. Including
them in this work is harmless to these models and their
representation is $(1,1,0)$.}

\vspace*{0.3cm}\begin{tabular}{c | c c c c c c}\hline\hline
Fields$/$Models&LR &LP&HP & FP &UN & NU \\
\hline $q_{\alpha L}\!\!=\!\!\left(\begin{array}{c}u_{\alpha
L}\\d_{\alpha L}\end{array}\right)$
&~~$(2,1,\frac{1}{6})$~~&~~$(2,1,\frac{1}{6})$~~&~~$(2,1,\frac{1}{6})$~~&~~$(2,1,\frac{1}{6})$~~&~~$(1,2,\frac{1}{6})$~~&~~
$(2,1,\frac{1}{6})\delta_{\alpha\alpha_1}\!\!\!+\!(1,2,\frac{1}{6})\delta_{\alpha\alpha_2}$\\
$q_{\alpha R}\!\!=\!\!\left(\begin{array}{c}u_{\alpha R}\\d_{\alpha
R}\end{array}\right)$
&$(1,2,\frac{1}{6})$&$(1,2,\frac{1}{6})$&$\begin{array}{c}(1,1,\frac{2}{3})\\(1,1,-\frac{1}{3})\end{array}$
&$\begin{array}{c}(1,1,\frac{2}{3})\\(1,1,-\frac{1}{3})\end{array}$
&$\begin{array}{c}(1,1,\frac{2}{3})\\(1,1,-\frac{1}{3})\end{array}$
&~~$\begin{array}{c}(1,1,\frac{2}{3})\\(1,1,-\frac{1}{3})\end{array}$\\
$l_{\alpha L}\!\!=\!\!\left(\begin{array}{c}\nu_{\alpha
L}\\e^-_{\alpha L}\end{array}\right)$
&$(2,1,-\frac{1}{2})$&$(2,1,-\frac{1}{2})$&$(2,1,-\frac{1}{2})$&$(2,1,-\frac{1}{2})$&
$(2,1,-\frac{1}{2})$&
$(2,1,-\frac{1}{2})\delta_{\alpha\alpha_1}\!\!\!+\!(1,2,-\frac{1}{2})\delta_{\alpha\alpha_2}$\\
$l_{\alpha R}\!\!=\!\!\left(\begin{array}{c}\nu_{\alpha
R}\\e^-_{\alpha R}\end{array}\right)$
&$(1,2,-\frac{1}{2})$&$\begin{array}{c}-\\(1,1,-1)\end{array}$&$(1,2,-\frac{1}{2})$&$\begin{array}{c}-\\
(1,1,-1)\end{array}$& $\begin{array}{c}-\\(1,1,-1)\end{array}$&
$\begin{array}{c}-\\(1,1,-1)\end{array}$\\
 \hline\hline
\end{tabular}
\end{table}

 Since above fermions can belong to different representations
for different underlying models,  an universal expression to cover
all these possible arrangements is needed. To reach this aim, we
introduce two goldstone operators $\hat{U}_L$ and $\hat{U}_R$ by
defining their arbitrary function
$f(\hat{U}_R,\hat{U}_L,D_\mu\hat{U}_R,D_\mu\hat{U}_L)$ action on
fermion field as
\begin{eqnarray}
&&\hspace{-1cm}f(\hat{U}_R,\hat{U}_L,D_\mu\hat{U}_R,D_\mu\hat{U}_L)q_{\alpha}=\left\{\begin{array}{ll}
f(U_2,U_1,D_{\mu}U_2,D_{\mu}U_1)q_{\alpha}&\mbox{LR}\\
f(U_2,U_1,D_{\mu}U_2,D_{\mu}U_1)q_{\alpha}&\mbox{LP}\\
f(1,U_1,0,D_{\mu}U_1)q_{\alpha}&\mbox{HP}\\
f(1,U_1,0,D_{\mu}U_1)q_{\alpha}&\mbox{FP}\\
f(1,U_2,0,D_{\mu}U_2)q_{\alpha}&\mbox{UN}\\
f(1,U_1,0,D_{\mu}U_1)q_{\alpha}\delta_{\alpha\alpha_1}+f(1,U_2,0,D_{\mu}U_2)q_{\alpha}\delta_{\alpha\alpha_2}&
\mbox{NU}
\end{array}\right.\nonumber\\
&&\\
&&\hspace{-1cm}f(\hat{U}_R,\hat{U}_L,D_\mu\hat{U}_R,D_\mu\hat{U}_L)l_{\alpha}=\left\{\begin{array}{ll}
f(U_2,U_1,D_{\mu}U_2,D_{\mu}U_1)l_{\alpha}&\mbox{LR}\\
f(1,U_1,0,D_{\mu}U_1)l_{\alpha}&\mbox{LP}\\
f(U_2,U_1,D_{\mu}U_2,D_{\mu}U_1)l_{\alpha}&\mbox{HP}\\
f(1,U_1,0,D_{\mu}U_1)l_{\alpha}&\mbox{FP}\\
f(1,U_1,0,D_{\mu}U_1)l_{\alpha}&\mbox{UN}\\
f(1,U_1,0,D_{\mu}U_1)l_{\alpha}\delta_{\alpha\alpha_1}+f(1,U_2,0,D_{\mu}U_2)l_{\alpha}\delta_{\alpha\alpha_2}&
\mbox{NU}
\end{array}\right.\nonumber
\end{eqnarray}
where in the case of "Non-universality generation", $\alpha_1$
denote the specified generation (typically first two generations)
which acts as doublet of $SU(2)_1$ and singlet of $SU(2)_2$;
$\alpha_2$ denote the remaining generation which acts as doublet of
$SU(2)_2$ and singlet of $SU(2)_1$.

With help of above representations, we now can write down the
universal dimension three Yukawa type interactions. For lepton part,
\begin{eqnarray}
\mathcal{L}_{\mathrm{Y,lepton}}\!\!=\overline{l}^I_{\alpha
L}[\hat{U}_L(y^{\alpha\beta}\!+\!y_3^{\alpha\beta}\tau^3)\hat{U}_R^\dagger]l^I_{\beta
R}+\frac{1}{2}[h_L^{\alpha\beta}\overline{l^{Ic}_{\alpha
L}}\hat{U}_L^*(1\!+\!\tau^3)\hat{U}_L^{\dagger}l^I_{\beta
L}+(L\rightarrow R)]+\mathrm{h.c.}~\label{LYldef}
\end{eqnarray}
where $h_{L,R}^{\alpha\beta}$ are hermitian functions of Higgs field
$h$. $l^c=C\overline{l}^T$ is charge conjugate field of $l$ with $C$
being the charge conjugation matrix. Symbol "$^I$" indicates that
they are gauge eigenstates. For quark part,
\begin{eqnarray}
&&\hspace{-1.5cm}\mathcal{L}_{\mathrm{Y,quark}}=\overline{q}^I_{\alpha
L}[\hat{U}_L(\tau^uy_u^{\alpha\beta}+\tau^dy_d^{\alpha\beta})\hat{U}_R^\dagger]q^I_{\beta
R} +\mathrm{h.c.}\label{LYqdef}
\end{eqnarray}
where $\tau^u=\frac{1+\tau^3}{2}$ and $\tau^d=\frac{1-\tau^3}{2}$.
 Coefficients $y_u^{\alpha\beta},y_d^{\alpha\beta}$ are functions of Higgs
 field.

The next is dimension four gauge interaction part Lagrangian
\begin{eqnarray}
&&\hspace{-1.5cm}\mathcal{L}_{f-4}=i{\displaystyle\sum_\alpha}\bigg\{\overline{q}^I_{\alpha
L}\slashed{D}q^I_{\alpha L}
+\delta_{L,1,\alpha}\overline{q}^I_{\alpha
L}\hat{U}_L(\slashed{D}\hat{U}_L)^{\dagger}q^I_{\alpha L}
+\delta_{L,2,\alpha}\overline{q}^I_{\alpha
R}\hat{U}_R\hat{U}_L^{\dagger}(\slashed{D}\hat{U}_L)\hat{U}_R^{\dagger}q^I_{\alpha
R}\nonumber\\
&&+\delta_{L,3,\alpha}\overline{q}^I_{\alpha
L}[(\slashed{D}\hat{U}_L)\tau^3\hat{U}^{\dagger}_L-\hat{U}_L\tau^3(\slashed{D}\hat{U}_L)^{\dagger}]q^I_{\alpha
L}+\delta_{L,4,\alpha}\overline{q}^I_{\alpha
L}\hat{U}_L\tau^3\hat{U}^{\dagger}_L(\slashed{D}\hat{U}_L)
\tau^3\hat{U}^{\dagger}_Lq^I_{\alpha L}\nonumber\\
&&+\delta_{L,5,\alpha}\overline{q}^I_{\alpha
R}\hat{U}_R[\tau^3\hat{U}^{\dagger}_L(\slashed{D}\hat{U}_L)-(\slashed{D}\hat{U}_L)^{\dagger}\hat{U}_L\tau^3]
\hat{U}^\dagger_Rq^I_{\alpha R}
+\delta_{L,6,\alpha}\overline{q}^I_{\alpha
R}\hat{U}_R\tau^3\hat{U}^{\dagger}_L(\slashed{D}\hat{U}_L)\tau^3
\hat{U}^\dagger_Rq^I_{\alpha R}\nonumber\\
&&+\delta_{L,7,\alpha}[\overline{q}^I_{\alpha
L}\hat{U}_L\tau^3\hat{U}^{\dagger}_L\slashed{D}q^I_{\alpha L}
-(\overline{q}^I_{\alpha
L}\slashed{D}^{\dagger})\hat{U}_L\tau^3\hat{U}^\dagger_Lq^I_{\alpha
L}]\bigg\}+q^I\rightarrow
l^I,\delta\rightarrow\delta^l+L\leftrightarrow R\;,\label{L4}
\end{eqnarray}
in which
\begin{eqnarray}
&&D_{\mu}q_{\alpha}= \left\{\begin{array}{ll}
(\partial_\mu\!+ig_1\frac{\tau^a}{2}W^a_{1,\mu}P_L\!+\!ig_2\frac{\tau^a}{2}W^a_{2,\mu}P_R
    +\!\frac{i}{6}gB_\mu)q_{\alpha}&\mbox{LR,~LP}\\
(\partial_\mu\!+ig_1\frac{\tau^a}{2}W^a_{1,\mu}P_L\!+\!ig\frac{\tau^3}{2}B_{\mu}P_R
    +\!\frac{i}{6}gB_\mu)q_{\alpha}
&\mbox{HP,~FP}\\
(\partial_\mu\!+ig_2\frac{\tau^a}{2}W^a_{2,\mu}P_L\!+\!ig\frac{\tau^3}{2}B_{\mu}P_R
    +\!\frac{i}{6}gB_\mu)q_{\alpha}
&\mbox{~UN}\\
(\partial_\mu\!+i\delta_{\alpha\alpha_1}g_1\frac{\tau^a}{2}W^a_{1,\mu}P_L\!
+i\delta_{\alpha\alpha_2}g_2\frac{\tau^a}{2}W^a_{2,\mu}P_L\!+\!ig\frac{\tau^3}{2}B_{\mu}P_R
+\!\frac{i}{6}gB_\mu)q_{\alpha} &\mbox{NU}
\end{array}\right.\nonumber\\
&&\\ D_{\mu}l_{\alpha}&=& \left\{\begin{array}{ll}
(\partial_\mu\!+ig_1\frac{\tau^a}{2}W^a_{1,\mu}P_L\!+\!ig_2\frac{\tau^a}{2}W^a_{2,\mu}P_R-\!\frac{i}{2}gB_\mu)l_{\alpha}&\mbox{LR,~HP}\\
(\partial_\mu\!+ig_1\frac{\tau^a}{2}W^a_{1,\mu}P_L\!+\!ig\frac{\tau^3}{2}B_{\mu}P_R-\!\frac{i}{2}gB_\mu)l_{\alpha}
&\mbox{LP,~FP,~UN}\\
(\partial_\mu\!+i\delta_{\alpha\alpha_1}g_1\frac{\tau^a}{2}W^a_{1,\mu}P_L\!
+i\delta_{\alpha\alpha_2}g_2\frac{\tau^a}{2}W^a_{2,\mu}P_L\!+\!ig\frac{\tau^3}{2}B_{\mu}P_R
-\!\frac{i}{2}gB_\mu)l_{\alpha} &\mbox{NU}
\end{array}\right.\nonumber
\end{eqnarray}
and $P_{^R_L}=(1\pm\gamma_5)/2$.
$(\slashed{D}\hat{U}_i)^\dag\equiv\gamma^{\mu}(D_{\mu}\hat{U}_i)^\dagger$
for $i=L,R$. In (\ref{L4}), coefficients $\delta$ and $\delta^l$ in
general  depend on generation indices which was not considered in
original LR case in Ref.\cite{LRSMEWCLmatter}.

%%%%%%%%%%%%%%%%%%%%%%%%%%%%%%%%%%%%%%%%%%%%%%%%%%
\section{EWCL in mass eigenstates}

EWCL presented in last section is on the basis of gauge eigenstates.
In this section, we diagonalize them to the basis of mass
eigenstates. We will find that this diagonalization is universal for
either boson sector or fermion sectors.

We first discuss boson sector. This part is the same as that in LR
case\cite{LRSMEWCLboson}, so we just list down the result. With
convention $\displaystyle
W^{\pm}_{i,\mu}=\frac{1}{\sqrt{2}}(W^1_{i,\mu}\mp iW^2_{i,\mu}),
~i=1, 2$, the mass terms in our bosonic part EWCL is
\begin{eqnarray}
 \hspace{-0.5cm}\mathcal{L}_M
&=&\frac{1}{4}f_1^2g_1^2W^+_{1,\mu}W^{-,\mu}_1+\frac{1}{4}f_2^2g_2^2W^+_{2,\mu}W^{-,\mu}_2
-\frac{1}{2}\kappa f_1f_2g_1g_2(W^+_{1,\mu}W^{-,\mu}_2+W^+_{2,\mu}W^{-,\mu}_1)\nonumber\\
&&+\frac{1}{8}(1-2\beta_{1,1})f_1^2(g_1W^3_{1,\mu}-gB_{\mu})^2
 +\frac{1}{8}(1-2\beta_{2,1})f_2^2(g_2W^3_{2,\mu}-gB_{\mu})^2\nonumber\\
&&-\frac{1}{4}(\kappa+2\tilde{\beta}_1)
f_1f_2(g_1W^3_{1,\mu}-gB_{\mu})(g_2W^{3,\mu}_2-gB^{\mu}).
\end{eqnarray}
The charged and neutral gauge bosons are diagonalized through
rotations
\begin{eqnarray}
\left(\begin{array}{c}W^{\pm}_1\\W^{\pm}_2\end{array}\right)=\left(\begin{array}{cc}\cos\zeta&-\sin\zeta\\
\sin\zeta&\cos\zeta
\end{array}\right)\left(\begin{array}{c}W^{\pm}\\W^{\prime\pm}\end{array}\right)\hspace{1.5cm}
\left(\begin{array}{c}W_{1,\mu}^3\\ W_{2,\mu}^3\\
B_\mu\end{array}\right)=\left(\begin{array}{ccc}
x_1&x_2&x_3\\y_1&y_2&y_3\\
v_1&v_2&v_3\end{array}\right)\left(\begin{array}{c}
Z_\mu\\
Z_\mu'\\
A_{\mu}\end{array}\right)\label{gaugebosondef}
\end{eqnarray}
with mixing parameters given by
\begin{eqnarray}
\tan 2\zeta=\frac{2\kappa
f_1f_2g_1g_2}{f_2^2g_2^2-f_1^2g_1^2}\hspace{2cm}
\left(\begin{array}{ccc}
x_1&x_2&x_3\\y_1&y_2&y_3\\
v_1&v_2&v_3\end{array}\right)=V\Lambda\tilde{V}\;,\label{GaugeMixing}
\end{eqnarray}
and
\begin{eqnarray}
&\hspace{-0.5cm}\left(\begin{array}{ccc}1-\alpha_{1,8}g_1^2&-\frac{1}{2}\tilde{\alpha}_8g_1g_2&-\alpha_{1,1}g_1g\\
-\frac{1}{2}\tilde{\alpha}_8g_1g_2&1-\alpha_{1,8}g_2^2&-\alpha_{2,1}g_2g\\-\alpha_{1,1}g_1g&-\alpha_{2,1}g_2g&1
\end{array}\right)
=V\left(\begin{array}{ccc}\lambda_+&0&0\\
0&\lambda_-&0\\0&0&1\end{array}\right)V^T\;,\nonumber\\
&\hspace{-0.5cm}\lambda_{\pm}=1\!-\frac{1}{2}\alpha_{1,8}g_1^2\!-\frac{1}{2}\alpha_{2,8}g_2^2
           \pm[\alpha_{1,1}^2g_1^2g^2\!+\!\alpha_{2,1}^2g_2^2g^2
           \!+\frac{1}{4}\tilde{\alpha}_8^2g_1^2g_2^2\!+\frac{1}{4}(\alpha_{1,8}g_1^2
           \!-\!\alpha_{2,8}g_2^2)^2]^{1/2}\;.\nonumber\\
&\hspace{-0.5cm}\Lambda
V^T\tilde{M}_0^2V\Lambda=\tilde{V}\left(\begin{array}{ccc}
M_Z^2&0&0\\0&M_{Z'}^2&0\\0&0&0\end{array}\right) \tilde{V}^T\;,
\hspace{1cm}\Lambda\equiv\left(\begin{array}{ccc}\frac{1}{\sqrt{\lambda_+}}&0&0\\
0&\frac{1}{\sqrt{\lambda_-}}&0\\0&0&1\end{array}\right),\nonumber\\
&\hspace{-1cm}\tilde{M}_0^2\!
=\!\left(\begin{array}{ccc}\!\!\frac{1}{4}(1-2\beta_{L,1})f_1^2g_1^2&
-\frac{1}{4}(\kappa\!+\!2\tilde{\beta}_1)f_1f_2g_1g_2&
\left[(2\beta_{1,1}\!-\!1)f_1\!\!+\!(\kappa\!+\!2\tilde{\beta}_1)f_2\right]
\!\frac{f_1g_1g}{4}\!\!\!\\
\!\!-\frac{1}{4}(\kappa\!+\!2\tilde{\beta}_1)
f_1f_2g_1g_2&\frac{1}{4}(1-2\beta_{2,1})f_2^2g_2^2&
\left[(2\beta_{2,1}\!-\!1)f_2\!\!+\!(\kappa\!+\!2\tilde{\beta}_1)f_1
\right]\!\frac{f_2g_2g}{4}\!\!\!\!\\
\!\!\mbox{\footnotesize$\left[(2\beta_{1,1}\!\!-\!\!1)\!f_1\!\!+\!(\kappa\!+\!\!2\tilde{\beta}_1)\!f_2\right]
\!\frac{f_1g_1g}{4}$}
&\!\!\mbox{\footnotesize$\left[(2\beta_{2,1}\!\!-\!\!1)\!f_2\!\!+\!(\kappa\!+\!\!2\tilde{\beta}_1)\!f_1
\right]\!\frac{f_2g_2g}{4}$}&
\!\!\mbox{\footnotesize$\left[\frac{1\!-\!2\beta_{1,1}}{2}\!f_1^2\!+\!\frac{1\!-\!2\beta_{2,1}}{2}\!f_2^2\!
-\!(\kappa\!+\!\!2\tilde{\beta}_1)
f_1f_2\right]\!\frac{g^2}{2}$}\!\!\!\end{array}\right)\;.\nonumber
\end{eqnarray}
The results of gauge boson masses become
\begin{eqnarray}
M^2_W&=&\frac{1}{4}[f_1^2g_1^2+f_2^2g_2^2-\sqrt{(f_1^2g_1^2-f_2^2g_2^2)^2+4\kappa^2
f_1^2f_2^2g_1^2g_2^2}]\;,\nonumber\\
M^2_{W'}&=&\frac{1}{4}[f_1^2g_1^2+f_2^2g_2^2+\sqrt{(f_1^2g_1^2-f_2^2g_2^2)^2+4\kappa^2
f_1^2f_2^2g_1^2g_2^2}]\,\label{Mresult}\\
M^2_{Z}&=&
\frac{f_1^2[(1-\beta_{1,1})(1-\beta_{2,1})-(\kappa+\tilde{\beta}_1)^2]}{2(1-\beta_{2,1})(g_2^2+g^2)}
  \times  [g_2^2g^2+g_1^2g_2^2+g_1^2g^2\nonumber\\
&&-2\alpha_{1,1}(g_2^2+g^2)g_1^2g_2^2g^2+2g_2^4g^4\alpha_{2,1}
    +\alpha_{1,8}g_1^4(g_2^2+g^2)^2+\alpha_{2,8}g_2^4g^4-\alpha_8(g_2^2+g^2)g_1^2g_2^2g^2]\nonumber\\
     M^2_{Z'}&=&
    \frac{1}{2}[(1-\beta_{2,1})f_2^2(g_2^2+g^2)-2f_1f_2g^2(\kappa+\beta_1)+(1-\beta_{1,1})f_1^2(g_1^2+g^2)]\nonumber\\
&&-\alpha_{1,1}g^2g_1^2[f_1^2(1-\beta_{1,1})-f_1f_2(\kappa+\beta_1)]
  -\alpha_{2,1}g^2g_2^2[f_2^2(1-\beta_{2,1})-f_1f_2(\kappa+\beta_1)]\nonumber\\
&&+\frac{1}{2}\alpha_{1,8}g_1^4f_1^2(1-\beta_{1,1})+\frac{1}{2}\alpha_{2,8}g_2^4f_2^2(1-\beta_{2,1})
  -\frac{1}{2}\alpha_8g_1^2g_2^2f_1f_2(\kappa+\beta_1)-M^2_{Z}\nonumber\;.
\end{eqnarray}
 For the gauge boson part, the most stringent constraint comes from $W-W'$ mixing
 which is characterized by the
mixing angle $\zeta$. Fortunately (\ref{GaugeMixing}) tells us that
this angle depends on two independent parameters
$x\equiv\frac{f_1g_1}{f_2g_2}$ and $\kappa$. While (\ref{Mresult})
indicates that the ratio of $W$ and $W'$ mass depends also on these
two parameters, we just have two parameters $x$ and $\kappa$ to
describe two physical quantities $\zeta$ and $M_W/M_{W'}$ at this
stage of effective Lagrangian. We can tune this two parameters
making the mixing angle $\zeta$ be small enough to match experiment
data and at the same time keeping the $W'$ mass be in arbitrary
values. This result implies the importance of parameter $\kappa$.
Since (\ref{GaugeMixing}) tells that to make mixing angle small, one
can either take very small $\kappa$ or small $x$. While from
(\ref{Mresult}), small $x$ will cause very big mass difference
between $W$ and $W'$. In order to avoid this big mass difference
between $W$ and $W'$, the only way is to have small $\kappa$. In any
of candidate models, only those with very small $\kappa$ value are
phenomenologically allowed.

Next, we discuss fermion sector which includes lepton and quark
parts. For lepton part, in unitary gauge, (\ref{LYldef}) become
\begin{eqnarray}
&&\mathcal{L}_{\mathrm{Y,lepton}}\bigg|_{\mbox{\tiny
Unitary~gauge}}=\mathcal{L}_{Me}+\mathcal{L}_{M\nu}\\
&&\mathcal{L}_{Me}=\overline{e^{-I}}_{\alpha
L}(y^{\alpha\beta}-y_3^{\alpha\beta})e^{-I}_{\beta R}+
\overline{e^{-I}}_{\alpha
R}(y^{\dag\alpha\beta}-y_3^{\dag\alpha\beta})e^{-I}_{\beta L}\;,
\label{LMedef1}\\
&&\mathcal{L}_{M\nu}=\overline{\nu^I}_{\alpha
L}(y^{\alpha\beta}+y_3^{\alpha\beta})\nu^I_{\beta R}
+h_L^{\alpha\beta}\overline{\nu^{Ic}_{\alpha L}}\nu^I_{\beta L}
+h_R^{\alpha\beta}\overline{\nu^{Ic}_{\alpha R}}\nu^I_{\beta R}
+\mbox{h.c.}\;, \label{LMmudef1}
\end{eqnarray}
For electron part, rotating the gauge eigenstates into the mass
eigenstates with unitary matrices $\tilde{V}^e$ by
$e^-_{L,R}=\tilde{V}^{e}_{L,R}e_{L,R}^{-I}$, we can reduce
(\ref{LMedef1}) to
\begin{eqnarray}
\mathcal{L}_{Me}=\overline{e^-}_L\mathbf{M}^{e}e_R^-+
\overline{e^-}_R\mathbf{M}^{e\dag}e_L^-
\end{eqnarray}
 with diagonal mass matrix
$\mathbf{M}^{e}=\tilde{V}^e_L(y-y_3)\tilde{V}^{e\dagger}_R$. For
neutrino part,  with help of relation
$\overline{\nu_1^c}\nu_2^c=\overline{\nu_2}\nu_1$, we find
coefficient matrices  $h_L^{\alpha\beta}$ and $h_R^{\alpha\beta}$
can be chosen to be symmetric, then neutrino part of Lagrangian can
be written as
\begin{eqnarray}
\mathcal{L}_{M\nu}&=&\frac{1}{2}\left(\begin{array}{cc}\overline{\nu^I_L}&
\overline{\nu^{Ic}_R}\end{array}\right) \left(\begin{array}{cc}2h_L&y+y_3\\
(y+y_3)^T&2h_R\end{array}\right)\left(\begin{array}{c}\nu^{Ic}_L\\
\nu^I_R\end{array}\right)+\mbox{h.c.}\;. \label{LMmudef2}
\end{eqnarray}
where $\nu^{Ic}_L=\left(\begin{array}{c}\nu_e^{Ic}\\
\nu_\nu^{Ic}\\ \nu_\tau^{Ic} \end{array}\right)_L$ is
left-handed neutrino gauge eigenstates and $\nu^I_R=\left(\begin{array}{c}\nu_e^I\\
\nu_\nu^I\\ \nu_\tau^I \end{array}\right)_R$ is right-handed
neutrino gauge eigenstates. The overall $6\times 6$ neutrino mass
matrix $\left(\begin{array}{cc}2h_L&y+y_3\\
(y+y_3)^T&2h_R\end{array}\right)$ is symmetric and can be
diagonalized by a unitary transformation,
\begin{eqnarray}
\left(\begin{array}{cc}V&R\\S&U\end{array}\right)^\dag
\left(\begin{array}{cc}2h_L&y+y_3\\
(y+y_3)^T&2h_R\end{array}\right)
\left(\begin{array}{cc}V&R\\S&U\end{array}\right)^*
=\left(\begin{array}{cc}\hat{M}_\nu&0\\0&\hat{M}_N\end{array}\right)\;,
\end{eqnarray}
where $\hat{M}_\nu=\mathrm{diag}\{m_1,m_2,m_3\}$ and
$\hat{M}_N=\mathrm{diag}\{M_1,M_2,M_3\}$ with $m_i$ and $M_i$ (for
$i=1,2,3$) the light and heavy neutrino masses, respectively. $V$ is
the $3\times 3$ Maki-Nakagawa-Sakata (MNS) neutrino mixing matrix
\cite{MNS} responsible for neutrino oscillations, and $R, S, U$ are
all $3\times 3$ matrices. After this diagonalization, one may
express the neutrino gauge eigenstates $\nu_\alpha^I$ (for
$\alpha=e,\mu,\tau$) in terms of the light and heavy neutrino mass
states $\nu_\alpha$ and $N_\alpha$:
\begin{eqnarray}
\left(\begin{array}{c}\nu_e^I\\
\nu_\nu^I\\ \nu_\tau^I \end{array}\right)_L=V\left(\begin{array}{c}\nu_e\\
\nu_\nu\\ \nu_\tau \end{array}\right)+R\left(\begin{array}{c}N_e\\
N_\nu\\ N_\tau \end{array}\right)\;.
\end{eqnarray}
Unitarity of $6\times 6$ rotation matrix leads to
$VV^\dag+RR^\dag=I$, which implies that the MNS matrix $V$ is not
unitary and the matrix $R$ characterize this non-unitarity of $V$.
By testing the non-unitarity of $V$ matrix, we can examine heavy
neutrino effects at low energy region \cite{XingPLB08}. In the case
that $h_R\gg h_L,y+y_3$, we can diagonalize the mass matrix
approximately by
\begin{eqnarray}
\left(\begin{array}{cc}V&R\\S&U\end{array}\right)=\left(\begin{array}{cc}1-\frac{1}{8}(y+y_3)h_R^{-2}(y+y_3)^T&\frac{1}{2}(y+y_3)h_R^{-1}\\
-\frac{1}{2}h_R^{-1}(y+y_3)^T&1-\frac{1}{8}h_R^{-1}(y+y_3)^T(y+y_3)h_R^{-1}\end{array}\right)\;.\nonumber
\end{eqnarray}
which will lead to
\begin{eqnarray}
&&\left(\begin{array}{cc}V&R\\S&U\end{array}\right)^\dag
\left(\begin{array}{cc}2h_L&y+y_3\\
(y+y_3)^T&2h_R\end{array}\right)
\left(\begin{array}{cc}V&R\\S&U\end{array}\right)^*\nonumber\\
&&=\left(\begin{array}{cc}2h_L-\frac{1}{2}(y+y_3)h_R^{-1}(y+y_3)^T+O(h_R^{-2})&O(h_R^{-1})\\
O(h_R^{-1})&2h_R+O(h_R^{-1})\end{array}\right)~~~~\label{Seesaw}
\end{eqnarray}
If $h_L=0$, (\ref{Seesaw}) leads to the standard type I seesaw
mechanism, otherwise we obtain type II seesaw mechanism for
neutrinos.

For quark part, (\ref{LYqdef}) in unitary gauge is
\begin{eqnarray}
&&\mathcal{L}_{\mathrm{Y,quark}}\bigg|_{\mbox{\tiny Unitary~gauge}}
=\overline{q}^I_{\alpha
L}(\tau^uy_u^{\alpha\beta}+\tau^dy_d^{\alpha\beta})q^I_{\beta R}
+\mathrm{h.c.}\label{LYqdef1}
\end{eqnarray}
We can explicitly expand coefficients
  $y_u^{\alpha\beta},y_d^{\alpha\beta}$ in terms of powers of quantum
  fluctuation Higgs field $\tilde{h}$
  \begin{eqnarray}
y_i=y_i^0+y_i^1\tilde{h}+O(\tilde{h}^2)\hspace{1cm}i=u,d\;,~~~~\label{Yexp}
\end{eqnarray}
where $y_i^0,y_i^1$ are matrices independent of Higgs field $h$.

The gauge eigenstates can be rotated into the mass eigenstates with
unitary matrices $V^{u, d}_{L,R}$,
\begin{eqnarray}
u_{L,R}=V^u_{L,R}u_{L,R}^I \hspace{3cm}d_{L,R}=V^d_{L,R}d_{L,R}^I.
\end{eqnarray}
The $y^0_{u, d}$ matrices defined in (\ref{Yexp}) are  diagonalized
as follows:
\begin{eqnarray}
V^u_{L}y^0_uV^{u\dag}_{R}=M_{diag}^u,~~~V^d_{L}y^0_dV^{d\dag}_{R}=M_{diag}^d,~~~\label{Mdef}
\end{eqnarray}
 where  $M_{diag}^{u,d}$ represent the diagonal up- and down-quark
mass matrices of physical quark masses.
\begin{eqnarray}
&q_{\alpha L,R}=\left(\begin{array}{c}u_{\alpha L,R}\\d_{\alpha
L,R}\end{array}\right)
=[(V^u_{L,R})_{\alpha\beta}\tau^u+(V^d_{L,R})_{\alpha\beta}\tau^d]
\left(\begin{array}{c}u^I_{\beta L,R}\\d^I_{\beta
L,R}\end{array}\right)\;.\\
&(V^u_{L}\tau^u+V^d_{L}\tau^d)(\tau^uy_u^0+\tau^dy_d^0)(V^{u\dag}_{R}\tau^u+V^{d\dag}_{L,R}\tau^d)
=(\tau^uM^u_{diag}+\tau^dM^d_{diag})
\end{eqnarray}
The usual Cabibbo-Kobayashi-Maskawa (CKM) matrix in the left sector,
and the corresponding matrix in the right sector, are given by
\begin{eqnarray}
V_{L,R}^{\mathrm{CKM}}=V_{L,R}^uV_{L,R}^{d\dag}.
\end{eqnarray}
Note that, {\it a priori}, there is no reason for
$V_{L}^{\mathrm{CKM}}$ to equal $V_{R}^{\mathrm{CKM}}$.

Any $n\times n$ unitary matrix has $n^2$ real parameters among which
$n(n-1)/2$ may be expressed in the form of
$\sin\theta_{\alpha\beta},~ \cos\theta_{\alpha\beta}$ with
$n^2-n(n-1)/2=n(n+1)/2$ phases left. Since each quark field can be
redefined through a phase transformation, $2n-1$ phases are not
physical. If $V_{L}^{\mathrm{CKM}}$ and $V_{R}^{\mathrm{CKM}}$ are
independent, the total number of physical phases is
$2\times\frac{n(n+1)}{2}-(2n-1)=n^2-n+1.$
 In our case of 3 generations of fermions, $V_{L}^{\mathrm{CKM}}$
can be taken as the standard form \cite{PDG06},
\begin{eqnarray}
V_L^{\mathrm{CKM}}=\left( \begin{array}{ccc}
V_L^{ud} & V_L^{us}  & V_L^{ub}\\
V_L^{cd} & V_L^{cs}  & V_L^{cb}\\
V_L^{td} & V_L^{ts}  & V_L^{tb}
\end{array} \right)
=\left( \begin{array}{ccc}
c_{12}c_{13} & s_{12}c_{13}  & s_{13}e^{-i\delta}\\
-s_{12}c_{23}-c_{12}s_{23}s_{13}e^{i\delta} & c_{12}c_{23}-s_{12}s_{23}s_{13}e^{i\delta}  & s_{23}c_{13}\\
s_{12}s_{23}-c_{12}c_{23}s_{13}e^{i\delta} &
-c_{12}s_{23}-s_{12}c_{23}s_{13}e^{i\delta}  & c_{23}c_{13}
\end{array} \right).~~~~\label{VLCKM}
\end{eqnarray}
Then the most general $V_{R}^{\mathrm{CKM}}$ may be in the form of
standard CKM matrix with 5 phases added:
\begin{eqnarray}
\label{VRCKM}
 V_{R}^{\mathrm{CKM}}=\left( \begin{array}{lll}
 \bar{V}_{R}^{ud}e^{2i\alpha_1} & \bar{V}_{R}^{us}e^{i(\alpha_1+\alpha_2+\beta_1)}  & \bar{V}_{R}^{ub}e^{i(\alpha_1+\alpha_3+\beta_1+\beta_2)}\\
  \bar{V}_{R}^{cd}e^{i(\alpha_1+\alpha_2-\beta_1)} & \bar{V}_{R}^{cs}e^{2i\alpha_2}  & \bar{V}_{R}^{cb}e^{i(\alpha_2+\alpha_3+\beta_2)}\\
  \bar{V}_{R}^{td}e^{i(\alpha_1+\alpha_3-\beta_1-\beta_2)} & \bar{V}_{R}^{ts}e^{i(\alpha_2+\alpha_3-\beta_2)}  & \bar{V}_{R}^{tb}e^{2i\alpha_3}
   \end{array} \right),
\end{eqnarray}
where
\begin{eqnarray}
\left( \begin{array}{ccc}
 \bar{V}_{R}^{ud} & \bar{V}_{R}^{us}  &\bar{V}_{R}^{ub}\\
  \bar{V}_{R}^{cd} & \bar{V}_{R}^{cs}  & \bar{V}_{R}^{cb}\\
   \bar{V}_{R}^{td} &\bar{V}_{R}^{ts}  & \bar{V}_{R}^{tb}
   \end{array} \right)
   =\left( \begin{array}{ccc}
 \bar{c}_{12}\bar{c}_{13} & \bar{s}_{12}\bar{c}_{13}  & \bar{s}_{13}e^{-i\bar{\delta}}\\
  -\bar{s}_{12}\bar{c}_{23}-\bar{c}_{12}\bar{s}_{23}\bar{s}_{13}e^{i\bar{\delta}} & \bar{c}_{12}\bar{c}_{23}-\bar{s}_{12}\bar{s}_{23}\bar{s}_{13}e^{i\bar{\delta}}  & \bar{s}_{23}\bar{c}_{13}\\
  \bar{s}_{12}\bar{s}_{23}-\bar{c}_{12}\bar{c}_{23}\bar{s}_{13}e^{i\bar{\delta}} & -\bar{c}_{12}\bar{s}_{23}-\bar{s}_{12}\bar{c}_{23}\bar{s}_{13}e^{i\bar{\delta}}  & \bar{c}_{23}\bar{c}_{13}
   \end{array} \right)
  \end{eqnarray}
  with $\bar{c}_{12}=\cos\bar{\theta}_{12},
  \bar{s}_{12}=\sin\bar{\theta}_{12}$, etc. In general,
  $\bar{\theta}_{\alpha\beta}$ do not equal to those in
  $V_{L}^{\mathrm{CKM}}$. If $\bar{V}_{R}^{\alpha\beta}=(V_{L}^{\alpha\beta})^*$ hold
   for $\alpha=u,c,t$ and  $\beta=d,s,b$, then $V_{R}^{\mathrm{CKM}}$ in (\ref{VRCKM}) coincides with
  that in \cite{Anatomy}\cite{YLWu07} which is called pseudo-manifest left-right
  symmetric and is originally proposed to construct left-right symmetric
models with spontaneously CP violation.

%%%%%%%%%%%%%%%%%%%%%%%%%%%%%%%
\section{Goldstone, Higgs and gauge couplings to quarks}

The discussions in last section are limited in unitary gauge without
the Goldstone contributions and Higgs contributions included. As a
compensation and preparation of next section computation, we now
focus our attention on quark-Goldstone-boson and quark-Higgs
couplings. We will find that, unlike the mixing terms dealt above,
these coupling are no longer universal. We explicitly expanded out
Goldstone fields by
\begin{eqnarray}
U_{1,2}=\textrm{exp}(\frac{ig_{1,2}}{\sqrt{2}M_{1,2}}\phi_{1,2})
\hspace{1cm}\phi_{1,2}=\left(\begin{array}{cc}\frac{\phi_{1,2}^{0}}{\sqrt{2}}&\phi_{1,2}^{+}\\
\phi^{-}_{1,2}&-\frac{\phi_{1,2}^{0}}{\sqrt{2}}\end{array}\right)\hspace{1cm}M_{1,2}=\frac{1}{2}f_{1,2}g_{1,2}\;.
\end{eqnarray}
in which we have taken $\zeta=0$. In terms of the masses
eigenstates, Lagrangian (\ref{LYqdef}) can be expanded according to
the goldstone and Higgs fields,
\begin{eqnarray}
\hspace{-1.5cm}\mathcal{L}_{\mathrm{Y,quark}}
&=&(1+\frac{ig_L}{2M_L}\phi_{L}^0-\frac{ig_R}{2M_R}\phi_R^0)\overline{u}_LM^u_{diag}u_R
+(1-\frac{ig_L}{2M_L}\phi_{L}^{0}+\frac{ig_R}{2M_R}\phi_R^0)\overline{d}_{L}M^d_{diag}d_{R}
\nonumber\\
&&+\mathrm{h.c.}+ \mathcal{L}_{\mathrm{Y,CC}}+
\mathcal{L}_{\mathrm{Y,}h}
+O(\overline{q}\phi^2q,\overline{q}\tilde{h}^2q,\overline{q}\phi\tilde{h}q)\;.\label{LYq}
\end{eqnarray}
where the charged Yukawa coupling $\mathcal{L}_{\mathrm{Y,CC}}$ in
Lagrangian (\ref{LYq}) is
\begin{eqnarray}
\label{LYq,CC}
 \mathcal{L}_{\mathrm{Y,CC}}
&=&-\frac{ig_{L}}{\sqrt{2}M_{L}}(\overline{u}_{R}M^u_{diag}\phi_{L}^{+}V^{\mathrm{CKM}}_{L}d_{L}
-\overline{u}_{L}V^{\mathrm{CKM}}_{L}\phi_{L}^{+}M^d_{diag}d_R)\nonumber\\
&&-\frac{ig_{R}}{\sqrt{2}M_{R}}(\overline{u}_{L}M^u_{diag}V^{\mathrm{CKM}}_{R}\phi_{R}^{+}d_R
-\overline{u}_{R}\phi_{R}^{+}V^{\mathrm{CKM}}_{R}M^d_{diag}d_{L})+\mathrm{h.c.}\nonumber\\
&\equiv&-\frac{i}{\sqrt{2}}\overline{u}_\alpha(A^{\alpha\beta}_L+B^{\alpha\beta}_L\gamma^5)d_\beta
\phi_{L}^{+}
-\frac{i}{\sqrt{2}}\overline{u}_\alpha(A^{\alpha\beta}_R+B^{\alpha\beta}_R\gamma^5)d_\beta\phi^+_{R}\nonumber\\
&&-\frac{i}{\sqrt{2}}\overline{d}_\beta(A^{\alpha\beta\dag}_L+B^{\alpha\beta\dag}_L\gamma^5)u_\alpha
\phi_{L}^{-}
-\frac{i}{\sqrt{2}}\overline{d}_\beta(A^{\alpha\beta\dag}_R+B^{\alpha\beta\dag}_R\gamma^5)u_\alpha\phi^-_{R}
\end{eqnarray}
 with
\begin{eqnarray}
\label{LYq,CCAB}
 && A^{\alpha\beta}_L=\frac{1}{2}\frac{g_L}{M_{L}}(m_{u_\alpha}-m_{d_\beta})V_L^{\alpha\beta},~~~
                       B^{\alpha\beta}_L=\frac{1}{2}\frac{g_L}{M_{L}}(-m_{u_\alpha}-m_{d_\beta})V_L^{\alpha\beta}\nonumber\\
&&A^{\alpha\beta}_R=\frac{1}{2}\frac{g_R}{M_{R}}(m_{u_\alpha}-m_{d_\beta})V_R^{\alpha\beta},~~~
                       B^{\alpha\beta}_R=\frac{1}{2}\frac{g_R}{M_{R}}(m_{u_\alpha}+m_{d_\beta})V_R^{\alpha\beta}\nonumber\\
&&A^{\alpha\beta\dag}_L=\frac{1}{2}\frac{g_L}{M_{L}}(m_{d_\beta}-m_{u_\alpha})V_L^{\alpha\beta*},~~~
                       B^{\alpha\beta\dag}_L=\frac{1}{2}\frac{g_L}{M_{L}}(-m_{d_\beta}-m_{u_\alpha})V_L^{\alpha\beta*}\nonumber\\
&&A^{\alpha\beta\dag}_R=\frac{1}{2}\frac{g_R}{M_{R}}(m_{d_\beta}-m_{u_\alpha})V_R^{\alpha\beta*},~~~
                       B^{\alpha\beta\dag}_R=\frac{1}{2}\frac{g_R}{M_{R}}(m_{d_\beta}+m_{u_\alpha})V_R^{\alpha\beta*}
\end{eqnarray}
To cover various models, we use symbol $\frac{g_L}{M_L}\phi_L$ and
$\frac{g_R}{M_R}\phi_R$ to represent Goldstone fields and
corresponding couplings. Their relations with Goldstone fields
$\phi_1$, $\phi_2$ and corresponding couplings are
\begin{eqnarray}
\frac{g_L}{M_L}\phi_L&=&\left\{\begin{array}{ll}
\frac{g_1}{M_1}\phi_1&\mbox{LR,LP,HP,FP}\\
\frac{g_2}{M_2}\phi_2&\mbox{UN}\\
\frac{g_1}{M_1}\phi_1\delta_{\alpha\alpha_1}+\frac{g_2}{M_2}\phi_2\delta_{\alpha\alpha_2}&
\mbox{NU}
\end{array}\right.\\
\frac{g_R}{M_R}\phi_R&=&\left\{\begin{array}{lll}
\frac{g_2}{M_2}\phi_2&\hspace*{2.7cm}&\mbox{LR,LP,}\\
0&&\mbox{HP,FP,UN,NU}
\end{array}\right.
\end{eqnarray}
The quark-Higgs-boson couplings $\mathcal{L}_{\mathrm{Y,}h}$ in
Lagrangian (\ref{LYq}) are
\begin{eqnarray}
\label{LYq,h}
 \mathcal{L}_{\mathrm{Y,}h}
&=&\tilde{h}(\overline{u}_LV^u_{L}y_u^1V^{u\dag}_{R}u_R
+\overline{d}_LV^d_{L}y_d^1V^{d\dag}_{R}d_R)+\mathrm{h.c.}\nonumber\\
&=&
\frac{1}{2}\overline{u}_\alpha(A^{\alpha\beta}_u+B^{\alpha\beta}_u\gamma^5)u_\beta\tilde{h}
+\frac{1}{2}\overline{d}_\alpha(A^{\alpha\beta}_d+B^{\alpha\beta}_d\gamma^5)d_j\beta\tilde{h}~~~~\label{Lh}
\end{eqnarray}
 where
\begin{eqnarray}
\label{LYq,hAB}
A^{\alpha\beta}_u=(\tilde{y}_u+\tilde{y}_u^{\dag})^{\alpha\beta},~~~~
                       B^{\alpha\beta}_u=(\tilde{y}_u-\tilde{y}_u^{\dag})^{\alpha\beta},~~~~
A^{\alpha\beta}_d=(\tilde{y}_d+\tilde{y}_d^{\dag})^{\alpha\beta},~~~~
                       B^{\alpha\beta}_d=(\tilde{y}_d-\tilde{y}_d^{\dag})^{\alpha\beta},
\end{eqnarray}
 where $\tilde{y}_u=V^u_{L}y_u^1V^{u\dag}_{R}$ and
 $\tilde{y}_d=V^d_{L}y_d^1V^{d\dag}_{R}$. Note that for neutral goldstones, there is no flavor-changing
$\overline{q}\phi q$ coupling. For charged goldstone bosons, the
non-diagonal CKM matrices and nontrivial mass difference of quarks
will yields flavor-changing couplings. If $y^1_i=y^0_i/v$ with $v$
the expectation value of $h$, the quark-Higgs couplings is in agree
with that of the SM. However, flavor-changing couplings for neutral
Higgs field $\tilde{h}$ can exist in general due to the fact that
matrices $V^i_{L}y_i^1V^{i\dag}_{R}$ may not be diagonal.

Now we come to discuss gauge couplings. In unitary gauge, Lagrangian
(\ref{L4}) become
\begin{eqnarray}
\mathcal{L}_{f-4}\bigg|_{\mbox{\tiny Unitary
gauge}}&=&{\displaystyle\sum_{\alpha}}~\overline{q}^I_\alpha[i\slashed{\partial}
-(~\Delta_{1,1,\alpha}g_1\frac{\tau^{a'}}{2}\slashed{W}^{a'}_1+\Delta_{1,2,\alpha}\frac{\tau^{a'}}{2}g_2
\slashed{W}^{a'}_2+\Delta^3_{1,1,\alpha}g_1\slashed{W}^3_1\label{L4unitary}\\
&&+\Delta^3_{1,2,\alpha}g_2\slashed{W}^3_2+\Delta_{1,\alpha}g\slashed{B}~)P_L
-(~\Delta_{2,1,\alpha}g_1\frac{\tau^{a'}}{2}\slashed{W}^{a'}_1+\Delta_{2,2,\alpha}\frac{\tau^{a'}}{2}g_2
\slashed{W}^{a'}_2\nonumber\\
&&+\Delta^3_{2,1,\alpha}g_1\slashed{W}^3_1+\Delta^3_{2,2,\alpha}g_2\slashed{W}^3_2
+\Delta_{2,\alpha}g\slashed{B}~)P_R]q^I_\alpha+q^I\rightarrow
l^I,\delta\rightarrow\delta^l,\Delta\rightarrow\Delta^l\;,\nonumber
\end{eqnarray}
where $a'=1,2$. The above anomalous gauge couplings $\Delta$'s can
be expressed by $\delta$'s introduced in (\ref{L4}) and the detailed
results are given in Appendix A. In terms of mass eigenstates for
gauge bosons, (\ref{L4unitary}) become
\begin{eqnarray}
\mathcal{L}_{f-4}\bigg|_{\mbox{\tiny Unitary
gauge}}&=&i\overline{q}^I_\alpha\gamma^\mu\partial_\mu
q^I_\alpha+\mathcal{L}_\mathrm{CC}+\mathcal{L}_\mathrm{NC}+\mathcal{L}_\mathrm{EM}
\end{eqnarray}
with charge current part $\mathcal{L}_\mathrm{CC}$
\begin{eqnarray}
\mathcal{L}_\mathrm{CC}&=&\frac{-1}{\sqrt{2}}{\displaystyle\sum_\alpha}\bigg[
\overline{q}^I_\alpha[(g_1\cos\zeta\Delta_{1,1,\alpha}\!+g_2\sin\zeta\Delta_{1,2,\alpha})\gamma^\mu
P_L+(g_2\sin\zeta\Delta_{2,1,\alpha}\!+g_1\cos\zeta\Delta_{2,2,\alpha})\gamma^\mu P_R]\nonumber\\
&&\times(\tau^+ W^+_{\mu}+\tau^-
W^-_{\mu})q^I_\alpha+\overline{q}^I_\alpha[(-g_1\sin\zeta\Delta_{1,1,\alpha}
+g_2\cos\zeta\Delta_{1,2,\alpha})\gamma^\mu P_L+(g_2\cos\zeta\Delta_{2,1,\alpha}\nonumber\\
&&-g_1\sin\zeta\Delta_{2,2,\alpha})\gamma^\mu P_R](\tau^+
W^{\prime+}_{\mu}+\tau^-
W^{\prime-}_{\mu})q^I_\alpha\bigg]+q^I\rightarrow
l^I,\Delta\rightarrow\Delta^l\;,\label{LCC}
\end{eqnarray}
neutral current part $\mathcal{L}_\mathrm{NC}$
\begin{eqnarray}
\mathcal{L}_\mathrm{NC}&=&
-\frac{1}{2}{\displaystyle\sum_\alpha}\bigg[\overline{q}^I_\alpha\{[g_1x_1(\Delta^3_{1,1,\alpha}+\Delta^3_{2,2,\alpha})
+g_2y_1(\Delta^3_{2,1,\alpha}+\Delta^3_{1,2,\alpha})+gv_1(\Delta_{1,\alpha}+\Delta_{2,\alpha})]\gamma^\mu
\nonumber\\
&&-[g_1x_1(\Delta^3_{1,1,\alpha}-\Delta^3_{2,2,\alpha})-g_2y_1(\Delta^3_{2,1,\alpha}-\Delta^3_{1,2,\alpha})
+gv_1(\Delta_{1,\alpha}-\Delta_{2,\alpha})]\gamma^\mu\gamma^5\}q^I_\alpha
Z_{\mu}\nonumber\\
&&+\overline{q}^I_\alpha\{[g_1x_2(\Delta^3_{1,1,\alpha}+\Delta^3_{2,2,\alpha})
+g_2y_2(\Delta^3_{2,1,\alpha}+\Delta^3_{1,2,\alpha})+gv_2(\Delta_{1,\alpha}+\Delta_{2,\alpha})]\gamma^\mu
\nonumber\\
&&-[g_1x_2(\Delta^3_{1,1,\alpha}-\Delta^3_{2,2,\alpha})-g_1y_2(\Delta^3_{2,1,\alpha}-\Delta^3_{1,2,\alpha})
+gv_2(\Delta_{1,\alpha}-\Delta_{2,\alpha})]\gamma^\mu\gamma^5\}q^I_\alpha
Z'_{\mu}\bigg]\nonumber\\
&&+q^I\rightarrow l^I,\Delta\rightarrow\Delta^l\;,\nonumber
\end{eqnarray}
electro-magnetic current part $\mathcal{L}_\mathrm{EM}$
\begin{eqnarray}
\mathcal{L}_\mathrm{EM}&=&
-\frac{1}{2}{\displaystyle\sum_{\alpha}}\overline{q}^I_\alpha\{[g_1x_3(\Delta^3_{1,1,\alpha}\!+\Delta^3_{2,2,\alpha})
+g_2y_3(\Delta^3_{2,1,\alpha}\!+\Delta^3_{1,2,\alpha})
+gv_3(\Delta_{1,\alpha}\!+\Delta_{2,\alpha})]\gamma^\mu\nonumber\\
&&-[g_1x_3(\Delta^3_{1,1,\alpha}\!-\Delta^3_{2,2,\alpha})-g_2y_3(\Delta^3_{2,1,\alpha}\!-\Delta^3_{1,2,\alpha})
+gv_3(\Delta_{1,\alpha}\!-\Delta_{2,\alpha})]\gamma^\mu\gamma^5\}q^I_\alpha
A_{\mu}\nonumber\\
&&+q^I\rightarrow l^I,\Delta\rightarrow\Delta^l\;,\nonumber
 \end{eqnarray}
Further in terms of fermion mass eigenstates,
$\mathcal{L}_\mathrm{NC}$ and $\mathcal{L}_\mathrm{EM}$ keep their
present form, but we must replace original summation over generation
indices
${\displaystyle\sum_{\alpha}}\overline{q}^I_\alpha\Delta_{i,\alpha}
q^I_\alpha$ with
${\displaystyle\sum_{\alpha\beta}}\overline{q}_\alpha\Delta_{i,\alpha\beta}'
q_\beta$ where
\begin{eqnarray}
\Delta_{i,\alpha\beta}'&\equiv&
[V_L^u\mathrm{diag}(\Delta_{i,1},\Delta_{i,2},\Delta_{i,3})V_L^{u\dag}+
V_R^u\mathrm{diag}(\Delta_{i,1},\Delta_{i,2},\Delta_{i,3})V_R^{u\dag}]_{\alpha\beta}\tau^u\nonumber\\
&&+[V_L^d\mathrm{diag}(\Delta_{i,1},\Delta_{i,2},\Delta_{i,3})V_L^{d\dag}+
V_R^d\mathrm{diag}(\Delta_{i,1},\Delta_{i,2},\Delta_{i,3})V_R^{d\dag}]_{\alpha\beta}\tau^d\label{Delta'def}
\end{eqnarray}
It is easy to see that if $\Delta_{i,\alpha}$ is universal in
generation, i.e. it is independent of index $\alpha$, then
$\Delta_{i,\alpha\beta}'=\Delta_{i,\alpha}\delta_{\alpha\beta}$
which leads $\mathcal{L}_\mathrm{NC}$ and $\mathcal{L}_\mathrm{EM}$
unchanged. In order to suppress the possible flavor changing neutral
and electro-magnetic currents, either non-universal effect of
$\Delta_{i,\alpha}$ appeared in $\mathcal{L}_\mathrm{NC}$ and
$\mathcal{L}_\mathrm{EM}$ is small or there is some cancelations
among different terms in (\ref{Delta'def}).

The charge current Lagrangian for quarks in mass eigenstates is
changed to
\begin{eqnarray}
\label{LRCCLagrangian}
 \mathcal{L}_{\mathrm{CC}}&=&-\frac{1}{\sqrt{2}}\overline{u}_\alpha\gamma^\mu(A^{\alpha\beta}_W
 +B^{\alpha\beta}_W\gamma^5)d_\beta
W^+_{\mu}-\frac{1}{\sqrt{2}}\overline{u}_\alpha\gamma^\mu(A^{\alpha\beta}_{W'}
+B^{\alpha\beta}_{W'}\gamma^5)d_\beta
W^{\prime+}_{\mu}+\mathrm{h.c.}\;,
 \end{eqnarray}
 where
\begin{eqnarray}
\label{LRCCAB}A^{\alpha\beta}_W&=&\frac{1}{2}\bigg[g_1\cos\zeta[V^u_L\mathrm{diag}(\Delta_{1,1,1},\Delta_{1,1,2},\Delta_{1,1,3})
V^{d\dag}_L+V^u_R\mathrm{diag}(\Delta_{2,2,1},\Delta_{2,2,2},\Delta_{2,2,3})V^{d\dag}_R]_{\alpha\beta}\nonumber\\
&&+g_2\sin\zeta[V^u_L\mathrm{diag}(\Delta_{1,2,1},\Delta_{1,2,2},\Delta_{1,2,3})V^{d\dag}_L
+V^u_R\mathrm{diag}(\Delta_{2,1,1},\Delta_{2,1,2},\Delta_{2,1,3})V^{d\dag}_R]_{\alpha\beta}\bigg]\nonumber\\
B^{\alpha\beta}_W&=&\frac{1}{2}\bigg[g_1\cos\zeta[-V^u_L\mathrm{diag}(\Delta_{1,1,1},\Delta_{1,1,2},\Delta_{1,1,3})
V^{d\dag}_L+V^u_R\mathrm{diag}(\Delta_{2,2,1},\Delta_{2,2,2},\Delta_{2,2,3})V^{d\dag}_R]_{\alpha\beta}\nonumber\\
&&+g_2\sin\zeta[-V^u_L\mathrm{diag}(\Delta_{1,2,1},\Delta_{1,2,2},\Delta_{1,2,3})V^{d\dag}_L
+V^u_R\mathrm{diag}(\Delta_{2,1,1},\Delta_{2,1,2},\Delta_{2,1,3})V^{d\dag}_R]_{\alpha\beta}\bigg]\nonumber\\
A^{\alpha\beta}_{W'}&=&\frac{1}{2}\bigg[g_2\cos\zeta[V^u_R\mathrm{diag}(\Delta_{2,1,1},\Delta_{2,1,2},\Delta_{2,1,3})
V^{d\dag}_R+V^u_L\mathrm{diag}(\Delta_{1,2,1},\Delta_{1,2,2},\Delta_{1,2,3})V^{d\dag}_L]_{\alpha\beta}\nonumber\\
&&-g_1\sin\zeta[V^u_R\mathrm{diag}(\Delta_{2,2,1},\Delta_{2,2,2},\Delta_{2,2,3})V^{d\dag}_R
+V^u_L\mathrm{diag}(\Delta_{1,1,1},\Delta_{1,1,2},\Delta_{1,1,3})V^{d\dag}_L]_{\alpha\beta}\bigg]\nonumber\\
B^{\alpha\beta}_{W'}&=&\frac{1}{2}\bigg[-g_2\cos\zeta[-V^u_R\mathrm{diag}(\Delta_{2,1,1},\Delta_{2,1,2},\Delta_{2,1,3})
V^{d\dag}_R+V^u_L\mathrm{diag}(\Delta_{1,2,1},\Delta_{1,2,2},\Delta_{1,2,3})V^{d\dag}_L]_{\alpha\beta}\nonumber\\
&&+g_1\sin\zeta[-V^u_R\mathrm{diag}(\Delta_{2,2,1},\Delta_{2,2,2},\Delta_{2,2,3})V^{d\dag}_R
+V^u_L\mathrm{diag}(\Delta_{1,1,1},\Delta_{1,1,2},\Delta_{1,1,3})V^{d\dag}_L]_{\alpha\beta}\bigg]\nonumber\\
&&\label{ABdef}
\end{eqnarray}
If $\Delta_{i,\alpha}$ is universal in generation index, then
rotation matrices appeared in above formulae will meet together
constituting CKM matrices.

If we only focus on gauge couplings to light gauge boson $A,W,Z$,
above $\Delta$'s cause a serious anomalous couplings. In
Ref.\cite{LRSMEWCLmatter}, we parameterized these anomalous
couplings in terms of ten coefficients in the case that
$\Delta_{i,\alpha}$ is universal in generation index, for which two
are in charged current, four in neutral current and four in
electro-magnetic current. The fact that SM is consistent with
experiment to very high precision implies these ten anomalous
couplings must be very small in values.
 %%%%%%%%%%%%%%%%%%%%%%%%%%%%%%%%%%%%%%%%%%%%%%%%%%%%%%%%%%%%%
\section{Effective Hamiltonian for neutral K and B system}

\begin{figure}[t] \caption{Box diagrams for $K^0-\bar{K}^0$ effective Hamiltonian $H^{WW}_\mathrm{eff}$. }
 \label{fig-WW-K-box}
\centering
\begin{minipage}[t]{\textwidth}
    \includegraphics[scale=0.7]{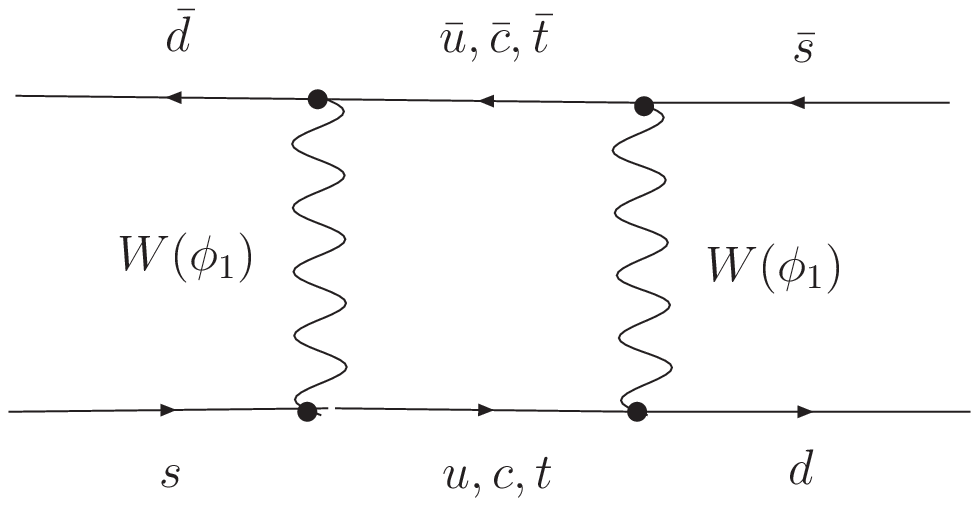}
    \includegraphics[scale=0.65]{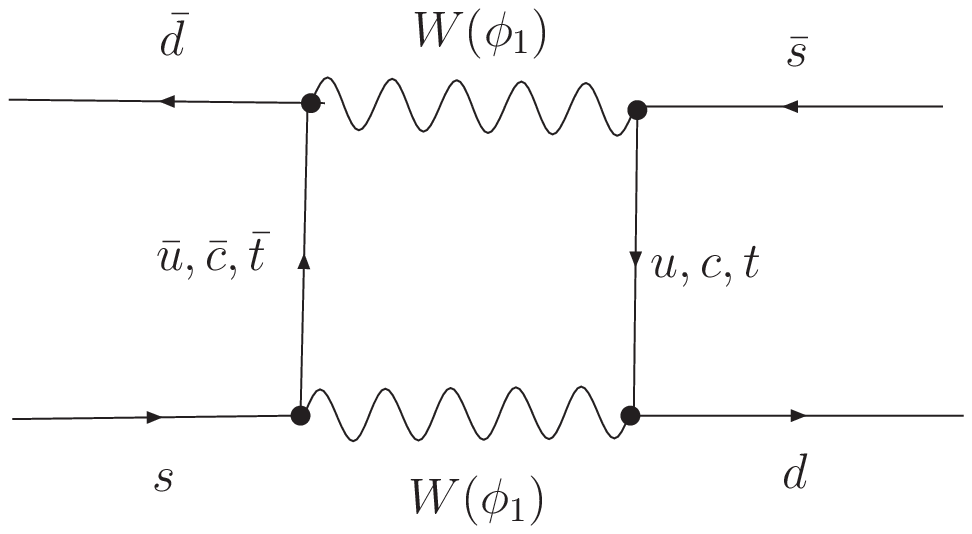}
\end{minipage}
\end{figure}

\begin{figure}[t] \caption{Higgs exchange diagrams for $K^0-\bar{K}^0$ effective Hamiltonian. } \label{fig-h-K-box}
\centering\begin{minipage}[t]{\textwidth}
    \includegraphics[scale=0.7]{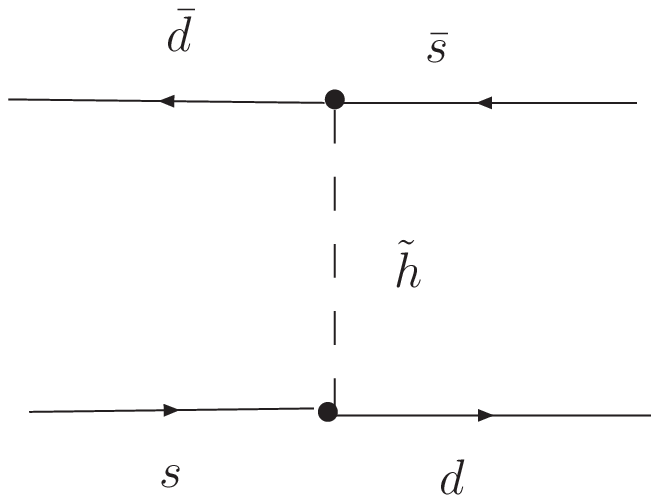}\hspace*{1cm}\includegraphics[scale=0.85]{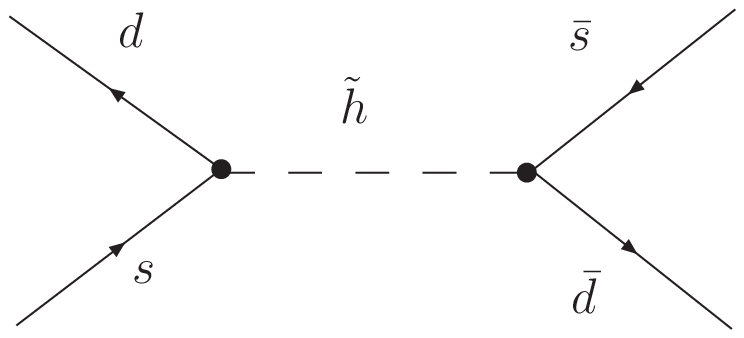}
\end{minipage}
\end{figure}

Once there exists $W'$ boson, there may be low energy
phenomenological constraints from $K^0-\bar{K}^0$,
$B_d^0-\bar{B}_d^0$ and $B_s^0-\bar{B}_s^0$ system. In most cases
$W'$ will generate extra Feynman box diagrams which contribute to
mass differences in $K^0-\bar{K}^0$, $B_d^0-\bar{B}_d^0$,
$B_s^0-\bar{B}_s^0$ system and corresponding CP violation
parameters. These mixings are described by a effective Hamiltonian
which is composed of four parts:
\begin{eqnarray}
H_\mathrm{eff}=H^{WW}_\mathrm{eff}+H^{W'W'}_\mathrm{eff}+H^{WW'}_\mathrm{eff}+H^{h^0}_\mathrm{eff}\label{HeffDef}
\end{eqnarray}
 The effective Hamilton $H_\mathrm{eff}$ is model dependent.
 We take the $K^0-\bar{K}^0$ system in LR and LP models as an example,
  other models and $B^0-\bar{B}^0$ system can be given in the similar way. The $WW$ box diagram for
$K^0-\bar{K}^0$ system is plotted in Fig.\ref{fig-WW-K-box}. Let
$M^{XY}$ be the amplitudes of the diagram mediated by particles $X$
and $Y$which may be gauge bosons $W,W'$ and goldstone bosons
$\phi_1,\phi_2$. $H^{WW}_\mathrm{eff}$ can be further decomposed
into $H^{WW}_\mathrm{eff}=\frac{1}{2}(M^{WW}+M^{W\phi_1}+M^{\phi_1W}
+M^{\phi_1\phi_1})+\mathrm{h.c.}$
$W'W'$ box diagram part $H^{W'W'}_\mathrm{eff}$ can be obtained from
$H^{WW}_\mathrm{eff}$ by
$H^{W'W'}_\mathrm{eff}=H^{WW}_\mathrm{eff}|_{W\rightarrow
W',~1\rightarrow 2}$. Similarly $WW'$ box diagram part
$H^{WW'}_\mathrm{eff}$ is
$H^{WW'}_\mathrm{eff}=M^{WW'}+M^{W\phi_2}+M^{\phi_1W'}
+M^{\phi_1\phi_2}+\mathrm{h.c.}$. $H^{h^0}_\mathrm{eff}$ is the part
of effective Hamiltonian arises from the flavor changing Yukawa
coupling via neutral Higgs exchange at tree level. The corresponding
Feynman diagrams is plot in Fig.\ref{fig-h-K-box} for
$K^0-\bar{K}^0$ system.

The $W'$ dependent part of $H_\mathrm{eff}$ introduced in
(\ref{HeffDef}) is also model dependent. LR and LP models are main
cases we are going to discuss in which $H^{W'W'}_\mathrm{eff}$ is
usually neglected due to existence of a suppression factor
$(M_W/M_{W'})^4$. SM calculation shows that just SM effect in
$H^{WW}_\mathrm{eff}$ itself can already match experiment data.
Therefore the constraints left is that either non SM effects in
$H^{WW}_\mathrm{eff}$, $H^{WW'}_\mathrm{eff}$ and
$H^{h^0}_\mathrm{eff}$ are all small in values separately or they
cancel each other. The cancelation will demand detailed model
arrangements which need fine tuning such as introducing in theory
the second bi-doublet higgs discussed in Ref.\cite{YLWu07}.  In this
work we do not consider this special fine tuning situation and only
limit us in the case that all non SM effects in
$H^{WW}_\mathrm{eff}$, $H^{WW'}_\mathrm{eff}$ and
$H^{h^0}_\mathrm{eff}$ are  small separately in values. This choice
is in accordance with the approximation that only dimension three
and four matter part operators are included in our calculation. If
we consider more higher dimension operators, dimension six four
quark operators such as $\bar{d_R}s_L\bar{d_L}s_R$  will contribute
to $H_\mathrm{eff}$ as a contact term.  This will raise the
possibility that using four quark operator contributions to cancel
 non SM effects in $H^{WW}_\mathrm{eff}$, $H^{WW'}_\mathrm{eff}$ and
$H^{h^0}_\mathrm{eff}$. This four quark operator can be seen as
remnant of exchanging some more heavier unknown particles and the
coupling of the operator is proportional to inverse of heavy
particle mass square, like traditional Fermi weak interaction theory
induced by exchanging electroweak gauge bosons.  In our treatment we
have ignored possible cancelations among operators of different
classes. If we generalize this treatment to higher dimension
operators,  the cancelations among contributions of four quark
operators and $W,W',h^0$ to $H_\mathrm{eff}$ are not allowed. This
implies the effective coupling in front of corresponding four quark
operator must be small which will improve the convergence of our
expansion and we can safely drop out four quark operator in our
first order approximation. This is the discussion for LR and LP
models. The situation in NU model is similar as in LR and LP models,
except there exists explicit non-universality term in (\ref{ABdef}).
For other models, HP and FP models are irrelevant, since in these
models $W'$ does not couple to light ordinary quarks if we ignore
small mixing between $W$ and $W'$. Then there are approximately no
$H^{W'W'}_\mathrm{eff}$ and $H^{WW'}_\mathrm{eff}$ terms in
(\ref{HeffDef}). The only constraint for these models is the value
of $H^{h^0}_\mathrm{eff}$ must be small. In the case of UN model,
$W$ does not couple to ordinary quark if we ignore  mixing between
$W$ and $W'$. The role of $W$ is replaced by $W'$. Considering the
facts that $H^{W'W'}_\mathrm{eff}$ is much smaller than
$H^{WW}_\mathrm{eff}$ in value due to suppression factor and there
is no $H^{WW}_\mathrm{eff}$ and $H^{WW'}_\mathrm{eff}$ terms, the
value of $H^{h^0}_\mathrm{eff}$ in this case can be larger than that
of HP and FP models. Since the constraints for UN model from mass
 differences in $K^0-\bar{K}^0$, $B_d^0-\bar{B}_d^0$,
$B_s^0-\bar{B}_s^0$ system and corresponding CP violation parameters
 are relatively weak, we skip the discussion of this situation.
 Combining above discussions together, for the $W'$ dependent part of
$H_\mathrm{eff}$, we only need to discuss two situations: one is LR
and LP models, the other is NU model.

 In performing detailed computations for box diagrams, we choose
Feynman gauge and take the masses and four-momenta of the external
legs to be zero ($m_d=m_s=0$) thus the internal lines carry the same
momentum. Detailed calculations give following amplitudes for the
diagram mediated by $X-Y, X, Y=W,W'$,
\begin{eqnarray}
M^{XY}&=&(\frac{\sqrt{2}g_1^2}{8M_{X}^2})^2\frac{M_{X}^2}{2\pi^2g_1^4}\beta\sum_{\alpha,\beta}
\sqrt{x_\alpha x_\beta}[4I_1(x_\alpha,x_\beta,\beta)]\bar{d}[(A_X^{\alpha s}A_Y^{\alpha d\dag}-B_X^{\alpha s}
B_Y^{\alpha d\dag})\nonumber\\
&&+\gamma_5(B_X^{\alpha s}A_Y^{\alpha d\dag}-A_X^{\alpha
s}B_Y^{\alpha d\dag})]s \otimes\bar{d}[(A_Y^{\beta s}A_X^{\beta
d\dag}-B_Y^{\beta s}B_X^{\beta d\dag})
+\gamma_5(B_Y^{\beta s}A_X^{\beta d\dag}-A_Y^{\beta s}B_X^{\beta d\dag})]s\nonumber\\
&&+(\frac{\sqrt{2}g_1^2}{8M_{X}^2})^2\frac{M_{X}^2}{2\pi^2g_1^4}\beta\sum_{\alpha,\beta}
\{[\frac{1}{4}I_2(x_\alpha,x_\beta,\beta)]\bigg[10\bar{d}\gamma_\mu[(A_X^{\alpha s}A_Y^{\alpha d\dag}
+B_X^{\alpha s}B_Y^{\alpha d\dag})\nonumber\\
&&+\gamma_5(A_X^{\alpha s}B_Y^{\alpha d\dag}+B_X^{\alpha
s}A_Y^{\alpha d\dag})]s \otimes\bar{d}\gamma^\mu[(A_Y^{\beta
s}A_X^{\beta d\dag}+B_Y^{\beta s}B_X^{\beta d\dag})
 +\gamma_5(A_Y^{\beta s}B_X^{\beta d\dag}+B_Y^{\beta s}A_X^{\beta d\dag})]s\nonumber\\
&&-6\bar{d}\gamma_\mu[\gamma_5(A_X^{\alpha s}A_Y^{\alpha
d\dag}+B_X^{\alpha s}B_Y^{\alpha d\dag})
+(A_X^{\alpha s}B_Y^{\alpha d\dag}+B_X^{\alpha s}A_Y^{\alpha d\dag})]s\nonumber\\
&&\otimes\bar{d}\gamma^\mu[\gamma_5(A_Y^{\beta s}A_X^{\beta
d\dag}+B_Y^{\beta s}B_X^{\beta d\dag})
 +(A_Y^{\beta s}B_X^{\beta d\dag}+B_Y^{\beta s}A_X^{\beta d\dag})]\bigg]s\}
\end{eqnarray}
where $x_\alpha=m_\alpha^2/M_{X}^2$ and $\beta=M_{X}^2/M_{Y}^2$, and
$A_W^{\alpha\beta}$, $A_{W'}^{\alpha\beta}$, $B_W^{\alpha\beta}$,
$B_{W'}^{\alpha\beta}$ are defined in (\ref{ABdef}),
\begin{eqnarray}
I_1(x_\alpha,x_\beta,\beta)&=&\frac{x_\alpha\ln
x_\alpha}{(1-x_\alpha)(1-x_\alpha\beta)(x_\alpha-x_\beta)}+(\alpha\leftrightarrow
\beta)
-\frac{\beta\ln \beta}{(1-\beta)(1-x_\alpha\beta)(1-x_\beta\beta)},\nonumber\\
I_2(x_\alpha,x_\beta,\beta)&=&\frac{x_\alpha^2\ln
x_\alpha}{(1-x_\alpha)(1-x_\alpha\beta)(x_\alpha-x_\beta)}+(\alpha\leftrightarrow
\beta) -\frac{\ln
\beta}{(1-\beta)(1-x_\alpha\beta)(1-x_\beta\beta)}\textcolor
 {red}{.}~~~
\end{eqnarray}
The amplitude of the diagram mediated by $X^+-\phi^-_n, X=W,W', n=1,
2$ is:
\begin{eqnarray}
\label{MWphi}
M^{X\phi_n}&=&(\frac{\sqrt{2}g_1^2}{8M_{X}^2})^2\frac{M_{X}^2}{2\pi^2g_1^4}\beta\sum_{\alpha,\beta}
\sqrt{x_\alpha
x_\beta}[I_1(x_\alpha,x_\beta,\beta)]\bar{d}\gamma_\nu[(A_X^{\alpha
s}A_n^{\alpha d\dag}-B_X^{\alpha s}B_n^{\alpha d\dag})
\\
&&+\gamma_5(B_X^{\alpha s}A_n^{\alpha d\dag}-A_X^{\alpha
s}B_n^{\alpha d\dag})]s \otimes\bar{d}\gamma^\nu[(A_n^{\beta
s}A_X^{\beta d\dag}+B_n^{\beta s}B_X^{\beta d\dag})
+\gamma_5(B_n^{\beta s}A_X^{\beta d\dag}+A_n^{\beta s}B_X^{\beta d\dag})]s\nonumber\\
&&+(\frac{\sqrt{2}g_1^2}{8M_{X}^2})^2\frac{M_{X}^2}{2\pi^2g_1^4}\beta\sum_{\alpha,\beta}
[I_2(x_\alpha,x_\beta,\beta)]\bar{d}[(A_X^{\alpha s}A_n^{\alpha
d\dag}+B_X^{\alpha s}B_n^{\alpha d\dag})
\nonumber\\
&&+\gamma_5(A_X^{\alpha s}B_n^{\alpha d\dag}+B_X^{\alpha
s}A_n^{\alpha d\dag})]s\otimes\bar{d}[(A_n^{\beta s}A_X^{\beta
d\dag}-B_n^{\beta s}B_X^{\beta d\dag})
 +\gamma_5(-A_n^{\beta s}B_X^{\beta d\dag}+B_n^{\beta s}A_X^{\beta d\dag})]s\nonumber
\end{eqnarray}
The amplitude of the diagram mediated by $\phi_m^+-Y^-, m=1, 2,
Y=W,W'$ is,
\begin{eqnarray}
\label{MphiW} M^{\phi_mY}
&=&(\frac{\sqrt{2}g_1^2}{8M_{X}^2})^2\frac{M_{X}^2}{2\pi^2g_1^4}\beta\sum_{\alpha,\beta}
\sqrt{x_\alpha
x_\beta}[I_1(x_\alpha,x_\beta,\beta)]\bar{d}\gamma_\mu[(A_m^{\alpha
s}A_Y^{\alpha d\dag}+B_m^{\alpha s}B_Y^{\alpha d\dag})
\\
&&+\gamma_5(B_m^{\alpha s}A_Y^{\alpha d\dag}+A_m^{\alpha
s}B_Y^{\alpha d\dag})]s \otimes\bar{d}[\gamma^\mu(A_Y^{\beta
s}A_m^{\beta d\dag}-B_Y^{\beta s}B_m^{\beta d\dag})
+\gamma_5(B_Y^{\beta s}A_m^{\beta d\dag}-A_Y^{\beta s}B_m^{\beta d\dag})]s\nonumber\\
&&+(\frac{\sqrt{2}g_1^2}{8M_{X}^2})^2\frac{M_{X}^2}{2\pi^2g_1^4}\beta\sum_{\alpha,\beta}
\{[\frac{1}{4}I_2(x_\alpha,x_\beta,\beta)]\bar{d}[(A_m^{\alpha
s}A_Y^{\alpha d\dag}-B_m^{\alpha s}B_Y^{\alpha d\dag})
\nonumber\\
&&+\gamma_5(-A_m^{\alpha s}B_Y^{\alpha d\dag}+B_m^{\alpha
s}A_Y^{\alpha d\dag})]s\otimes\bar{d}[(A_Y^{\beta s}A_m^{\beta
d\dag}+B_Y^{\beta s}B_m^{\beta d\dag})
 +\gamma_5(A_Y^{\beta s}B_m^{\beta d\dag}+B_Y^{\beta s}A_m^{\beta d\dag})]s\}\nonumber
\end{eqnarray}
The amplitude of the diagram mediated by $\phi_m^+-\phi^-_n, m, n=1,
2$ is,
\begin{eqnarray}
\label{phiphi}
M^{\phi_m\phi_n}&=&(\frac{\sqrt{2}g_1^2}{8M_{X}^2})^2\frac{M_{X}^2}{2\pi^2g_1^4}\beta\sum_{\alpha,\beta}
\sqrt{x_\alpha
x_\beta}[I_1(x_\alpha,x_\beta,\beta)]\bar{d}[(A_m^{\alpha
s}A_n^{\alpha d\dag}+B_m^{\alpha s})B_n^{\alpha d\dag})
\\
&&+\gamma_5(B_m^{\alpha s}A_n^{\alpha d\dag}+A_m^{\alpha
s}B_n^{\alpha d\dag})]s\otimes\bar{d}[(A_n^{\beta s}A_m^{\beta
d\dag}+B_n^{\beta s}B_m^{\beta d\dag})
+\gamma_5(B_n^{\beta s}A_m^{\beta d\dag}+A_n^{\beta s}B_m^{\beta d\dag})]s\nonumber\\
&&+(\frac{\sqrt{2}g_1^2}{8M_X^2})^2\frac{M_{X}^2}{2\pi^2g_1^4}\beta\sum_{\alpha,\beta}
\{[\frac{1}{4}I_2(x_\alpha,x_\beta,\beta)]\bar{d}\gamma_\rho[(A_m^{\alpha
s}A_n^{\alpha d\dag}-B_m^{\alpha s}B_n^{\alpha d\dag})
+\gamma_5(-A_m^{\alpha s}B_n^{\alpha d\dag}\nonumber\\
&&+B_m^{\alpha s}A_n^{\alpha
d\dag})]s\otimes\bar{d}\gamma^\rho[(A_n^{\beta s}A_m^{\beta
d\dag}-B_n^{\beta s}B_m^{\beta d\dag})
 +\gamma_5(-A_n^{\beta s}B_m^{\beta d\dag}+B_n^{\beta s}A_m^{\beta d\dag})]s)\}\nonumber
\end{eqnarray}
Since the mixing angle $\zeta$ is expected to be small, for
simplicity in the following we take $\zeta=0$.

For LR and LP models, we can ignore the generation index $\alpha$
dependence in all $\Delta$'s appeared in (\ref{ABdef}), then the
rotation matrices $V^u$ and $V^{d\dag}$ can meet together forming
CKM matrices. We introduce CKM factors
$\lambda_\alpha^{LR}(K)=V^{\mathrm{CKM},u_\alpha
s}_LV^{\mathrm{CKM},u_\alpha d*}_{R}$ for $K^0-\bar{K}^0$ system,
$\lambda_\alpha^{LR}(B_q)=V^{\mathrm{CKM},u_\alpha
b}_LV^{\mathrm{CKM},u_\alpha q*}_{R}$ for $B_q^0-\bar{B}_q^0$
system, etc. By taking $m_u=0$ and using the relation
$\lambda_u+\lambda_c+\lambda_t=0$, ignoring the higher order of
$\Delta_{2,2}, \Delta_{1,2}$ (we have dropped out their generation
indices) and accurate to the order linear in
$\beta=M_{W}^2/M_{W'}^2$, we obtain
 \begin{eqnarray}
H^{WW}_\mathrm{eff}\!&=&\frac{G_F^2M_W^2}{16\pi^2}\Delta_{1,1}^2\times
\left\{\begin{array}{ll}f_{LL}(K)\bar{d}\gamma^\mu(1\!-\!\gamma_5)s
\otimes\bar{d}\gamma_\mu(1\!-\!\gamma_5)s&~~~K^0-\bar{K}^0~\mathrm{system}\\
f_{LL}(B_q)\bar{q}\gamma^\mu(1\!-\!\gamma_5)b
\otimes\bar{q}\gamma_\mu(1\!+\!\gamma_5)b&~~~B^0_q-B^0_{\bar{q}}~\mathrm{system}\end{array}\right.+\textrm{h.c.}~~~~~\\
H^{WW'}_\mathrm{eff}\!&=&\frac{G_F^2M_W^2}{16\pi^2}2\beta\Delta_{2,1}^2\frac{g_2^2}{g_1^2}\times
\left\{\begin{array}{ll} f_{LR}(K)\bar{d}(1\!-\!\gamma_5)s
\otimes\bar{d}(1\!+\!\gamma_5)s&~~~K^0-\bar{K}^0~\mathrm{system}\\
f_{LR}(B_q)\bar{q}(1\!-\!\gamma_5)b
\otimes\bar{q}(1\!+\!\gamma_5)b&~~~B^0_q-B^0_{\bar{q}}~\mathrm{system}\end{array}\right.+\textrm{h.c.}~~~~~
\end{eqnarray}
where $q=d,s$ and
\begin{eqnarray}
f_{LL}(K)&=&(\lambda_c^{LL}(K))^2\eta_{cc}\tilde{S}_0(x_c)+(\lambda_t^{LL}(K))^2\eta_{tt}\tilde{S}_0(x_t)+
2\lambda_c^{LL}(K)\lambda_t^{LL}(K)\eta_{ct}\tilde{S}_0(x_c, x_t)\label{fLLK}\\
f_{LL}(B_q)&=&(\lambda_t^{LL}(B_q))^2\eta_{B_q}\tilde{S}_0(x_t)\label{fLLB}\\
\tilde{S}_0(x)&=&
\frac{x}{(1-x)^2}\bigg[\Delta_{1,1}^4+\frac{4\Delta_{1,1}^4-16\Delta_{L,1}^2+1}{4}x+\frac{x^2}{4}
+\frac{2x\ln
x}{1-x}(\Delta_{1,1}^4-\Delta_{1,1}^2-\frac{4\Delta_{1,1}^2-1}{4}x)\bigg]\\
\tilde{S}_0(x_c, x_t)&=&
x_cx_t\bigg[\frac{1}{(1-x_c)(1-x_t)}(\Delta_{1,1}^4-2\Delta_{1,1}^2+\frac{1}{4})+\frac{\ln
x_t}{(x_t-x_c)(1-x_t)^2}
\big(\Delta_{1,1}^4-2\Delta_{1,1}^2x_t+\frac{x_t^2}{4}\big)\nonumber\\
&&+\frac{\ln x_c}{(x_c-x_t)(1-x_c)^2}
\big(\Delta_{1,1}^4-2\Delta_{1,1}^2x_c+\frac{x_c^2}{4}\big)\bigg].\\
f_{LR}(K)&=&\lambda_c^{LR}(K)\lambda_c^{RL}(K)S_{cc}(K)+\lambda_t^{LR}(K)\lambda_t^{RL}(K)S_{tt}(K)\nonumber\\
&&+(\lambda_c^{LR}(K)\lambda_t^{RL}(K)
+\lambda_t^{LR}(K)\lambda_c^{RL}(K))S_{ct}(K)\label{fLRK}\\
f_{LR}(B_q)&=&f_{LR}(K)\bigg|_{K\rightarrow
B_q}\label{fLRB}\\
S_{cc}(K)&=&\frac{x_c}{(1-x_c)^2}[(4\Delta_{1,1}^2\eta_1^{LR}(K)-
x_c\eta_2^{LR}(K))(1-x_c)\nonumber\\
&&+(4\Delta_{1,1}^2\eta_1^{LR}(K)-2x_c\eta_2^{LR}(K)
+x_c^2\eta_2^{LR}(K))\ln x_c+\eta_2^{LR}(K)(1-x_c)^2\ln \beta]\\
S_{tt}(K)&=&S_{cc}(K)\bigg|_{c\rightarrow t}\\
S_{ct}(K)&=&\frac{\sqrt{x_cx_t}}{(1-x_c)(1-x_t)(x_t-x_c)}
[x_t(4\Delta_{1,1}^2\eta_1^{LR}(K)-x_t\eta_2^{LR}(K))(1-x_c) \ln x_t\\
&&-x_c(4\Delta_{1,1}^2\eta_1^{LR}(K)-x_c\eta_2^{LR}(K))(1-x_t) \ln
x_c +\eta_2^{LR}(K)(1-x_c)(1-x_t)(x_t-x_c)\ln
\beta]\nonumber\\
S_{cc}(B_q)&=&S_{cc}(K)\bigg|_{K\rightarrow B_q},~~
S_{tt}(B_q)=S_{tt}(K)\bigg|_{K\rightarrow B_q},~~
S_{ct}(B_q)=S_{ct}(K)\bigg|_{K\rightarrow B_q}
\end{eqnarray}
with $x_c=m_c^2/M_W^2$,~$x_t=m_t^2/M_W^2$, $\beta=M_W^2/M_{W'}^{2}$.
The next-to-leading-order QCD short-distance corrections are
$\eta_{cc}=1.38\pm0.20, \eta_{ct}= 0.47\pm0.04
,\eta_{tt}=0.57\pm0.01$\cite{eta}\cite{eta1}, $\eta_{B_d}=0.551,
\eta_{B_s}=0.837$\cite{etaB}. The QCD corrections are
$\eta_1^{LR}(K)=1.4, \eta_2^{LR}(K)= 1.17$ for
$\Lambda_{QCD}=0.2\mathrm{GeV}$\cite{etaQCD} and
$\eta_1(B_q)\simeq1.8, \eta_2(B_q)\simeq1.7$ at scale
$m_{b}$\cite{etaBQCD}.

The matrix elements are given by
\begin{eqnarray}
&&\langle K^0|\bar{d}\gamma^\mu(1\pm\gamma_5)s\otimes
\bar{d}\gamma_\mu(1\pm\gamma_5)s|\bar{K}^0\rangle=\frac{4}{3}f_K^2m_KB_K,\\
&&\langle B_q^0|\bar{d}\gamma^\mu(1\pm\gamma_5)s\otimes
\bar{d}\gamma_\mu(1\pm\gamma_5)s|\bar{B}_q^0\rangle=\frac{4}{3}f_{B_q}^2m_{B_q}B_{B_q},\\
&&\langle K^0|\bar{d}(1-\gamma_5)s\otimes
\bar{d}(1+\gamma_5)s|\bar{K}^0\rangle=\frac{1}{2m_K}[\frac{1}{3}+\frac{2m_K^2}{(m_s+m_d)^2}]f_K^2m_K^2B_K^S,\\
&&\langle B^0|\bar{q}(1-\gamma_5)b\otimes
\bar{q}(1+\gamma_5)b|\bar{B}^0\rangle=\frac{1}{2m_{B_q}}[\frac{1}{3}+\frac{2m_{B_q}^2}{m_b^2}]
f_{B_q}^2m_{B_q}^2B_{B_q}^S\hspace{1cm}q=d,s.
\end{eqnarray}
The decay constant for neutral K meson is given by
$f_K/f_\pi=1.198\pm0.003 $\cite{fK1,fK2} with $f_\pi=(130\pm5)
\times10^{-3}\mathrm{GeV}$\cite{PDG06} and
 the bag parameter is $B_K=0.79\pm0.04\pm0.09$\cite{latQCDBK}. For $B_d$ and
$B_s$ mesons,  $f_{B_d}\sqrt{B_{B_d}}=0.220\pm
0.040\mathrm{GeV}$\cite{etaB} and
$f_{B_s}\sqrt{B_{B_s}}=0.221\mathrm{GeV}$\cite{bagBd2} . The bag
parameter from QCD sum rule gives
$B_{B_q}^S/B_{B_q}=1.2\pm0.2$\cite{Anatomy}.

For NU models, we must consider the generation dependence $\alpha$
in all $\Delta$'s appeared in (\ref{ABdef}). To simplify the
expressions, we denote the CKM factors as $V^{\alpha \beta}_L\equiv
V^{\mathrm{CKM},u_\alpha b_\beta}_L, $ and
\begin{eqnarray}
V_{L,11}^{\alpha\beta}=\sum_{\alpha'}V_L^{u,\alpha\alpha'}\Delta_{1,1,\alpha'}V_L^{d\dag,\alpha'\beta},~~
V_{L,12}^{\alpha\beta}=\sum_{\alpha'}V_L^{u,\alpha\alpha'}\Delta_{1,2,\alpha'}V_L^{d\dag,\alpha'\beta}\label{CKMfactor}
\end{eqnarray} etc.  Ignoring the higher order of $\Delta_{2,1,\alpha}$
and $\Delta_{2,2,\alpha}$, After tedious calculations, we get the
effective Hamilton for $K^0-\bar{K}^0$ system in NU models as
follows:
 \begin{eqnarray}
H^{WW}_\mathrm{eff}\!&=&\frac{G_F^2M_W^2}{16\pi^2}\bigg\{
[\sum_{\alpha,\beta=u,c,t}(V_{L,11}^{\alpha s}V_{L,11}^{\alpha
d*})(V_{L,11}^{\beta s}V_{L,11}^{\beta d*})
+\frac{1}{4}\sum_{\alpha,\beta=u,c}x_\alpha x_\beta
(V_{L,11}^{\alpha s}V_{L}^{\alpha d*})
(V_{L}^{\beta s}V_{L,11}^{\beta d*})]\nonumber\\
&&\times I_2(x_\alpha,x_\beta,1)-2\sum_{\alpha,\beta=u,c}x_\alpha
x_\beta (V_{L}^{\alpha s}V_{L}^{\alpha d*})(V_{L}^{\beta
s}V_{L}^{\beta
d*})I_1(x_\alpha,x_\beta,1)\bigg\}\nonumber\\
&&\times\bar{d}\gamma_\mu(1-\gamma_5)s
\otimes\bar{d}\gamma^\mu(1-\gamma_5)s+\textrm{h.c.}~~~~~\\
H^{W'W'}_\mathrm{eff}\!&=&\frac{G_F^2M_W^2}{16\pi^2}\bigg\{
\sum_{\alpha,\beta=u,c,t}(V_{L,12}^{\alpha s}V_{L,12}^{\alpha
d*})(V_{L,12}^{\beta s}V_{L,12}^{\beta
d*})I_2(x_\alpha,x_\beta,1)\nonumber\\
&&+\frac{1}{4}\beta x_t^2 (V_{L,12}^{t s}V_{L}^{t d*})(V_{L}^{t
s}V_{L,12}^{t d*})I_2(x_t,1) -2\beta^2x_t^2 (V_{L}^{t s}V_{L}^{t
d*})^2I_1(x_t,1)\bigg\}\nonumber\\
&&\times\bar{d}\gamma_\mu(1-\gamma_5)s
\otimes\bar{d}\gamma^\mu(1-\gamma_5)s+\textrm{h.c.}~~~~~\\
H^{WW'}_\mathrm{eff}\!&=&\frac{G_F^2M_W^2}{8\pi^2(g_1^2/g_2^2)}\beta\bigg\{\sum_{\alpha,\beta=u,c,t}[(V_{L,11}^{\alpha
s}V_{L,12}^{\alpha d*}+V_{L,12}^{\alpha s}V_{L,11}^{\alpha
d*})(V_{L,12}^{\beta s}V_{L,11}^{\beta d*}
+V_{L,11}^{\beta s}V_{L,12}^{\beta d*})]I_2(x_\alpha,x_\beta,\beta)\nonumber\\
 &&-x_t^2\beta[(V_{L,11}^{ts}V_{L}^{td*})(V_{L}^{t
s}V_{L,11}^{td*})]I_1(x_t,\beta) - \sum_{\alpha,\beta=u,c}(x_\alpha
x_\beta)^{3/2}(V_{L}^{\alpha s}V_{L,12}^{\alpha d*})
(V_{L,12}^{\beta s}V_{L}^{\beta d*})]\nonumber\\
&&\times
I_1(x_\alpha,x_\beta,\beta)\bigg\}\bar{d}\gamma_\mu(1-\gamma_5)s
\otimes\bar{d}\gamma^\mu(1-\gamma_5)s+\textrm{h.c.}~~~~~.
\end{eqnarray}
The effective Hamilton for $B^0-\bar{B}^0$ system can be obtained
through the same procedure.

 %%%%%%%%%%%%%%%%%%%%%%%%%%%%%%%%%%%%%%%%%%%%%%%%%%%%%%%%%%%%%
\section{Constraints from neutral K and B system for LR and LP models}

In this section, we will concentrate on the constraints on our EWCL
from mass differences in $K^0-\bar{K}^0$, $B_d^0-\bar{B}_d^0$,
$B_s^0-\bar{B}_s^0$ systems and indirect CP violation parameter
$|\epsilon_K|$, mainly for LR and LP models. Due to complexity of
CKM factors introduced in (\ref{CKMfactor}) for NU models, we will
leave the investigation for NU model elsewhere.

\begin{figure}[t]
\caption{Pure $W$ contribution to $K^0-\bar{K}^0$,
$B_d^0-\bar{B}_d^0$, $B_s^0-\bar{B}_s^0$ systems and indirect CP
violation in K mesons. Anomalous coupling $\Delta_{1,1}=1$
corresponds to SM results which are explicitly written down in
brackets. } \label{fig-WW} \centering
\begin{minipage}[t]{\textwidth}
    \includegraphics[scale=1]{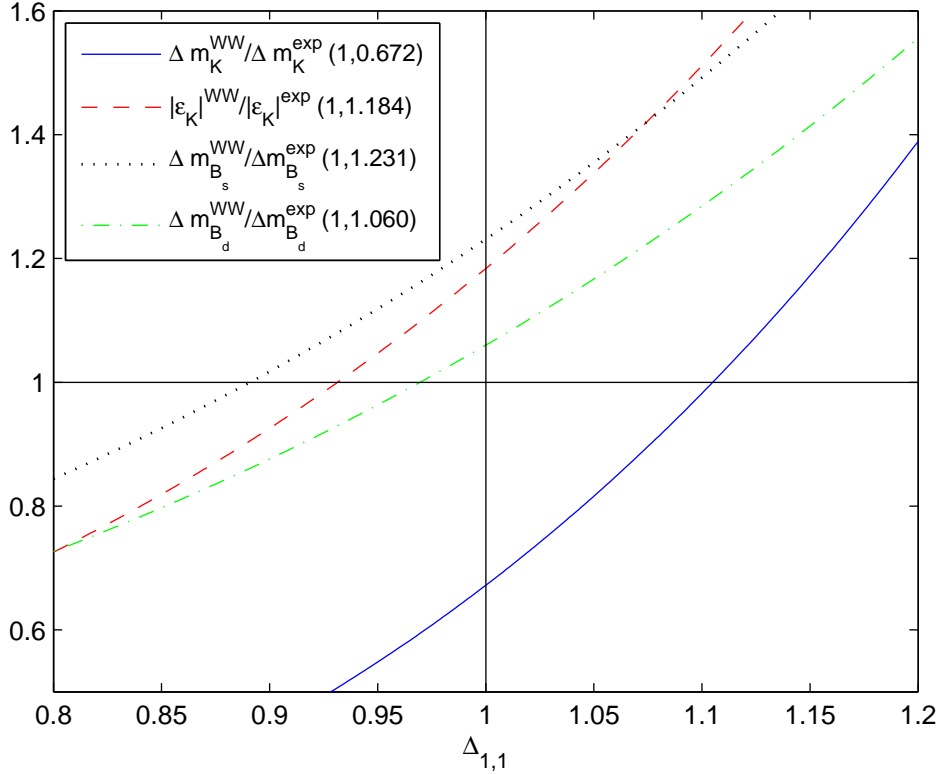}
\end{minipage}
\end{figure}

The mass differences in $K^0-\bar{K}^0$, $B_d^0-\bar{B}_d^0$ and
$B_s^0-\bar{B}_s^0$ systems are determined by
\begin{eqnarray}
\Delta m_{K}=2\mathrm{Re}\langle
K^0|H_\mathrm{eff}|\bar{K}^0\rangle\hspace{2cm}\Delta
m_{B_q}=2|\langle
B^0_q|H_\mathrm{eff}|\bar{B}_q^0\rangle|\hspace{0.5cm}q=d,s~~~~
\end{eqnarray}
and the indirect CP violation in K mesons can be expressed as
\begin{eqnarray}
|\epsilon_{K}|=\frac{1}{2\sqrt{2}} \bigg(\frac{\mathrm{Im}\langle
K^0|H_\mathrm{eff}|\bar{K}^0\rangle}{\mathrm{Re}\langle
K^0|H_\mathrm{eff}|\bar{K}^0\rangle}+2\xi_0\bigg)
\approx\frac{\mathrm{Im}\langle
K^0|H_\mathrm{eff}|\bar{K}^0\rangle}{\sqrt{2}\Delta m_{K}}
\end{eqnarray}
where $\xi_0$ is the weak phase of $K\rightarrow\pi\pi$ decay
amplitude with isospin zero. The pure $W$ contribution to mass
differences in $K^0-\bar{K}^0$, $B_d^0-\bar{B}_d^0$ and
$B_s^0-\bar{B}_s^0$ systems and indirect CP violation in K mesons as
functions of anomalous coupling $\Delta_{1,1}$ introduced in
(\ref{Delta11def}) is shown in Fig.\ref{fig-WW}. From (\ref{LCC}),
we know that $\Delta_{1,1}$ characterize the anomalous coupling for
charge current,  it can deviate from $1$ very much and therefore we
choose region [0.8,1.2] for $\Delta_{1,1}$ as horizontal coordinate
in Fig.\ref{fig-WW}.
In numerical calculation, the input parameters are taken from
particle data group \cite{PDG06} except those explicitly labeled.
$$\begin{array}{ll} G_F=1.16637(1)\times
10^{-5}\mathrm{GeV}^{-2},&M_W=80.403\pm0.029\mathrm{GeV},\\
m_{K}=(497.648\pm0.022)\times10^{-3}\mathrm{GeV},&\Delta
m_{K}^{exp}=(3.483\pm0.006)\times10^{-15}\mathrm{GeV},\\
m_{d}=5\times10^{-3}\mathrm{GeV},&m_{s}=95\times10^{-3}\mathrm{GeV},\\
m_{c}=1.25\pm0.09\mathrm{GeV},&m_t=174.2\pm3.3 \mathrm{GeV},\\
m_{B_d}=(5279.4\pm0.5)\times10^{-3}\mathrm{GeV},&\Delta
 m_{B_d}^{exp}=(3.337\pm0.003)\times10^{-13}\mathrm{GeV},\\
 m_{B_s}=(5367.5\pm1.8)\times10^{-3}\mathrm{GeV},&\Delta
 m_{B_s}^{exp}=(1.17\pm0.003)\times10^{-11}\mathrm{GeV}.
\end{array}
$$
The CKM elements are given in terms of Wolfenstein parameterization
\cite{PDG06}:
$$\lambda=0.2272, ~A=0.818,~
\bar{\rho}=0.221, ~\bar{\eta}=0.340,$$ with the relations
$\displaystyle s_{13}e^{i\delta}=(V_L^{ub})^*=A\lambda^3(\rho+i\eta)
=\frac{A\lambda^3(\bar{\rho}+i\bar{\eta})\sqrt{1-A^2\lambda^4}}{\sqrt{1-\lambda^2}[1-A^2\lambda^4(\bar{\rho}+i\bar{\eta})]}$.

 We find that for $\Delta m_K$, $|\epsilon_K|$, $\Delta m_{B_d}$ and $\Delta
 m_{B_s}$, SM theoretical results ($\Delta_{1,1}=1$) match to experiment values
 with error $33\%$, $18\%$, $6\%$ and $23\%$ respectively. These
 errors are expected from uncertainty of matrix
elements and long distance contributions \cite{SMcontribution}. New
physics contributions must hide in these errors.

 Up to the order liner in $\beta$, $W'$ contribution to
mass differences in $K^0-\bar{K}^0$, $B_d^0-\bar{B}_d^0$,
$B_s^0-\bar{B}_s^0$ systems and indirect CP violation in K mesons
are
\begin{eqnarray}
&&\hspace{-0.5cm}\Delta m_{K}^{WW'}\!\!\!\!=2\mathrm{Re}\!\langle
K^0|H^{WW'}_\mathrm{eff}|\bar{K}^0\rangle =\frac{G_F^2M_W^2f_K^2m_K
B_K^S\beta}{4\pi^2}\Delta_{2,1}^2\frac{g_2^2}{g_1^2}\mathrm{Re}(f_{LR}(K))
[\frac{1}{6}\!+\!\frac{m_K^2}{(m_s\!+\!m_d)^2}]~~~\label{DeltamK}\\
&&\hspace{-0.5cm}\Delta m_{B_q}^{WW'}\!\!\!\!=2|\langle
B^0_q|H^{WW'}_\mathrm{eff}|\bar{B}_q^0\rangle|
=\frac{G_F^2M_W^2f_B^2m_{B_q}
B_B^S\beta}{4\pi^2}\Delta_{2,1}^2\frac{g_2^2}{g_1^2}|f_{LR}(B_q)|
[\frac{1}{6}\!+\!\frac{m_{B_q}^2}{m_b^2}]~~~~\label{DeltamB}\\
&&\hspace{-0.5cm}|\epsilon_{K}|^{WW'}\!\!\!\!\approx
\frac{\mathrm{Im}\!\langle
K^0|H^{WW'}_\mathrm{eff}|\bar{K}^0\rangle}{\sqrt{2}\Delta m_{K}}=
\frac{G_F^2M_W^2f_K^2m_K B_K^S\beta}{8\sqrt{2}\pi^2\Delta
m_{K}}\Delta_{2,1}^2\frac{g_2^2}{g_1^2}\mathrm{Im}(f_{LR}(K))
[\frac{1}{6}\!+\!\frac{m_K^2}{(m_s\!+\!m_d)^2}]~~~\label{epsilonK}
\end{eqnarray}

For $W'$ contributions, we discuss $K^0-\bar{K}^0$, $B^0-\bar{B}^0$
systems separately.
%%%%%%%%%%%%%%%%%%%%%%%%%%%%%%%%%%%%%%%%
\subsection{$K^0-\bar{K}^0$ system}

\begin{figure}[t] \caption{Ratio of $tt$ loop  to experiment data for $W'$ contribution to $K_L-K_S$ mass
difference, $cc$ to $tt$ loop and $ct$ to $tt$ loop for $W'$
contribution to $K_L-K_S$ mass difference in $K^0-\bar{K}^0$ system
with solid blue line for $M_{W'}=10M_W$, dash red line for
$M_{W'}=15M_W$, dash-dot pink line for $M_{W'}=20M_W$ and dot black
line for $M_{W'}=25M_W$, respectively.\\~} \label{fig-WW'-mK-cc-ct}
\centering \hspace*{-2.5cm}\begin{minipage}[t]{\textwidth}
\includegraphics[scale=0.9]{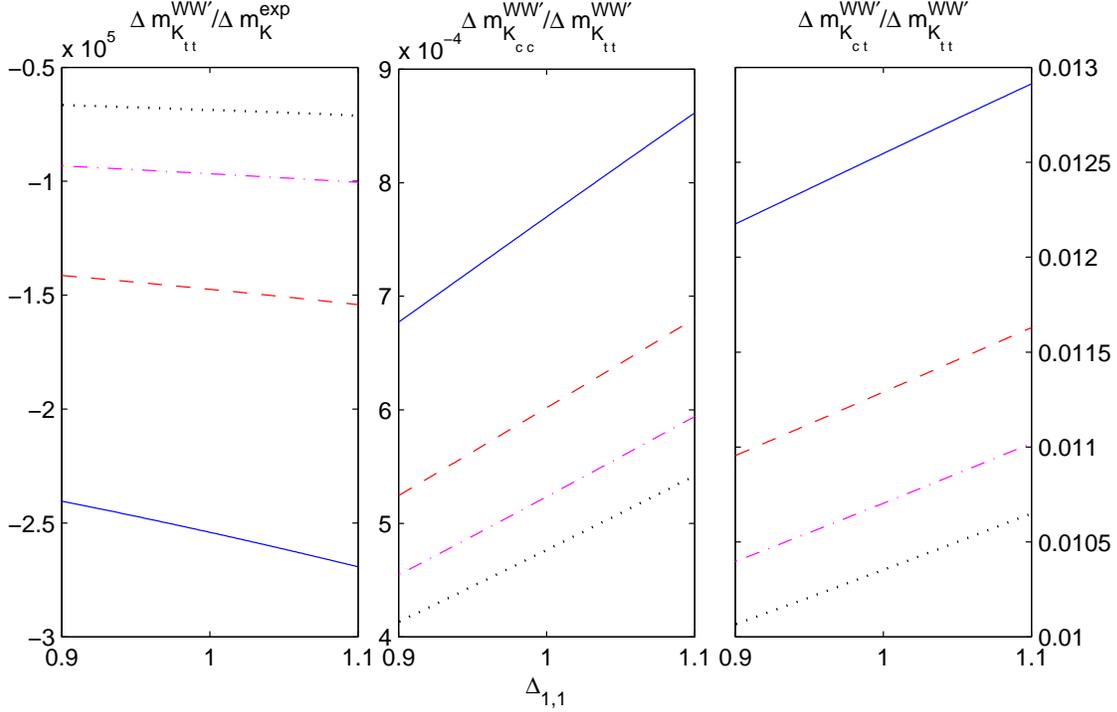}
\end{minipage}
\end{figure}

 According to types of inner quark lines in the box diagrams, $W'$ contributions
 to $\Delta m_{K}$ in (\ref{DeltamK}) can be decomposed into $tt$, $cc$ and $ct$ quark loop contributions,
\begin{eqnarray} \Delta
m_{K}^{WW'}&=&\Delta_{2,1}^2\frac{g_2^2}{g_1^2}\Delta
m_{K_{tt}}^{WW'}\bigg[~\mathrm{Re}[\lambda_t^{LR}(K)\lambda_t^{RL}(K)]+\mathrm{Re}[\lambda_c^{LR}(K)
\lambda_c^{RL}(K)]\frac{\Delta
m_{K_{cc}}^{WW'}}{\Delta m_{K_{tt}}^{WW'}}\nonumber\\
&&+\mathrm{Re}[\lambda_c^{LR}(K)\lambda_t^{RL}(K)+\lambda_t^{LR}(K)\lambda_c^{RL}(K)]\frac{\Delta
m_{K_{ct}}^{WW'}}{\Delta m_{K_{tt}}^{WW'}}~\bigg]\label{mK-WW'}
\end{eqnarray}
in which the CKM matrices are
\begin{eqnarray}
&&\hspace{-1cm}\lambda_x^{LR}(K)\lambda_x^{RL}(K)
=|V^{xs}_LV^{xd*}_{L}\bar{V}^{xs}_R\bar{V}^{xd*}_{R}|e^{-i(\alpha_1-\alpha_2-\beta_1-\phi_{xs}-\bar{\phi}_{xs}
+\phi_{xd}+\bar{\phi}_{xd})}\hspace{1.4cm}x=c,t
\nonumber\\
&&\hspace{-1cm}\lambda_c^{LR}(K)\lambda_t^{RL}(K)
=|V^{cs}_L\bar{V}^{cd*}_{R}\bar{V}^{ts}_RV^{td*}_{L}
|e^{-i(\alpha_1-\alpha_3-\beta_1+\beta_2-\phi_{cs}+\bar{\phi}_{cd}-\bar{\phi}_{ts}+\phi_{td})}\hspace{1.2cm}
\mathrm{arg}(V^{\alpha\beta}_L)\!=\!\phi_{\alpha\beta}\nonumber\\
&&\hspace{-1cm}\lambda_t^{LR}(K)\lambda_c^{RL}(K)
=|V^{ts}_L\bar{V}^{td*}_{R}\bar{V}^{cs}_RV^{cd*}_{L}
|e^{-i(\alpha_1-2\alpha_2+\alpha_3-\beta_1-\beta_2-\phi_{ts}+\bar{\phi}_{td}-\bar{\phi}_{cs}+\phi_{cd})}
\hspace{0.5cm}{\color{black}\mathrm{arg}(\bar{V}^{\alpha\beta}_R)\!=\!\bar{\phi}_{\alpha\beta}}~~~\label{phasedef}
\end{eqnarray}
In Fig.\ref{fig-WW'-mK-cc-ct}, we plot $\frac{\Delta
m_{K_{tt}}^{WW'}}{\Delta m_K^\mathrm{exp}}$, $\frac{\Delta
m_{K_{cc}}^{WW'}}{\Delta m_{K_{tt}}^{WW'}}$ and $\frac{\Delta
m_{K_{ct}}^{WW'}}{\Delta m_{K_{tt}}^{WW'}}$  separately,
From Fig.\ref{fig-WW'-mK-cc-ct}, we find that $\frac{\Delta
m_{K_{tt}}^{WW'}}{\Delta m_K^\mathrm{exp}}$ is of order $10^5$,
$\frac{\Delta m_{K_{cc}}^{WW'}}{\Delta m_{K_{tt}}^{WW'}}$ is of
order $10^{-3}$ and $\frac{\Delta m_{K_{ct}}^{WW'}}{\Delta
m_{K_{tt}}^{WW'}}$ is of order $10^{-2}$. Therefore to reduce total
contributions of $\Delta m_{K}^{WW'}$, we have following four
different kind of mechanisms
\begin{itemize}
\item Large $M_{W'}$:~Take very large $W'$ mass. This is traditional naive constraints to $W'$
mass.
\item Small $g_2$:~Take very small $W'$ gauge coupling $g_2\ll g_1$.
This can only happens if $f_2\gg g_1f_1/g_2$ to make large enough
$W'$ mass.  Since two gauge couplings are not equal to each other,
this situation is parity explicitly broken.
\item Small $\Delta_{2,1}$:~Take very small $\Delta_{2,1}$.
This is the situation pointed out in our previous work
\cite{LRSMEWCLmatter}. Although realization of this situation in
detail model is still lacking.
\item Specific $V^{\mathrm{CKM}}_R$:~Choose special right hand CKM
matrix elements to make $f_{LR}(K)$ in (\ref{fLRK}) small.
Numerically
\begin{eqnarray}
&&\mathrm{Re}[\lambda_t^{LR}(K)\lambda_t^{RL}(K)]+\mathrm{Re}[\lambda_c^{LR}(K)\lambda_c^{RL}(K)]\frac{\Delta
m_{K_{cc}}^{WW'}}{\Delta
m_{K_{tt}}^{WW'}}\nonumber\\
&&+\mathrm{Re}[\lambda_c^{LR}(K)\lambda_t^{RL}(K)+\lambda_t^{LR}(K)\lambda_c^{RL}(K)]\frac{\Delta
m_{K_{ct}}^{WW'}}{\Delta m_{K_{tt}}^{WW'}}\ll
10^{-5}~~~\label{ReCKMconstrain}
\end{eqnarray}
\end{itemize}

\begin{figure}[t] \caption{Ratio of $tt$ loop to
experiment data for $W'$ contribution to $|\epsilon_K|$ , $cc$ to
$tt$ loop and $ct$ to $tt$ loop for $W'$ contribution to indirect CP
violation in K mesons $|\epsilon_K|$ in $K^0-\bar{K}^0$ system with
solid blue line for $M_{W'}=10M_W$, dash red line for
$M_{W'}=15M_W$, dash-dot pink line for $M_{W'}=20M_W$ and dot black
line for $M_{W'}=25M_W$, respectively.\\~ }
\label{fig-WW'-epsilonK-cc-ct} \centering
\hspace*{-2.5cm}\begin{minipage}[t]{\textwidth}
\includegraphics[scale=0.9]{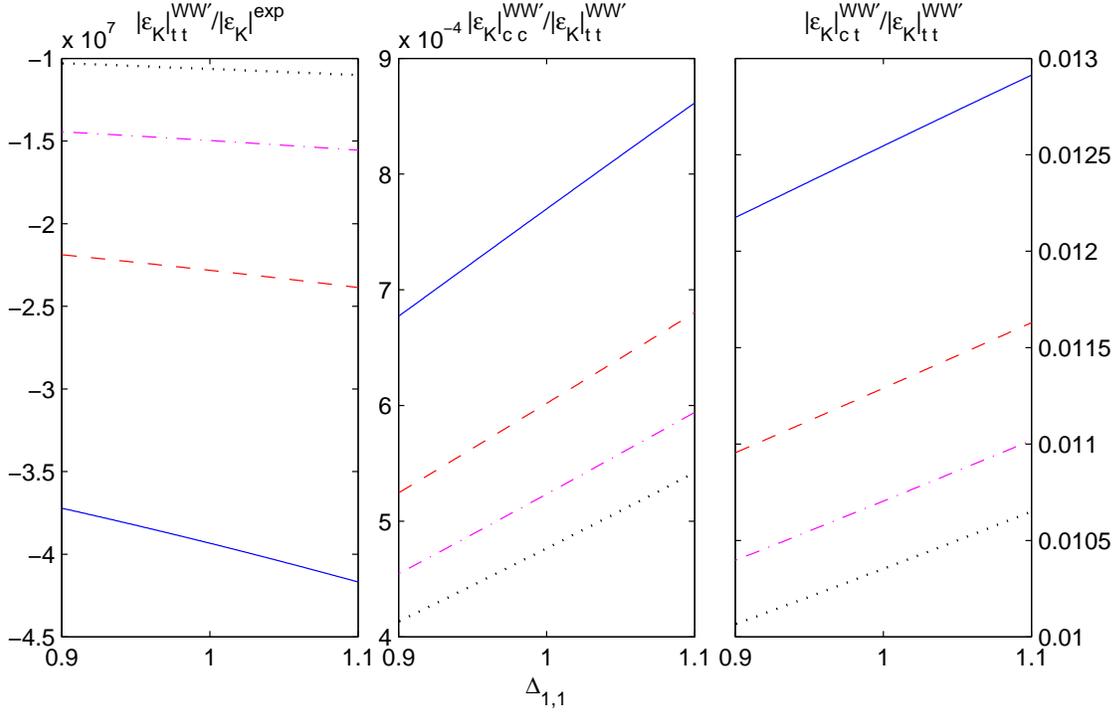}
\end{minipage}
\end{figure}

\begin{figure}[t] \caption{$\frac{\Delta m_{K}^{WW'}}{\Delta
m_{K}^\mathrm{exp}}/(\Delta_{2,1}^2\frac{g_2^2}{g_1^2})$ as function
of $\cos(\alpha_2-\alpha_3+\beta_2)$ with anomalous coupling
$\Delta_{1,1}=1$. } \label{fig-WW'-mK-manifest-cosdelta} \centering
\begin{minipage}[t]{\textwidth}
    \includegraphics[scale=1]{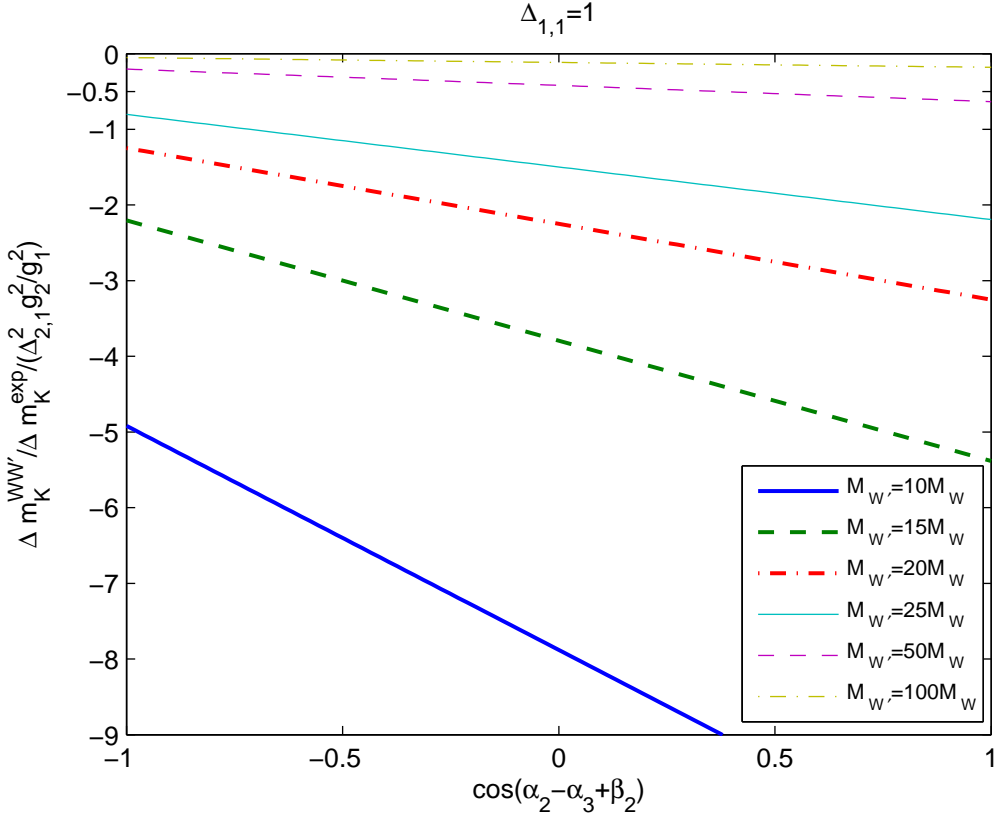}
\end{minipage}
\end{figure}

 Similar to $K_L-K_S$ mass difference, we can also decompose
indirect CP violation parameter $|\epsilon_K|$ in K system as
\begin{eqnarray}
|\epsilon_K|^{WW'}&=&\Delta_{2,1}^2\frac{g_2^2}{g_1^2}|\epsilon_K|_{tt}^{WW'}\bigg|~
\mathrm{Im}[\lambda_t^{LR}(K)\lambda_t^{RL}(K)]+\mathrm{Im}[\lambda_c^{LR}(K)\lambda_c^{RL}(K)]
\frac{|\epsilon_K|_{cc}^{WW'}}{|\epsilon_K|_{tt}^{WW'}}\nonumber\\
&&+\mathrm{Im}[\lambda_c^{LR}(K)\lambda_t^{RL}(K)+\lambda_t^{LR}(K)\lambda_c^{RL}(K)]
\frac{|\epsilon_K|_{ct}^{WW'}}{|\epsilon_K|_{tt}^{WW'}}~\bigg|\label{epsilonK-WW'}
\end{eqnarray}
In Fig.\ref{fig-WW'-epsilonK-cc-ct}, we plot
$\frac{|\epsilon_K|_{tt}^{WW'}}{|\epsilon_K|^\mathrm{exp}}$,
$\frac{|\epsilon_K|_{cc}^{WW'}}{|\epsilon_K|_{tt}^{WW'}}$ and
$\frac{|\epsilon_K|_{ct}^{WW'}}{|\epsilon_K|_{tt}^{WW'}}$
 separately,
From Fig.\ref{fig-WW'-epsilonK-cc-ct}, we find that
$\frac{|\epsilon_K|_{tt}^{WW'}}{|\epsilon_K|^\mathrm{exp}}$
 is of order $10^7$, $\frac{|\epsilon_K|_{cc}^{WW'}}{|\epsilon_K|_{tt}^{WW'}}$ is of
order $10^{-3}$ and
$\frac{|\epsilon_K|_{ct}^{WW'}}{|\epsilon_K|_{tt}^{WW'}}$ is of
order $10^{-2}$. To reduce total contributions of
 $|\epsilon_K|^{WW'}$, we also can take either
large $M_{W'}$; or small $g_2$; or small $\Delta_{2,1}$; or specific
$V^{\mathrm{CKM}}_R$ satisfying
\begin{eqnarray}
&&\mathrm{Im}[\lambda_t^{LR}(K)\lambda_t^{RL}(K)+\mathrm{Im}[\lambda_c^{LR}(K)\lambda_c^{RL}(K)]
\frac{|\epsilon_K|_{cc}^{WW'}}{|\epsilon_K|_{tt}^{WW'}}\nonumber\\
&&+\mathrm{Im}[\lambda_c^{LR}(K)\lambda_t^{RL}(K)+\lambda_t^{LR}(K)\lambda_c^{RL}(K)]
\frac{|\epsilon_K|_{ct}^{WW'}}{|\epsilon_K|_{tt}^{WW'}}\ll
10^{-7}~~~~\label{ImCKMconstrain}
\end{eqnarray}
The relation (\ref{ReCKMconstrain}) and (\ref{ImCKMconstrain}) offer
constraints for right hand CKM matrix elements, as long as they
really take the role of suppressing contribution from $W'$ boson. If
constraints (\ref{ReCKMconstrain}) and (\ref{ImCKMconstrain}) can
not be satisfied, we must adjust $M_{W'}$, $g_2$ and $\Delta_{2,1}$
to suppress contribution of $W'$. To quantitatively estimate
constraints for $M_{W'}$, $g_2$ and $\Delta_{2,1}$, we take a
special pseudo-manifest left-right symmetric situation as an
example. In this situation,
$\bar{V}_{R}^{\alpha\beta}=(V_{L}^{\alpha\beta})^*$, which implies
the relations $\phi_{\alpha\beta}=-\bar{\phi}_{\alpha\beta}$ between
phases defined in (\ref{phasedef}). Then CKM factors appeared in
(\ref{mK-WW'}) and (\ref{epsilonK-WW'}) can be simplified as
\begin{eqnarray}
&&\hspace{-1cm}\lambda_c^{LR}(K)\lambda_c^{RL}(K)=|V^{cs}_LV^{cd}_{L}|^2e^{-i(\alpha_1-\alpha_2-\beta_1)}\hspace{1cm}
\lambda_t^{LR}(K)\lambda_t^{RL}(K)=|V^{ts}_LV^{td}_{L}|^2e^{-i(\alpha_1-\alpha_2-\beta_1)}\label{CKMmanifest}\\
&&\hspace{-1cm}\lambda_c^{LR}(K)\lambda_t^{RL}(K)+\lambda_t^{LR}(K)\lambda_c^{RL}(K)=2|V^{cs}_LV^{cd}_{L}V^{ts}_LV^{td}_{L}|[\cos(\alpha_1-\alpha_2-\beta_1)\cos(\alpha_2-\alpha_3+\beta_2)\nonumber\\
&&\hspace{5.7cm}-i\sin(\alpha_1-\alpha_2-\beta_1)\cos(\alpha_2-\alpha_3+\beta_2)]\nonumber
\end{eqnarray}
Notice that constraint from $|\epsilon_K|$ demands imaginary part of
above CKM matrix elements must at least two order of magnitude
smaller than their real part, this leads us to take following choice
of phase angle
\begin{eqnarray}
\alpha_1-\alpha_2-\beta_1=0.
\end{eqnarray}
Then the imaginary part of all
 CKM matrix elements in (\ref{CKMmanifest}) will vanish and the cc, tt and ct loops do not
 contribute to $|\epsilon_K|^{WW'}$
separately. This special choice of phase angle is originally
proposed in Ref.\cite{YLWu07} which directly leads to
\begin{eqnarray}
|\epsilon_K|^{WW'}=0.
\end{eqnarray}
 The values of  CKM matrix factors in (\ref{CKMmanifest}) now can be
worked out in terms of left hand CKM matrix in \cite{PDG06},
\begin{eqnarray}
&&\hspace{-1cm}\lambda_c^{LR}(K)\lambda_c^{RL}(K)\bigg|_\mathrm{manifest}
\hspace*{-1cm}\stackrel{\alpha_1-\alpha_2-\beta_1=0}{=======}0.0488
\hspace{2cm}
\lambda_t^{LR}(K)\lambda_t^{RL}(K)\bigg|_\mathrm{manifest}
\hspace*{-1cm}\stackrel{\alpha_1-\alpha_2-\beta_1=0}{=======}5.89\times 10^{-6}\nonumber\\
&&\hspace{-1cm}\lambda_c^{LR}(K)\lambda_t^{RL}(K)+\lambda_t^{LR}(K)\lambda_c^{RL}(K)
\bigg|_\mathrm{manifest}
\hspace*{-1cm}\stackrel{\alpha_1-\alpha_2-\beta_1=0}{=======}0.00107\cos(\alpha_2-\alpha_3+\beta_2)\label{CKMmanifest1}
\end{eqnarray}
Now, except an overall factor $\Delta_{2,1}^2\frac{g_2^2}{g_1^2}$,
$\frac{\Delta m_{K}^{WW'}}{\Delta m_{K}^\mathrm{exp}}$ depends on
two other parameters, $\Delta_{1,1}$ and
$\cos(\alpha_2-\alpha_3+\beta_2)$. Considering that anomalous
coupling $\Delta_{1,1}$ can not deviate from $1$ very much, in
Fig.\ref{fig-WW'-mK-manifest-cosdelta}, we plot $\frac{\Delta
m_{K}^{WW'}}{\Delta
m_{K}^\mathrm{exp}}/(\Delta_{2,1}^2\frac{g_2^2}{g_1^2})$ as function
of $\cos(\alpha_2-\alpha_3+\beta_2)$ with anomalous coupling
$\Delta_{1,1}=1$.
From Fig.\ref{fig-WW'-mK-manifest-cosdelta}, we find if
$\Delta_{2,1}\sim 1$ and $g_2\sim g_1$, then to make $\frac{\Delta
m_{K}^{WW'}}{\Delta m_{K}^\mathrm{exp}}\ll 1$, we must have
$M_{W'}\sim$ several TeVs. This is the naive prediction of $W'$ mass
in traditional left-right symmetric models. While if $M_{W'}$ is at
order of 1TeV, to make $\frac{\Delta m_{K}^{WW'}}{\Delta
m_{K}^\mathrm{exp}}\ll 1$, we must have
$\Delta_{2,1}^2\frac{g_2^2}{g_1^2}\ll 10^{-1}$ which demands either
very small anomalous coupling $\Delta_{2,1}$ or small gauge coupling
$g_2$. Note that from (\ref{LRCCLagrangian}), unlike $\Delta_{1,1}$
which roughly is 1 since it is anomalous coupling of charged current
for $W$ boson, $\Delta_{2,1}$ is anomalous coupling of charged
current for $W'$ boson and there is no experiment constraint on its
value. This provides us an alternative way to reduce $W'$
contribution. This possibility was first pointed out in our previous
work \cite{LRSMEWCLmatter} where $\Delta_{2,1}$ is denoted by
$\Delta_{R,1}$.
%%%%%%%%%%%%%%%%%%%%%%%%%%%%%%%%%%%%%%%%
\subsection{$B^0-\bar{B}^0$ system}

\begin{figure}[t] \caption{Ratio of $tt$ loop to experiment data for $W'$ contribution to
$\Delta m_{B_d}$, $cc$ to $tt$ loop and $ct$ to $tt$ loop for $W'$
contribution to mass difference in $B_d^0-\bar{B}_d^0$ system with
solid blue line for $M_{W'}=10M_W$, dash red line for
$M_{W'}=15M_W$, dash-dot pink line for $M_{W'}=20M_W$ and dot black
line for $M_{W'}=25M_W$, respectively.\\~ }
\label{fig-WW'-mBd-cc-ct} \centering
\hspace*{-2.5cm}\begin{minipage}[t]{\textwidth}
\includegraphics[scale=0.9]{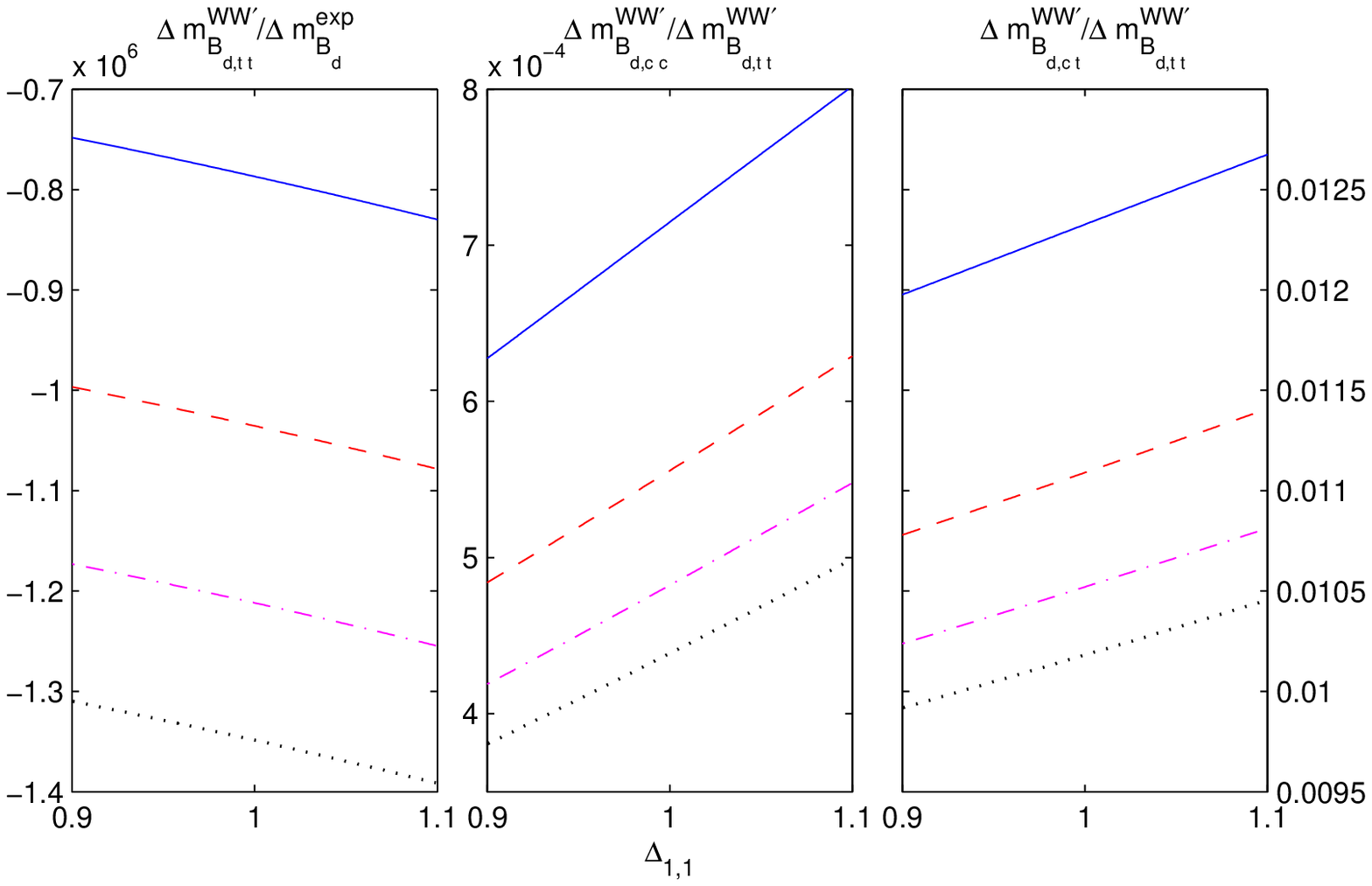}
\end{minipage}
\end{figure}

For $B_d^0-\bar{B}_d^0$ system, similar to $K^0-\bar{K}^0$ system,
we can decompose corresponding effective Hamiltonian as,
\begin{eqnarray}
\Delta
m_{B_d}^{WW'}&=&\Delta_{2,1}^2\frac{g_2^2}{g_1^2}\Delta
m_{B_dtt}^{WW'}\bigg|\lambda_t^{LR}(B_d)\lambda_t^{RL}(B_d)+\lambda_c^{LR}(B_d)\lambda_c^{RL}(B_d)
\frac{\Delta m_{B_dcc}^{WW'}}{\Delta m_{B_dtt}^{WW'}}\nonumber\\
&&+[\lambda_c^{LR}(B_d)\lambda_t^{RL}(B_d)+\lambda_t^{LR}(B_d)\lambda_c^{RL}(B_d)]\frac{\Delta
m_{B_dct}^{WW'}}{\Delta m_{B_dtt}^{WW'}}\bigg|\\
&&\hspace{-2cm}\lambda_x^{LR}(B_d)\lambda_x^{RL}(B_d)
=|V^{xb}_LV^{xd*}_{L}\bar{V}^{xb}_R\bar{V}^{xd*}_{R}|e^{-i(\alpha_1-\alpha_3-\beta_1-\beta_2-\phi_{xb}-\bar{\phi}_{xb}
+\phi_{xd}+\bar{\phi}_{xd})}\hspace{2cm}x=c,t
\nonumber\\
&&\hspace{-2cm}\lambda_c^{LR}(B_d)\lambda_t^{RL}(B_d)
=|V^{cb}_L\bar{V}^{cd*}_{R}\bar{V}^{tb}_RV^{td*}_{L}
|e^{-i(\alpha_1+\alpha_2-2\alpha_3-\beta_1-\phi_{cb}+\bar{\phi}_{cd}-\bar{\phi}_{tb}+\phi_{td})}\hspace{1cm}
\mathrm{arg}(V^{\alpha\beta}_L)\!=\!\phi_{\alpha\beta}\nonumber\\
&&\hspace{-2cm}\lambda_t^{LR}(B_d)\lambda_c^{RL}(B_d)
=|V^{tb}_L\bar{V}^{td*}_{R}\bar{V}^{cb}_RV^{cd*}_{L}
|e^{-i(\alpha_1-\alpha_2-\beta_1-2\beta_2-\phi_{tb}+\bar{\phi}_{td}-\bar{\phi}_{cb}+\phi_{cd})}
\hspace{1cm}\mathrm{arg}(\bar{V}^{\alpha\beta}_R)\!=\!\bar{\phi}_{\alpha\beta}\nonumber
\end{eqnarray}
In Fig.\ref{fig-WW'-mBd-cc-ct}, we plot $\frac{\Delta
m_{B_dtt}^{WW'}}{\Delta m_{B_d}^\mathrm{exp}}$, $\frac{\Delta
m_{B_dcc}^{WW'}}{\Delta m_{B_dtt}^{WW'}}$ and $\frac{\Delta
m_{B_dct}^{WW'}}{\Delta m_{B_dtt}^{WW'}}$ separately,
From Fig.\ref{fig-WW'-mBd-cc-ct}, we find that $\frac{\Delta
m_{B_dtt}^{WW'}}{\Delta m_{B_d}^\mathrm{exp}}$ is of order $10^5$,
$\frac{\Delta m_{B_dcc}^{WW'}}{\Delta m_{B_dtt}^{WW'}}$ is of order
$10^{-3}$ and $\frac{\Delta m_{B_dct}^{WW'}}{\Delta
m_{B_dtt}^{WW'}}$ is of order $10^{-2}$. To reduce total
contributions of
 $\frac{\Delta
m_{B_d}^{WW'}}{\Delta m_{B_d}^\mathrm{exp}}$, we can take either
large $M_{W'}$; or small $g_2$; or small $\Delta_{2,1}$; or specific
$V^{\mathrm{CKM}}_R$ which satisfy
\begin{eqnarray}
&&\bigg|\lambda_t^{LR}(B_d)\lambda_t^{RL}(B_d)+\lambda_c^{LR}(B_d)\lambda_c^{RL}(B_d)
\frac{\Delta m_{B_dcc}^{WW'}}{\Delta m_{B_dtt}^{WW'}}\nonumber\\
&&+[\lambda_c^{LR}(B_d)\lambda_t^{RL}(B_d)+\lambda_t^{LR}(B_d)\lambda_c^{RL}(B_d)]\frac{\Delta
m_{B_dct}^{WW'}}{\Delta m_{B_dtt}^{WW'}}\bigg|\ll 10^{-5}
\end{eqnarray}

\begin{figure}[t] \caption{Ratio of $tt$ loop to experiment data for $W'$ contribution to
$\Delta m_{B_s}$, $cc$ to $tt$ loop and $ct$ to $tt$ loop for $W'$
contribution to mass difference in $B_s^0-\bar{B}_s^0$ system with
solid blue line for $M_{W'}=10M_W$, dash red line for
$M_{W'}=15M_W$, dash-dot pink line for $M_{W'}=20M_W$ and dot black
line for $M_{W'}=25M_W$, respectively.\\~} \label{fig-WW'-mBs-cc-ct}
%\centering
\hspace*{-2.5cm}\begin{minipage}[t]{\textwidth}
\includegraphics[scale=0.9]{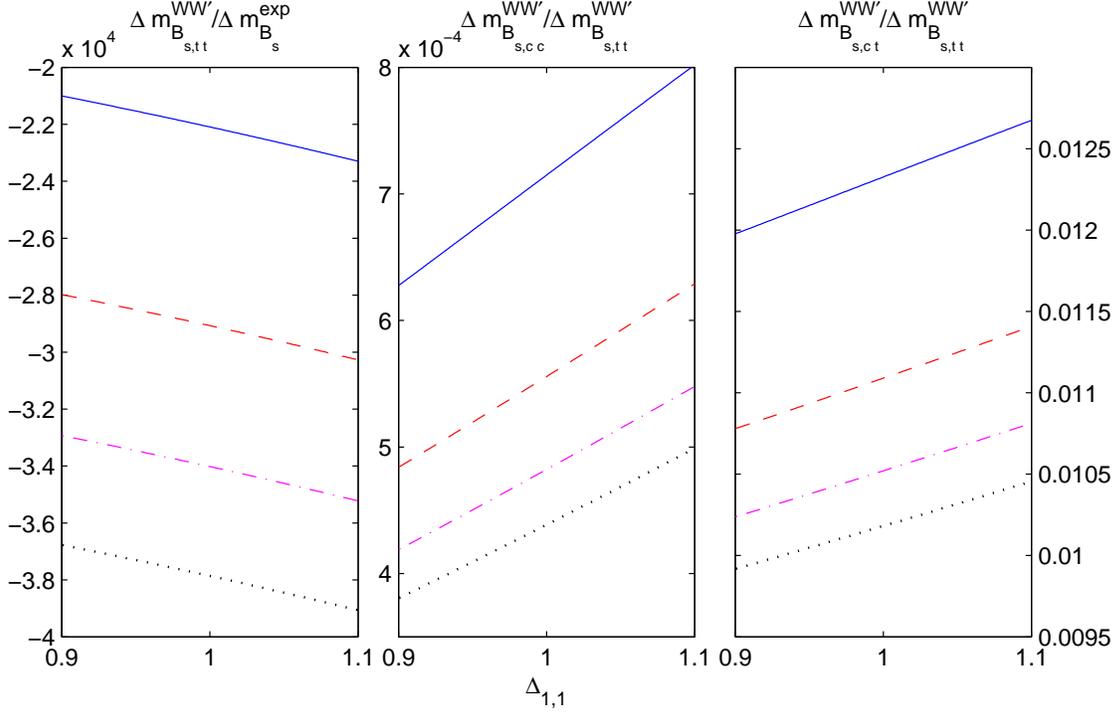}
\end{minipage}
\end{figure}

For $B^0_s-\bar{B}^0_s$ system,
\begin{eqnarray}
 \Delta
m_{B_s}^{WW'}&=&\Delta_{2,1}^2\frac{g_2^2}{g_1^2}\Delta
m_{B_stt}^{WW'}\bigg|\lambda_t^{LR}(B_s)\lambda_t^{RL}(B_s)+\lambda_c^{LR}(B_s)\lambda_c^{RL}(B_s)
\frac{\Delta m_{B_scc}^{WW'}}{\Delta m_{B_stt}^{WW'}}\nonumber\\
&&+[\lambda_c^{LR}(B_s)\lambda_t^{RL}(B_s)+\lambda_t^{LR}(B_s)\lambda_c^{RL}(B_s)]\frac{\Delta
m_{B_sct}^{WW'}}{\Delta m_{B_stt}^{WW'}}\bigg|\\
&&\hspace{-2cm}\lambda_x^{LR}(B_s)\lambda_x^{RL}(B_s)
=|V^{xb}_LV^{xs*}_{L}\bar{V}^{xb}_R\bar{V}^{xs*}_{R}|e^{-i(\alpha_2-\alpha_3-\beta_2-\phi_{xb}-\bar{\phi}_{xb}
+\phi_{xs}+\bar{\phi}_{xs})}\hspace{2cm}x=c,t
\nonumber\\
&&\hspace{-2cm}\lambda_c^{LR}(B_s)\lambda_t^{RL}(B_s)
=|V^{cb}_L\bar{V}^{cs*}_{R}\bar{V}^{tb}_RV^{ts*}_{L}
|e^{-i(2\alpha_2-2\alpha_3-\phi_{cb}+\bar{\phi}_{cs}-\bar{\phi}_{tb}+\phi_{ts})}\hspace{2.5cm}
\mathrm{arg}(V^{\alpha\beta}_L)\!=\!\phi_{\alpha\beta}\nonumber\\
&&\hspace{-2cm}\lambda_t^{LR}(B_s)\lambda_c^{RL}(B_s)
=|V^{tb}_L\bar{V}^{ts*}_{R}\bar{V}^{cb}_RV^{cs*}_{L}
|e^{-i(-2\beta_2-\phi_{tb}+\bar{\phi}_{ts}-\bar{\phi}_{cb}+\phi_{cs})}
\hspace{3cm}\mathrm{arg}(\bar{V}^{\alpha\beta}_R)\!=\!\bar{\phi}_{\alpha\beta}\nonumber
\end{eqnarray}
In Fig.\ref{fig-WW'-mBs-cc-ct}, we plot $\frac{\Delta
m_{B_stt}^{WW'}}{\Delta m_{B_s}^\mathrm{exp}}$, $\frac{\Delta
m_{B_scc}^{WW'}}{\Delta m_{B_stt}^{WW'}}$ and $\frac{\Delta
m_{B_sct}^{WW'}}{\Delta m_{B_stt}^{WW'}}$ separately,
From Fig.\ref{fig-WW'-mBs-cc-ct}, we find that $\frac{\Delta
m_{B_stt}^{WW'}}{\Delta m_{B_s}^\mathrm{exp}}$ is of order $10^4$,
$\frac{\Delta m_{B_scc}^{WW'}}{\Delta m_{B_stt}^{WW'}}$ is of order
$10^{-3}$ and $\frac{\Delta m_{B_sct}^{WW'}}{\Delta
m_{B_stt}^{WW'}}$ is of order $10^{-2}$. To reduce total
contributions of
 $\frac{\Delta
m_{B_s}^{WW'}}{\Delta m_{B_s}^\mathrm{exp}}$, we can also take
either large $M_{W'}$; or small $g_2$,; or small $\Delta_{2,1}$; or
specific $V^{\mathrm{CKM}}_R$ which satisfy
\begin{eqnarray}
&&\bigg|\lambda_t^{LR}(B_s)\lambda_t^{RL}(B_s)+\lambda_c^{LR}(B_s)\lambda_c^{RL}(B_s)
\frac{\Delta m_{B_scc}^{WW'}}{\Delta m_{B_stt}^{WW'}}\nonumber\\
&&+[\lambda_c^{LR}(B_s)\lambda_t^{RL}(B_s)+\lambda_t^{LR}(B_s)\lambda_c^{RL}(B_s)]\frac{\Delta
m_{B_sct}^{WW'}}{\Delta m_{B_stt}^{WW'}}\bigg|\ll 10^{-4}
\end{eqnarray}

Now we come to Higgs contribution to $H_\mathrm{eff}$.  This part is
universal and from vertices given by (\ref{Lh}), the amplitude of
the diagram mediated by neutral Higgs $\tilde{h}$ is
\begin{eqnarray}
\label{phiphi} M^{\tilde{h}}
&=&\frac{1}{4m_h^2}\left\{\begin{array}{ll}
\bar{d}(A_d^{ds}+B_d^{ds}\gamma^5)s\otimes\bar{d}(A_d^{ds}+B_d^{ds}\gamma^5)s&K^0-\bar{K}^0~\mathrm{system}\\
\bar{q}(A_d^{qb}+B_d^{qb}\gamma^5)b\otimes\bar{q}(A_d^{qb}+B_d^{qb}\gamma^5)b&B^0_q-\bar{B}_q^0~\mathrm{system}~~~
q=d,s
\end{array}\right.~~~~
\end{eqnarray}
and Higgs contribution to $H_\mathrm{eff}$ is related to the
amplitude by $H^{\tilde{h}}_\mathrm{eff}=\frac{1}{2}M^{\tilde{h}}$,
it's matrix element is
\begin{eqnarray}
\langle K^0|H^{\tilde{h}}_\mathrm{eff}|\bar{K}^0\rangle&=&
\frac{1}{8m_h^2}\langle
K^0|\bar{d}(A_d^{ds}+B_d^{ds}\gamma^5)s\otimes\bar{d}(A_d^{ds}+B_d^{ds}\gamma^5)s|\bar{K}^0\rangle
\\
&=&[(\tilde{y}_d^{ds}\!\!-\!\tilde{y}_d^{\dag
ds})^2\!\!-\!(\tilde{y}_d^{ds}\!\!+\!\tilde{y}_d^{\dag
ds})^2\!\!+\!\frac{((\tilde{y}_d^{ds}\!\!+\!\tilde{y}_d^{\dag
ds})^2\!\!-\!11(\tilde{y}_d^{ds}\!\!-\!\tilde{y}_d^{\dag
ds})^2)m_K^2}{(m_s\!+\!m_d)^2}]\frac{f_K^2m_KB_K^S}{96m_h^2}\nonumber\\
\langle B^0_q|H^{\tilde{h}}_\mathrm{eff}|B^0_{\bar{q}}\rangle&=&
\frac{1}{8m_h^2}\langle
B^0_q|\bar{q}(A_d^{qb}+B_d^{qb}\gamma^5)b\otimes\bar{q}(A_d^{qb}+B_d^{qb}\gamma^5)b|B^0_{\bar{q}}\rangle
\hspace{1cm}q=d,s\\
&=&[(\tilde{y}_d^{qb}\!\!-\!\tilde{y}_d^{\dag
qb})^2\!\!-\!(\tilde{y}_d^{qb}\!\!+\!\tilde{y}_d^{\dag
qb})^2\!\!+\!\frac{((\tilde{y}_d^{qb}\!\!+\!\tilde{y}_d^{\dag
qb})^2\!\!-\!11(\tilde{y}_d^{qb}\!\!-\!\tilde{y}_d^{\dag qb})^2)
m_{B_q}^2}{(m_s\!+\!m_d)^2}]\frac{f_{B_q}^2m_{B_q}B_{B_q}^S}{96m_h^2}\nonumber
\end{eqnarray}
In the case of SM, above matrix elements vanishes due to
non-existence of  flavor changing coulings. Then in SM, there is no
Higgs contribution to $\Delta m_{K}^{h}$, $\Delta m_{B_q}^{h}$ and
$|\epsilon_{K}|^{h}$. Beyond SM Higgs,
 demanding Higgs contributions to $\Delta m_{K}$ $\Delta
m_{B}$ are much smaller than their experimental value,  we find
 constraints
\begin{eqnarray}
&&\mathrm{Re}[(\tilde{y}_d^{ds}\!\!+\!\tilde{y}_d^{\dag
ds})^2\!-\!11.4(\tilde{y}_d^{ds}\!\!-\!\tilde{y}_d^{\dag ds})^2]\ll
6.45\!\times\!10^{-7}\left(\frac{m_h}{1\mathrm{TeV}}\right)^2\label{HconstraintK}\\
&& |(\tilde{y}_d^{db}\!\!+\!\tilde{y}_d^{\dag
db})^2\!-\!11.4(\tilde{y}_d^{db}\!\!-\!\tilde{y}_d^{\dag db})^2|\ll
1.74\!\times\!10^{-6}\left(\frac{m_h}{1\mathrm{TeV}}\right)^2\label{HconstraintBd}\\
&&|(\tilde{y}_d^{sb}\!\!+\!\tilde{y}_d^{\dag
sb})^2\!-\!11.4(\tilde{y}_d^{sb}\!\!-\!\tilde{y}_d^{\dag sb})^2|\ll
6.10\!\times\!10^{-5}\left(\frac{m_h}{1\mathrm{TeV}}\right)^2
\label{HconstraintBs}
\end{eqnarray}
While demanding Higgs contributions to $|\epsilon_{K}|$ is much
smaller than its experimental value,  we find constraint:
\begin{eqnarray}
\mathrm{Im}[(\tilde{y}_d^{ds}\!\!+\!\tilde{y}_d^{\dag
ds})^2\!-\!11.4(\tilde{y}_d^{ds}\!\!-\!\tilde{y}_d^{\dag ds})^2]\ll
0.0065\!\times\! \mathrm{Re}[(\tilde{y}_d^{ds}+\tilde{y}_d^{\dag
ds})^2-11.4(\tilde{y}_d^{ds}-\tilde{y}_d^{\dag ds})^2]
\end{eqnarray}
which imply term $(\tilde{y}_d^{ds}+\tilde{y}_d^{\dag
ds})^2-11.4(\tilde{y}_d^{ds}-\tilde{y}_d^{\dag ds})^2$ is
approximately real. Further combing constraint
$(\ref{HconstraintK})$,
\begin{eqnarray}
(\tilde{y}_d^{ds}\!\!+\!\tilde{y}_d^{\dag
ds})^2\!-\!11.4(\tilde{y}_d^{ds}\!\!-\!\tilde{y}_d^{\dag ds})^2\ll
6.45\!\times\!10^{-7}\left(\frac{m_h}{1\mathrm{TeV}}\right)^2\label{Hconstraint1}
\end{eqnarray}
%%%%%%%%%%%%%%%%%%%%%%%%%%%%%%%%%%%%%%%%%%%%%%%%%%%%%%%%
\section{Summary}

 In this paper, we have presented the complete list of electroweak chiral Lagrangian for $W'$, $Z'$,
 a neutral light higgs and those discovered SM particles with symmetry $SU(2)_1\otimes SU(2)_2\otimes
U(1)$. The bosonic part is accurate up to order of $p^4$. The matter
part involving various fermions representation arrangements such as
LR, LP, HP, FP, UN, NU includes dimension three Yukawa type and
dimension four gauge type operators. The gauge boson and fermion
mixings and masses are universal. For $W-W'$ mixing, there exists
two independent parameters just accounting two physical quantities,
mixing angle $\zeta$ and mass ratio $M_W/M_{W'}$. For neutrino
mixing, existence of three heavy neutrinos will violate unitarity of
rotation matrix among three light neutrinos. The left and right hand
quark mixings lead to left and right hand CKM matrices with right
hand CKM matrix parameterized in the same structure of left hand one
multiplying five extra phase angles. We express Goldstone, Higgs and
gauge couplings to quarks in gauge boson and fermion mass
eigenstates. We build up effective Hamiltonian for neutral $K$ and
$B$ systems and perform detail calculations for LR and LP models to
mass differences for $K^0-\bar{K}^0$, $B_d^0-\bar{B}_d^0$ and
$B_s^0-\bar{B}_s^0$ systems and indirect CP violation parameter
$\epsilon_K$ for $K$ mesons. We show that just $W$ itself with
anomalous coupling $\Delta_{1,1}$ near to 1 is already account for
experiment data. Except the case of mutual cancelations and very
heavy $W'$ up to order of $100M_W$, there are other three ways to
suppress $W'$ contributions: small right hand gauge coupling $g_2$,
 small anomalous coupling $\Delta_{2,1}$,
 or special combination of CKM matrix elements. The
smallness of higgs contribution leads  to some constraints on
$h\bar{q}q$ couplings.

%%%%%%%%%%%%%%%%%%%%%%%%%%%%%%%%%%%%%%%%%%%%%
\section*{Acknowledgment}
This work was  supported by National  Science Foundation of China
(NSFC) under Grant No. 10435040 and Specialized Research Fund for
the Doctoral Program of Higher Education.

%%%%%%%%%%%%%%%%%%%%%%%%%%%%%%%%%%%%%%%%%%%%%%%
\appendix
\section{Anomalous Gauge Couplings $\Delta$}

For quark,
\begin{eqnarray}
&&\hspace{-0.5cm}\Delta_{1,1,\alpha}=\left\{\begin{array}{ll}1-\delta_{L,1,\alpha}-\delta_{L,4,\alpha}&\mbox{LR,LP}\\
   1-\delta_{L,1,\alpha}-\delta_{L,4,\alpha}&\mbox{HP,FP}\\
   0&\mbox{UN}\\
   (1-\delta_{L,1,\alpha}-\delta_{L,4,\alpha})\delta_{\alpha\alpha_1}
   +0\delta_{\alpha\alpha_2}&\mbox{NU}\end{array}\right.\label{Delta11def}\\
&&\hspace{-0.5cm}\Delta_{1,2,\alpha}=\left\{\begin{array}{ll}\delta_{R,2,\alpha}-\delta_{R,6,\alpha}&\mbox{LR,LP}\\
   0&\mbox{HP,FP}\\
   1-\delta_{L,1,\alpha}-\delta_{L,4,\alpha}&\mbox{UN}\\
  0\delta_{\alpha\alpha_1}
   +(1-\delta_{L,1,\alpha}-\delta_{L,4,\alpha})\delta_{\alpha\alpha_2}&\mbox{NU}\end{array}\right.\nonumber\\
&&\hspace{-0.5cm} \Delta^3_{1,1,\alpha}=\left\{\begin{array}{ll}
   (1-\delta_{L,1,\alpha}+\delta_{L,4,\alpha})\frac{\tau^3}{2}+\delta_{L,3,\alpha}+\delta_{L,7,\alpha}&\mbox{LR,LP}\\
   (1-\delta_{L,1,\alpha}+\delta_{L,4,\alpha})\frac{\tau^3}{2}+\delta_{L,3,\alpha}+\delta_{L,7,\alpha}&\mbox{HP,FP}\\
   0&\mbox{UN}\\
   ((1-\delta_{L,1,\alpha}+\delta_{L,4,\alpha})\frac{\tau^3}{2}
   +\delta_{L,3,\alpha}+\delta_{L,7,\alpha})\delta_{\alpha\alpha_1}
   +0\delta_{\alpha\alpha_2}&\mbox{NU}\end{array}\right.\nonumber\\
&&\hspace{-0.5cm}\Delta^3_{1,2,\alpha}=\left\{\begin{array}{ll}
   (\delta_{R,2,\alpha}+\delta_{R,6,\alpha})\frac{\tau^3}{2}+\delta_{R,5,\alpha}&\mbox{LR,LP}\\
   0&\mbox{HP,FP}\\
   (1-\delta_{L,1,\alpha}+\delta_{L,4,\alpha})\frac{\tau^3}{2}+\delta_{L,3,\alpha}+\delta_{L,7,\alpha}&\mbox{UN}\\
   0\delta_{\alpha\alpha_1}
   +[(1-\delta_{L,1,\alpha}+\delta_{L,4,\alpha})\frac{\tau^3}{2}+\delta_{L,3,\alpha}+\delta_{L,7,\alpha}]\delta_{\alpha\alpha_2}&\mbox{NU}\end{array}\right.\nonumber\\
&&\hspace{-0.5cm} \Delta_{1,\alpha}=\left\{\begin{array}{ll}
   (\delta_{L,1,\alpha}\!-\delta_{R,2,\alpha}\!-\delta_{L,4,\alpha}\!-\delta_{R,6,\alpha}\!+\frac{2}{3}\delta_{L,
   7,\alpha})\frac{\tau_3}{2}+\frac{1}{6}-\delta_{L,3,\alpha}\!-\delta_{R,5,\alpha}&\mbox{LR,LP}\\
   (\delta_{L,1,\alpha}\!-\delta_{L,4,\alpha}\!+\frac{2}{3}\delta_{L,
   7,\alpha})\frac{\tau_3}{2}+\frac{1}{6}-\delta_{L,3,\alpha}\!&\mbox{HP,FP}\\
   (\delta_{L,1,\alpha}\!-\delta_{L,4,\alpha}\!+\frac{2}{3}\delta_{L,
   7,\alpha})\frac{\tau_3}{2}+\frac{1}{6}-\delta_{L,3,\alpha}&\mbox{UN}\\
   (\delta_{L,1,\alpha}\!-\delta_{L,4,\alpha}\!+\frac{2}{3}\delta_{L,
   7,\alpha})\frac{\tau_3}{2}+\frac{1}{6}-\delta_{L,3,\alpha}\!&\mbox{NU}\end{array}\right.\nonumber\\
&&\label{Delta}\\
&&\hspace{-0.5cm}\Delta_{2,1,\alpha}=\left\{\begin{array}{ll}1-\delta_{R,1,\alpha}-\delta_{R,4,\alpha}&\mbox{LR,LP}\\
   \delta_{L,2,\alpha}-\delta_{L,6,\alpha}&\mbox{HP,FP}\\
   0&\mbox{UN}\\
    (\delta_{L,2,\alpha}-\delta_{L,6,\alpha})\delta_{\alpha\alpha_1}
   +0\delta_{\alpha\alpha_2}&\mbox{NU}\end{array}\right.\nonumber\\
&&\hspace{-0.5cm}\Delta_{2,2,\alpha}=\left\{\begin{array}{ll}\delta_{L,2,\alpha}-\delta_{L,6,\alpha}&\mbox{LR,LP}\\
   0&\mbox{HP,FP}\\
   \delta_{L,2,\alpha}-\delta_{L,6,\alpha}&\mbox{UN}\\
   0\delta_{\alpha\alpha_1}
+(\delta_{L,2,\alpha}-\delta_{L,6,\alpha})\delta_{\alpha\alpha_2}&\mbox{NU}\end{array}\right.\nonumber\\
&&\hspace{-0.5cm} \Delta^3_{2,1,\alpha}=\left\{\begin{array}{ll}
   (1-\delta_{R,1,\alpha}+\delta_{R,4,\alpha})\frac{\tau^3}{2}+\delta_{R,3,\alpha}+\delta_{R,7,\alpha}&\mbox{LR,LP}\\
   (\delta_{L,2,\alpha}-\delta_{L,6,\alpha})\frac{\tau^3}{2}+\delta_{L,5,\alpha}+\delta_{L,6,\alpha}&\mbox{HP,FP}\\
   0&\mbox{UN}\\
   ((\delta_{L,2,\alpha}-\delta_{L,6,\alpha})\frac{\tau^3}{2}+\delta_{L,5,\alpha}+\delta_{L,6,\alpha})\delta_{\alpha\alpha_1}
   +0\delta_{\alpha\alpha_2}&\mbox{NU}\end{array}\right.\nonumber\\
&&\hspace{-0.5cm}\Delta^3_{2,2,\alpha}=\left\{\begin{array}{ll}
   (\delta_{L,2,\alpha}+\delta_{L,6,\alpha})\frac{\tau^3}{2}+\delta_{L,5,\alpha}&\mbox{LR,LP}\\
   0&\mbox{HP,FP}\\
   (\delta_{L,2,\alpha}-\delta_{L,6,\alpha})\frac{\tau^3}{2}+\delta_{L,5,\alpha}+\delta_{L,6,\alpha}&\mbox{UN}\\
   0\delta_{\alpha\alpha_1}
   +((\delta_{L,2,\alpha}-\delta_{L,6,\alpha})\frac{\tau^3}{2}+\delta_{L,5,\alpha}+\delta_{L,6,\alpha})\delta_{\alpha\alpha_2}&\mbox{NU}\end{array}\right.\nonumber\\
&&\hspace{-0.5cm} \Delta_{2,\alpha}=\left\{\begin{array}{ll}
   (\delta_{R,1,\alpha}\!-\delta_{L,2,\alpha}\!-\delta_{R,4,\alpha}\!-\delta_{L,6,\alpha}\!+\frac{2}{3}\delta_{R,
   7,\alpha})\frac{\tau_3}{2}+\frac{1}{6}-\delta_{R,3,\alpha}\!-\delta_{L,5,\alpha}&\mbox{LR,LP}\\
   (1-\delta_{L,2,\alpha}-\delta_{L,6,\alpha}+\frac{2}{3}\delta_{R,
   7,\alpha})\frac{\tau_3}{2}+\frac{1}{6}-\delta_{L,5,\alpha}+\delta_{R,
   7,\alpha}   &\mbox{HP,FP}\\
   (1-\delta_{L,2,\alpha}-\delta_{L,6,\alpha}+\frac{2}{3}\delta_{R,
   7,\alpha})\frac{\tau_3}{2}+\frac{1}{6}-\delta_{L,5,\alpha}+\delta_{R,
   7,\alpha}  &\mbox{UN}\\
   (1-\delta_{L,2,\alpha}-\delta_{L,6,\alpha}+\frac{2}{3}\delta_{R,
   7,\alpha})\frac{\tau_3}{2}+\frac{1}{6}-\delta_{L,5,\alpha}+\delta_{R,
   7,\alpha}&\mbox{NU}\end{array}\right.\nonumber
\end{eqnarray}
For lepton:
\begin{eqnarray}
&&\hspace{-0.5cm}\Delta_{1,1,\alpha}^l=\left\{\begin{array}{ll}
   1-\delta^l_{L,1,\alpha}-\delta^l_{L,4,\alpha}&\mbox{LR,HP}\\
   1-\delta^l_{L,1,\alpha}-\delta^l_{L,4,\alpha}&\mbox{LP,FP,UN}\\
   (1-\delta^l_{L,1,\alpha}-\delta^l_{L,4,\alpha})\delta_{\alpha\alpha_1}+0\delta_{\alpha\alpha_2}&\mbox{NU}\end{array}\right.\nonumber\\
&&\hspace{-0.5cm}\Delta_{1,2,\alpha}^l=
   \left\{\begin{array}{ll}\delta^l_{R,2,\alpha}-\delta^l_{R,6,\alpha}&\mbox{LR,HP}\\
   0&\mbox{LP,FP,UN}\\
   0\delta_{\alpha\alpha_1}+(1-\delta^l_{L,1,\alpha}-\delta^l_{L,4,\alpha})\delta_{\alpha\alpha_2}&\mbox{NU}\end{array}\right.\nonumber\\
&&\hspace{-0.5cm} \Delta^{l3}_{1,1,\alpha}=\left\{\begin{array}{ll}
   (1-\delta^l_{L,1,\alpha}+\delta^l_{L,4,\alpha})\frac{\tau^3}{2}+\delta^l_{L,3,\alpha}
   +\delta^l_{L,7,\alpha}&\mbox{LR,HP}\\
   (1-\delta^l_{L,1,\alpha}+\delta^l_{L,4,\alpha})\frac{\tau^3}{2}
     +\delta^l_{L,3,\alpha}+\delta^l_{L,7,\alpha}&\mbox{LP,FP,UN}\\
   ((1-\delta^l_{L,1,\alpha}+\delta^l_{L,4,\alpha})\frac{\tau^3}{2}
     +\delta^l_{L,3,\alpha}+\delta^l_{L,7,\alpha})\delta_{\alpha\alpha_1}+0\delta_{\alpha\alpha_2}&\mbox{NU}\end{array}\right.\nonumber\\
&&\hspace{-0.5cm}\Delta^{l3}_{1,2,\alpha}=\left\{\begin{array}{ll}
   (\delta^l_{R,2,\alpha}+\delta^l_{R,6,\alpha})\frac{\tau^3}{2}+\delta^l_{R,5,\alpha}&\mbox{LR,HP}\\
   0&\mbox{LP,FP,UN}\\
   0\delta_{\alpha\alpha_1}+((1-\delta^l_{L,1,\alpha}+\delta^l_{L,4,\alpha})\frac{\tau^3}{2}
     +\delta^l_{L,3,\alpha}+\delta^l_{L,7,\alpha})\delta_{\alpha\alpha_2}&\mbox{NU}\end{array}\right.\nonumber\\
&&\hspace{-0.5cm} \Delta^l_{1,\alpha}=\left\{\begin{array}{ll}
   (\delta^l_{L,1,\alpha}\!-\delta^l_{R,2,\alpha}\!-\delta^l_{L,4,\alpha}\!-\delta^l_{R,6,\alpha}\!-2\delta^l_{L,
   7,\alpha})\frac{\tau_3}{2}-\frac{1}{2}-\delta^l_{L,3,\alpha}\!-\delta^l_{R,5,\alpha}&\mbox{LR,HP}\\
   (\delta^l_{L,1,\alpha}\!-\delta^l_{L,4,\alpha}\!-2\delta^l_{L,
   7,\alpha})\frac{\tau_3}{2}-\frac{1}{2}-\delta^l_{L,3,\alpha}\!&\mbox{LP,FP,UN}\\
   (\delta^l_{L,1,\alpha}\!-\delta^l_{L,4,\alpha}\!-2\delta^l_{L,
   7,\alpha})\frac{\tau_3}{2}-\frac{1}{2}-\delta^l_{L,3,\alpha}\!&\mbox{NU}\end{array}\right.\nonumber\\
&&\label{Delta}\\
&&\hspace{-0.5cm}\Delta_{2,1,\alpha}^l=\left\{\begin{array}{ll}
   1-\delta^l_{R,1,\alpha}-\delta^l_{R,4,\alpha}&\mbox{LR,HP}\\
   \delta^l_{L,2,\alpha}-\delta^l_{L,6,\alpha}&\mbox{LP,FP,UN}\\
   ( \delta^l_{L,2,\alpha}-\delta^l_{L,6,\alpha})\delta_{\alpha\alpha_1}+0\delta_{\alpha\alpha_2}&\mbox{NU}\end{array}\right.\nonumber\\
&&\hspace{-0.5cm}\Delta_{2,2,\alpha}^l=\left\{\begin{array}{ll}
   \delta^l_{L,2,\alpha}-\delta^l_{L,6,\alpha}&\mbox{LR,HP}\\
   0&\mbox{LP,FP,UN}\\
   0\delta_{\alpha\alpha_1}+( \delta^l_{L,2,\alpha}-\delta^l_{L,6,\alpha})\delta_{\alpha\alpha_2}&\mbox{NU}\end{array}\right.\nonumber\\
&&\hspace{-0.5cm} \Delta^{l3}_{2,1,\alpha}=\left\{\begin{array}{ll}
   (1-\delta^l_{R,1,\alpha}+\delta^l_{R,4,\alpha})\frac{\tau^3}{2}+\delta^l_{R,3,\alpha}
   +\delta^l_{R,7,\alpha}&\mbox{LR,HP}\\
   (\delta^l_{L,2,\alpha}-\delta^l_{L,6,\alpha})\frac{\tau^3}{2}
     +\delta^l_{L,5,\alpha}+\delta^l_{L,6,\alpha}&\mbox{LP,FP,UN}\\
   ((\delta^l_{L,2,\alpha}-\delta^l_{L,6,\alpha})\frac{\tau^3}{2}
     +\delta^l_{L,5,\alpha}+\delta^l_{L,6,\alpha})\delta_{\alpha\alpha_1}+0\delta_{\alpha\alpha_2}&\mbox{NU}\end{array}\right.\nonumber\\
&&\hspace{-0.5cm}\Delta^{l3}_{2,2,\alpha}=\left\{\begin{array}{ll}
   (\delta^l_{L,2,\alpha}+\delta^l_{L,6,\alpha})\frac{\tau^3}{2}+\delta^l_{L,5,\alpha}&\mbox{LR,HP}\\
   0&\mbox{LP,FP,UN}\\
   0\delta_{\alpha\alpha_1}+((\delta^l_{L,2,\alpha}-\delta^l_{L,6,\alpha})\frac{\tau^3}{2}
     +\delta^l_{L,5,\alpha}+\delta^l_{L,6,\alpha})\delta_{\alpha\alpha_2}&\mbox{NU}\end{array}\right.\nonumber\\
&&\hspace{-0.5cm} \Delta^l_{2,\alpha}=\left\{\begin{array}{ll}
   (\delta^l_{R,1,\alpha}\!-\delta^l_{L,2,\alpha}\!-\delta^l_{R,4,\alpha}\!-\delta^l_{L,6,\alpha}\!-2\delta^l_{R,
   7,\alpha})\frac{\tau_3}{2}-\frac{1}{2}-\delta^l_{R,3,\alpha}\!-\delta^l_{L,5,\alpha}&\mbox{LR,HP}\\
   (1-\delta^l_{L,2,\alpha}-\delta^l_{L,6,\alpha}-2\delta^l_{R,
   7,\alpha})\frac{\tau_3}{2}-\frac{1}{2}-\delta^l_{L,5,\alpha}+\delta^l_{R,
   7,\alpha} &\mbox{LP,FP,UN}\\
   (1-\delta^l_{L,2,\alpha}-\delta^l_{L,6,\alpha}-2\delta^l_{R,
   7,\alpha})\frac{\tau_3}{2}-\frac{1}{2}-\delta^l_{L,5,\alpha}+\delta^l_{R,
   7,\alpha} &\mbox{NU}\end{array}\right.\nonumber
\end{eqnarray}

%%%%%%%%%%%%%%%%%%%%%%%%%%%%%%%%%%%%%%%%%%%%%%%


\begin{thebibliography}{1}

\bibitem{LRSM1}
J.C.Pati, A.Salam, Phys. Rev. {\bf D10}, 275(1974).

\bibitem{LRSM2}
R.N.Mohapatra, J.C.Pati, Phys. Rev. {\bf D11}, 566(1975);\\
R.N.Mohapatra, J.C.Pati, Phys. Rev. {\bf D11}, 2558(1975);\\
G.Senjanovic, R.N.Mohapatra, Phys. Rev. {\bf D12}, 1502(1975).

\bibitem{AlternateLRSM}
K.S.Babu, X.G.He, E.Ma, Phys. Rev. {\bf D36}, 878(1987).

\bibitem{UnunifiedSM}
H.Geogi, E.E.Jenkins, E.H.Simmons, Phys. Rev. Lett. {\bf
62},2789(1989); {\bf 63},1540(E)(1989); Nucl. Phys. {\bf B331},
541(1990)\\
E.Ma, S.Rajpoot, Mod. Phys. Lett. {\bf A5},979(1990);\\
V.Barger, T.Rizzo, Phys. Rev. {\bf D41}, 946(1990);\\
L.Randall, Phys. Lett. {B234}, 508(1990);\\
T.G.Rizzo, Int. J. Mod. Phys. {\bf A7}, 91(1992);\\
R.S.Chivukula, E.H.Simmons, J.Terning, Phys. Lett. {B346},
284(1995).

\bibitem{Non-CommutingExtendedTechnicolor}R.S.Chivukula,
E.H.Simmons, J.Terning, Phys. Rev. {\bf D53},5258(1996).

\bibitem{LH1}
N.Arkani-Hamed, A.G.Cohen, H.Geogi, Phys. Lett. {\bf B513},
232(2001).

\bibitem{LH2}
D.E.Kaplan, M.Schmaltz, JHEP {\bf 0310}(2003)039.

\bibitem{LH3}
M.Schmaltz, D.Tucker-Smith, Annu. Rev. Nucl. Part. Sci. {\bf
55},229(2005).

\bibitem{Composite4}
U.Baur, et al., Phys. Rev. {\bf D35}, 297(1987);\\
M.Kuroda, et al., Nucl. Phys. {\bf B261}, 432(1985).

\bibitem{SSTopflavor}
P.Batra, A.Delgado, D.E.Kaplan, T.M.P.Tait, JHEP {\bf
0402}(2004)043.

\bibitem{GUT}
R.W.Robinett, Phys. Rev. {\bf D26}, 2388(1982);\\
R.W.Robinett, J.L.Rosner, Phys. Rev. {\bf D26}, 2396(1982);\\
P.Langacker, R.W.Robinett, J.L.Rpsner, Phys. Rev. {\bf D30},
1470(1984).

\bibitem{Sstring1}
M.Cvetic, P.Langacker, Mod. Phys. Lett. {\bf A11}, 1247(1996).

\bibitem{Sstring2}
J.C.Pati, hep-ph/0606089.

\bibitem{Sstring3}
M.Green, J.Schwartz, Phys. Lett. {\bf B149}, 117(1984);\\
D.Gross, et al., Phys. Rev. Lett. {\bf 54}, 502(1985);\\
E.Witten, Phys. Lett. {\bf B155}, 1551(1985);\\
P.Candelas, et al., Nucl. Phys. {\bf B258}, 46(1985);\\
M.Dine, et al., Nucl. Phys. {\bf B259}, 549(1985).

\bibitem{ED1}
A.Datta, P.J.O'Donnell, Z.H.Lin, X.Zhang, T.Huang, Phys. Lett. {\bf
B483}, 203(2000).

\bibitem{ED2}
H.C.Cheng, et al., Phys. Rev. {\bf D64}, 065007(2001).

\bibitem{Unitarity0}
J.M.Cornwall, D.N.Levin, G.Tiktopoulos, Phys. Rev. Lett. {\bf 30},
1268(1973); {\bf 31}, 572(E)(1973); Phys. Rev. {\bf D10},
1145(1974); {\bf D11}, 972(E)(1975).

\bibitem{TCrho}
M.Bando, T.Kugo, K.Yamawaki, Phys. Rept. {\bf 164}, 217(1988).

\bibitem{ED3}
R.Sundrum, hep-th/0508134;\\
C.Csaki, J.Hubisz, P.Meade, hep-ph/0510275;\\
G.Kribs, hep-ph/0605325.

\bibitem{Unitarity}
D.A. Dicus and V.S. Mathur. Phys. Rev. \textbf{D7}, 3111(1973);~
B.W. Lee, C. Quigg and H.B. Thacker. Phys. Rev. \textbf{D16},
1519(1977), Phys. Rev. Lett. \textbf{38}, 883(1977);~ M.J.G.
Veltman. Acta Phys. Polon. \textbf{B8}, 475(1977).

\bibitem{WangLiMing}
L.-M.Wang, Q.Wang, hep-ph/0605104;\\
S.-Z.Wang, Q.Wang, to be published in Chin. Phys. Lett.

\bibitem{longhitano80}
A.Longhitano, Phys. Rev. {\bf D22},  1166 (1980).

\bibitem{Appelquist80}
T.Appelquist, C.Bernard Phys. Rev. {\bf D22}, 200(1980).

\bibitem{Appelquist93}
T.Appelquist, G.-H.Wu, Phys. Rev. {\bf D48}, 3235(1993).

\bibitem{Z'}
Y.Zhang, S.-Z.Wang, Q.Wang, JHEP {\bf 03}, (2008)047

\bibitem{LRSMEWCLboson}
Y.Zhang, S.-Z.Wang, F.-J.Ge and Q.Wang, Phys. Lett. {\bf B653},
259(2007).

\bibitem{LRSMEWCLmatter}
S.-Z.Wang, S.-Z.Jiang and Q.Wang, Phys. Lett. {\bf B662}, 375(2008).

\bibitem{FP0}
A.Donini, F.Feruglio, J.Matias, F.Zwirner, Nucl. Phys. {\bf B507},
51(1997).

\bibitem{LH4}
G.Burdman, M.Perelstein, A.Pierce, Phys. Rev. Lett. {\bf 90},
241802(2003), Erratum-ibid.{\bf 92}, 049903(2004).

\bibitem{FP}
V.Barger, W.Y.Keung, E.Ma, Phys. Rev. {\bf D22}727(1980); Phys.
Lett. {\bf B94},377;\\
V.Barger, E.Ma, K.Whisnant, Phys. Rev. Lett. {\bf 46},1501(1981);\\
J.L.Kneur, D.Schildknecht, Nucl. Phys. {\bf B357}, 357(1991).

\bibitem{NU}
X.-Y.Li, E.Ma, J. Phys. {\bf G19},1265(1993)\\
D.J.Muller and S.Nandi, Phys.Lett. {\bf B383},345(1996)\\
E.Malkawi, T.Tait, C.P.Yuan, Phys. Lett. {\bf B385}, 304(1996).

\bibitem{MNS}
Z.Maki, M.Nakagawa, S.Sakata, Prog. Theor. Phys. {\bf 28},
870(1962).

\bibitem{XingPLB08}
Z.Z.Xing, Phys. Lett. {\bf B660},515(2008).

\bibitem{PDG06}
W.-M.Yao, et al., Particla Data Group, J.Phys. {\bf G33}, 1(2006)

\bibitem{Anatomy} P. Ball, J.-M.Fr\'{e}re, J. Matas, Nucl. Phys. {\bf
B572}, 3(2000).

\bibitem{YLWu07}
Yue-Liang Wu, Yu-Feng Zhou, hep-ph/0709.0042.

\bibitem{eta}
S. Herrlich U. Nierste, Nucl. Phys.  {\bf B476}, 27(1996).

\bibitem{eta1}
A. J. Buras ,1997 ICHEP96, 28th Int. Conf. on High Energy Physics
(Warsaw, Poland) ad Z Ajduk and A K Wroblewski (Singapore: World
Scientific)p243.

\bibitem{etaB}
A. J. Buras ,M. Jamin and P. H. Weisz,
%Leading and next-to-leading order QCD corrections to epsilon paraters and $B^0-\bar{B}^0$
%mixing in the presence of a heavy top quark,
 Nucl. Phys.  {\bf
B347}, 491(1990).

\bibitem{etaQCD}
J. M. Frere, J. Galand, A. Le Yaouanc, L. Oliver, O. Pene,and J. C.
Raynal , Phys. Rev. {\bf D46}, 337(1992).

\bibitem{etaBQCD}
G. Ecker and W. Grimus, Nucl. Phys.  {\bf B258}, 27(1985).

\bibitem{fK1}
C. Aubin et al.,[MILC Collab.],  Phys.  Rev. {\bf D70},
114501(2004)[hep-lat/0407028].

\bibitem{fK2}
C. Bernard et al.,[MILC Collab.],  PoS LAT
2005,025(2005)[hep-lat/0509137].


\bibitem{latQCDBK}
C. Dawson,  PoS LAT 2005,007(2005).

%\bibitem{bagBd1}
%A. Gray et al.,[HPQCD Collab.],  Phys.  Rev. Lett. {\bf 95},
%212001(2005)[hep-lat/0507015].

\bibitem{bagBd2}
S. Aoki et al.,[JLQCD Collab.],  Phys.  Rev. Lett. {\bf 91},
212001(2003)[hep-lat/0307039].

\bibitem{SMcontribution}
G.Buchalla, A.J.Buras, M.E.Lautenbacher, Rev. Mod. Phys., {\bf 68},
1125(1996).
\end{thebibliography}
\end{document}